\documentclass[useAMS,usenatbib]{mn2e}
\usepackage{epsfig}
\usepackage{subfig}
\usepackage{times}
\usepackage{color}

\newcommand{\teff}{$T_{\rm eff}$} 
\newcommand{\kms}{km s$^{-1}$}
\newcommand{\logg}{$\log g$} 
\newcommand\rgc{$R_{\rm G}$}

\title[ARGOS bulge survey]{ARGOS III: Stellar Populations in the Galactic Bulge of the Milky Way}
\author[M. Ness et al.]
  {M.~Ness$^1$\thanks{E--mail:mkness@mso.anu.edu.au.},
  K.~Freeman$^1$,  E.~Athanassoula$^2$, E.~Wylie--de--Boer$^1$, 
  J.~Bland--Hawthorn$^3$, 
  \newauthor 
 M.~Asplund$^1$,
 G.F.~Lewis$^3$,
  D.~Yong$^1$,
 R.R. ~Lane$^4$,
  and L.L.~Kiss$^{3,5,6}$, \\
  $^1$Research School of Astronomy \& Astrophysics, Australian National University, Cotter Rd., Weston, ACT 2611, Australia\\
 $^2$ Aix Marseille Universit\'e, CNRS, LAM (Laboratoire d'Astrophysique de
Marseille) UMR 7326, 13388, Marseille, France.   \\
   $^3$Sydney Institute for Astronomy, University of Sydney, School of Physics A28, NSW 2006, Australia\\
  $^4$Departamento de Astronom\'{i}a Universidad de Concepci\'{o}n, Casilla 160 C, Concepci\'{o}n, Chile\\ 
    $^5$Konkoly Observatory, MTA Research Centre for Astronomy and Earth Sciences, Budapest, Hungary\\
$^6$ELTE Gothard-Lend\"ulet Research Group, H-9700, Szombathely, Hungary.\\
 } 
\date{Released 2002 Xxxxx XX}

\begin{document}

\date{Accepted 2012 TBD. Received 2012. As soon as possible; in original form 2012 February}

\pagerange{\pageref{firstpage}----\pageref{lastpage}} \pubyear{2012}

\maketitle

\label{firstpage}

\begin{abstract}

We present the metallicity results from the ARGOS spectroscopic survey of the Galactic bulge. Our aim is to understand the formation of the Galactic bulge: did it form via mergers, as expected from $\Lambda$CDM theory, or from disk instabilities, as suggested by its boxy/peanut shape, or both? Our stars are mostly red clump giants, which have a well defined absolute magnitude from which distances can be determined. We have obtained spectra for 28,000 stars at a spectral resolution of R = 11,000. From these spectra, we have determined stellar parameters and distances to an accuracy of $<1.5$ kpc.  The stars in the inner Galaxy span a large range in  [Fe/H],  --2.8 $\le$ [Fe/H] $\le$ +0.6. From the spatial distribution of the red clump stars as a function of [Fe/H] (Ness et al. 2012a), we propose that the stars with [Fe/H] $> -0.5$ are part of the boxy/peanut bar/bulge. We associate the lower metallicity stars ([Fe/H] $< -0.5$) with the thick disk, which may be puffed up in the inner region, and with the inner regions of the metal-weak thick disk and inner halo. For the bulge stars with [Fe/H] $> -0.5$, we find two discrete populations;  (i) stars with [Fe/H] $\approx -0.25$ which provide a roughly constant fraction of the stars in the latitude interval $b = -5^\circ$ to $-10^\circ$, and (ii) a kinematically colder, more metal-rich population with mean [Fe/H] $\approx +0.15$ which is more prominent closer to the plane. The changing ratio of these components with latitude appears as a vertical abundance gradient of the bulge. We attribute both of these bulge components to instability-driven bar/bulge formation from the thin disk. We associate the thicker component with the stars of the early less metal-rich thin disk, and associate the more metal-rich population concentrated to the plane with the colder more metal-rich stars of the early thin disk, similar to the colder and younger more metal-rich stars seen in the thin disk in the solar neighborhood today. We do not exclude a weak underlying classical merger--generated bulge component, but see no obvious kinematic association of any of our bulge stars with such a classical bulge component.  The clear spatial and kinematic separation of the two bulge populations (i) and (ii) makes it unlikely that any significant merger event could have affected the inner regions of the Galaxy since the time when the bulge-forming instabilities occurred.

\end{abstract}


\section{Introduction}
 
Recent observations show that the stars in the Galactic bulge region comprise several kinematic and chemical components spanning a range in $\alpha- $enhancement \citep{Babusiaux2010, Bensby2010}. Although the bulge region can no longer be described as a single stellar population, it appears to be a mostly old, $\alpha$-enhanced system whose stars formed over a short period of time early in the life of the Galaxy \citep{Zoccali2003, Melendez2008, McWilliam2010, Lecureur2007, AlvesBrito2010}. 

Near--infrared imaging (\citet{Okuda1977}, COBE/DIRBE \citep{Dwek1995, Smith2004} revealed the boxy/peanut morphology of the Milky Way bulge. \citet{Blitz1991} established the bar-like structure of the inner bulge region that is now understood to go with the boxy/peanut bulge structure.  The major axis of this bar points into the first Galactic quadrant, at about 20 degrees to the Sun--center line.  Its semi--length is about $3.1 - 3.5$ kpc \citep{Gerhard2002} and its axial ratios are about 1: 0.33: 0.23  \citep{Dwek1995}.  Near--IR star-counts \citep[e.g.][]{Lopez2005} provide evidence for a longer flatter component of the bar, but the nature of this long bar is uncertain. \citet{Inma2011} and \citet{Romero2011} argue that the apparent long bar is an artefact associated with leading spiral features at the end of the shorter primary bar. 

In Cold Dark Matter ($\Lambda$CDM) simulations of Galaxy formation \citep[e.g.][]{Abadi2003, Kobayashi2011}, bulges are built up primarily through hierarchical mergers. However, it is now widely believed from simulations and observations that the boxy/peanut shaped bulges, like the bulge of the Milky Way, are not merger products but formed via instability of the inner disk: see e.g. \citet{Combes1981, Raha1991, Bureau1999}. The early simulations showed that, after a few revolutions, the flat disk develops a bar which then becomes vertically unstable and puffs up into a boxy/peanut-bulge structure. More recent N--body simulations of bulge formation via  disk instability \citep{Athanassoula2008} help to understand the timescales and the spatial, kinematic and chemical abundance tracers of this instability process.  For example, the simulations can be used to interpret the mapping of the early disk into the boxy/peanut-bulge via the instabilities.

The simulations of bar/bulges produced by disk instability also show that cylindrical rotation is expected, and these are supported by observations of edge--on boxy/peanut bulges \citep[e.g.][]{FalconBarroso2006}.  If the Galactic boxy/peanut bulge is a disk instability product, then we would expect it to show cylindrical rotation. From their survey of M giants in the inner Galaxy and based  on such kinematical criteria, \citet{Howard2009} constrain any merger-generated component of the bulge to be less than 8\%.  Several authors continue to argue in favour of  the importance of dissipational collapse or mergers in the formation of the Milky Way bulge \citep{Zoccali2008, Babusiaux2010}, based on their interpretation of new chemical and kinematic data.

We have completed a large spectroscopic survey of about 28,000 stars in the bulge and surrounding disk in order to measure kinematics and chemistry and so to constrain the formation process of the bulge. We aim to   establish whether it is primarily a merger or primarily an instability product.  We used models of classical bulges to estimate that 28,000 stars is sufficient to detect a 5\% merger--generated bulge underlying an instability--generated bar/bulge. We note that this estimate is contingent upon the classical bulge being a slowly rotating and dispersion-supported component. We would not be able to identify a classical bulge component which has been spun up into cylindrical rotation as in the simulations of \citet{Saha2012}.  

This is the third paper in the ARGOS series and concentrates primarily on the metallicity distribution function (MDF) for the ARGOS stars in the inner Galaxy.  Paper I \citep{Ness2012a} examined the nature of the bimodal distribution of apparent magnitudes of the red clump stars along the minor axis of the bulge. This split clump appears to be a generic feature of boxy/peanut bulges and provides a useful guide to the interpretation of the metallicity components in the bulge.  Paper II (Freeman et al., 2012) gave a description of the survey, our choice of fields, and the details of our observations, sample selection and analysis, including estimation of stellar parameters and distances.  Paper IV \citep[][in preparation]{Ness2012c} will describe the kinematical properties of the bulge as evaluated from the ARGOS stars.  In Paper V, we will further interpret the origin of the MDF components by using our dynamical models to discuss how the instability of the disk generates a phase-space-dependent mapping of the disk into the bulge.

In Section 2 of this paper, we examine the spatial and metallicity distributions of stars in our sample, identify the components in the metallicity distribution in the inner Galaxy, and discuss the abundance gradients of these components in latitude and galactocentric cylindrical radius \rgc. In Section 3 we present the components that comprise our MDF and discuss their properties in Section 4. We present our results for the $\alpha-$enhancement of the bulge as a function of latitude in Section 5, and show the relationship between $\alpha-$enhancement and [Fe/H] as a function of \rgc. In Sections 6, 7 and 8 we discuss the interpretation and origin of the MDF components that we have identified and compare our results to other studies. Section 9 gives our conclusions. 

\begin{figure}
\centering
\includegraphics[scale=0.2]{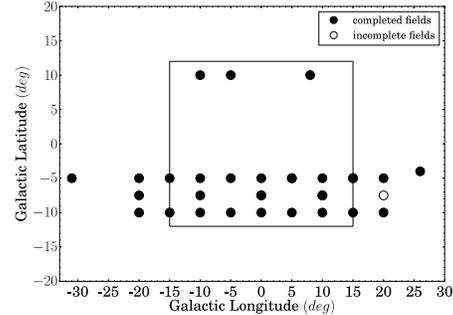}
\caption{The 28 x 2$^\circ$ fields in our survey, in Galactic latitude and longitude. Filled circles indicate fields for which we have complete data. One field at $(l,b) = (20^\circ, -7.5^\circ)$ is incomplete; only 600 stars were observed for this field. The rectangle denotes the approximate extent of the boxy/peanut-bulge}
\label{fig:donefields}
\end{figure}

\section{Results}

\subsection{Overview}

\begin{figure}
        \includegraphics[scale=0.3]{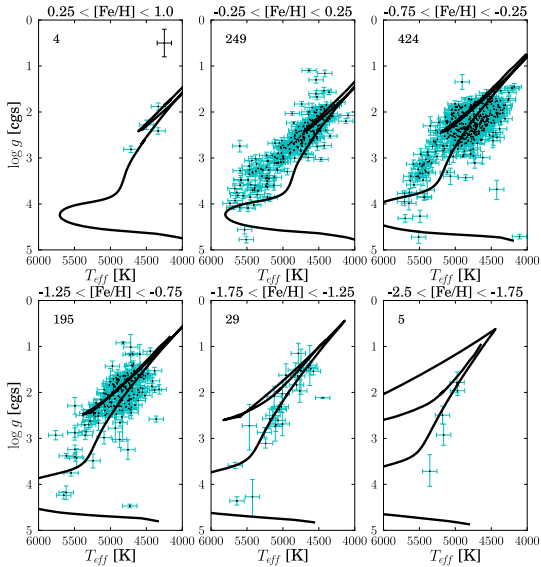} \\ 
\caption{(\teff, \logg) values for stars in the field $(l,b) = (0^\circ,-10^\circ)$, overlaid with 10 Gyr isochrones \citep{Cassisi2006}.The number of stars is given in the left hand corner of each panel. The $\chi^2$ fitting errors for \logg\ are shown for each star and an error bar representing the calibration errors in \logg\ along the $y$-axis and in temperature along the $x$-axis is shown in the right hand corner of the upper left panel.}
\label{fig:stellarparams}
\end{figure}

\begin{figure}
      \includegraphics[scale=0.3]{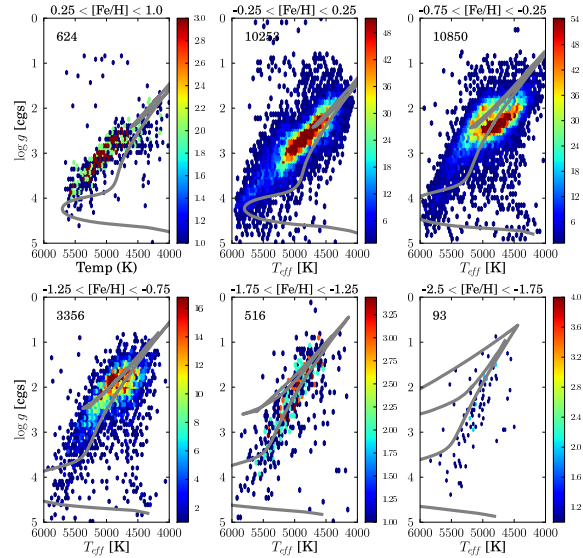} \\
\caption{Density images in (\teff, \logg) space of 25,500 stars in our survey overlaid with 10 Gyr isochrones \citep{Cassisi2006}. The number of stars is given in the left hand corner of each panel. The colorbars show the number of stars in each bin.}
\label{fig:stellarparams3}
\end{figure}

For reference, Figure \ref{fig:donefields} shows the location of the 28 two--degree fields of the ARGOS survey. Each field has about 1000 stars, mostly red clump giants, as discussed in detail in Freeman et al. (2012). 

As part of the survey, we have obtained radial velocities for each star with an error $<  0.9$ \kms\ and determined the stellar parameters \teff, \logg, [Fe/H] and [$\alpha$/Fe] for each of our stars. The [Fe/H] values are determined from the Fe lines in the Ca-triplet region of the spectrum, using a masked $\chi^2$ technique against a grid of model spectra. We have excluded stars with poor quality or low signal-to-noise (S/N) spectra, giving a final sample of about 25,500 stars. The resolution of the spectra is about 11,000 and the typical S/N for our stars is 75 per resolution element. From calibration against stars in well-observed globular and open clusters, the typical accuracies of the [Fe/H] and \logg\ values are about $0.09$ and $0.30$ dex respectively. Figure \ref{fig:stellarparams} shows an example of the (\teff, \logg) distribution of stars for our minor axis fields at $(l,b) = (0^\circ,-10^\circ)$ with BaSTI isochrones \citep{Cassisi2006} of age $10$ Gyr superimposed. The $6$ panels show the distributions and isochrones for $6$ intervals of [Fe/H] from $-2.5$ to $+1.0$. The isochrones have the mean abundance for the panel range in [Fe/H], and [$\alpha$/Fe] $= 0.4$ for the isochrones at [Fe/H] $<$ --0.25. The scatter of the stars about the isochrones is as expected. For some stars, the stellar parameters returned by our $\chi^2$ process were too close to the edge of the model grid ( i.e. \logg\ $< 0.5$) and these stars were excluded from the analysis. 

In order not to exclude metal-poor giants from the survey, we selected stars with $(J-K)_\circ > 0.38$ and therefore expect to find some nearby dwarfs with spectral types later than about G2, as seen in Figure \ref{fig:stellarparams}. These dwarfs are readily identified from their position relative to the isochrones in the (\teff,\logg) plane. We see that most of the stars are clustered around the red clump (\logg $= 2$ to $3$ and \teff $= 4500$ to $5300$ K), as we would expect \citep{Zhao2001,Valentini2010, Saguner2011}. \citet{Zhao2001} studied Hipparcos clump giants at high resolution and found two populations; a metal rich clump for which the gravity is higher ([Fe/H] $> 0$, \logg $= 2.5$ to $3.1$) plus a few more metal-poor stars with lower gravity ([Fe/H] $< -0.3$, \logg\ $ = 1.9$ to $2.4$). We also find that the gravity of the main clump of stars is shifted towards lower values for lower [Fe/H], and we expect that the clump is not contributing significantly to the stellar population for [Fe/H] $< -1.0$. 

The clustering around the red clump region is seen across all of the fields in our survey. Figure \ref{fig:stellarparams3} shows the density distribution in the (\teff, \logg) plane for the 25,500 stars in our survey, plotted in bins of metallicity against the isochrones. While the more metal-poor stars lie close to the isochrones, the more metal-rich stars deviate progressively more and more from the isochrones. The deviations are larger for the giants that lie above the red clump, and are most marked for stars in the most metal-rich bin, [Fe/H] $> 0.25$.  We need to discuss these observed deviations from the isochrones, because we will be deriving isochrone distances for all of the survey stars. The apparent deviations of our stars to the left may be due to several potential sources of error. Before discussing these error sources, we briefly summarise the relevant details of our procedure for deriving stellar parameters \citep[see][for further discussion]{Freeman2012}.

While our $\chi^2$ technique is in principle able to estimate \teff, \logg\ and [Fe/H] from the AAOmega spectra in the Ca-triplet region, we find in practice that there are too many degrees of freedom at our resolution, which leads to larger errors in the stellar parameters. It is better to estimate \teff\ independently. This is consistent with the findings of \citet{Kirby2008} at their slightly lower resolution of $8000$. After correction for the reddening \citep*{Schlegel}, stellar temperatures are derived from $(J-K)_{\circ}$ using a metallicity-dependent calibration due to \citet{Bessell1998}\footnote{http://wwwuser.oat.ts.astro.it/castelli/colors/bcp.html}. The procedure is iterative: starting with an initial estimate of [Fe/H] to get an initial estimate of temperature, new estimates of gravity and [Fe/H] are then determined from the $\chi^2$ technique against a grid of models. Typically four cycles are needed for convergence.  The mismatch between isochrones and stellar parameters seen in Figure \ref{fig:stellarparams3} for the more metal-rich stars could possibly be due to problems with the stellar models, the adopted reddening, or our temperature calibration.  It seems unlikely that our abundance scale or our \logg\ scale are the problem: they have both been carefully checked from observations of stars in previously well-observed star clusters covering a wide range of [Fe/H] and \logg.  

The BaSTI isochrones appear to work well for high metallicity giants \citep{Cassisi2010}.  Although we note that there is some evidence for a spread in age for the metal-rich dwarfs in the bulge \citep{Bensby2010}, which could lead to some displacement towards higher temperatures for the corresponding metal-rich giants,  we will assume that the 10 Gyr BaSTI isochrones are appropriate for our bulge stars.

\begin{figure}
\centering
\subfloat[]{\label{fig:onesmall}\includegraphics[scale=0.2]{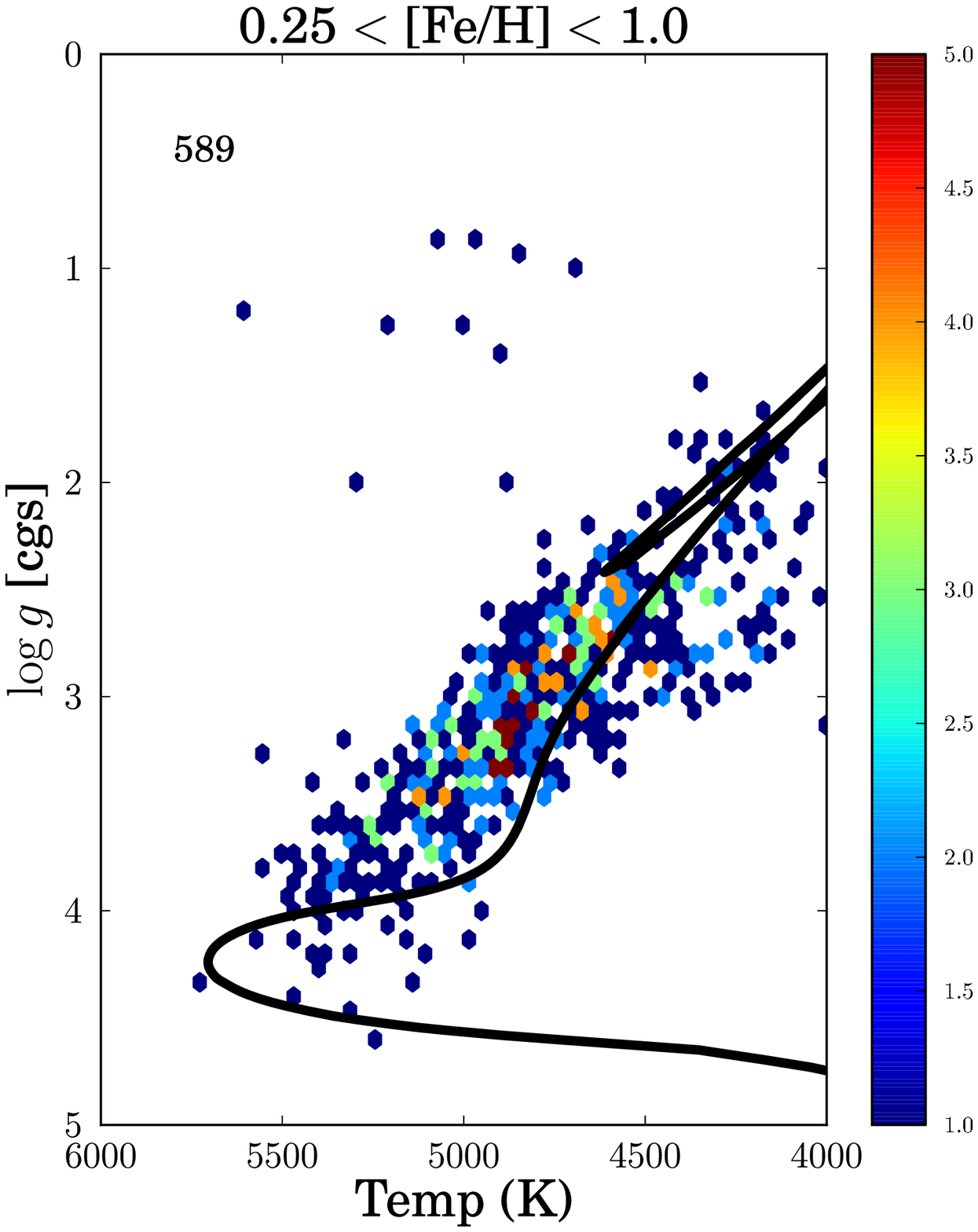}}  
\subfloat[]{\label{fig:twosmall}\includegraphics[scale=0.2]{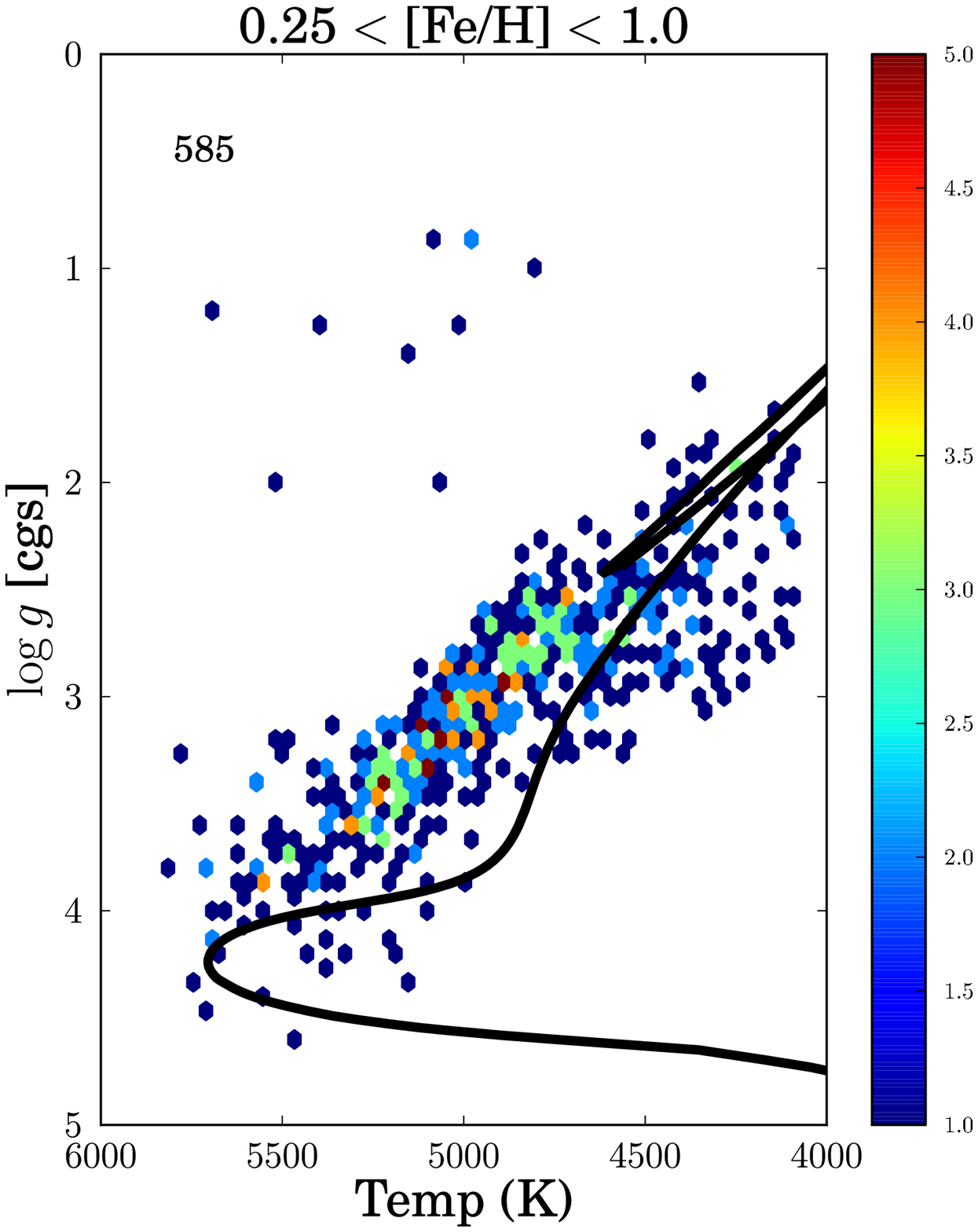}}  \\ 
\caption[The metal-rich stars on the isochrone for 90\% reddening and alpha-enhanced temperatures]{The most metal-rich stars on the isochrone from \citet{Cassisi2006}, showing the effect of adopting a lower interstellar reddening (a) and an alpha-enhanced temperature calibration (b).}
\label{fig:offset1}
\end{figure}

Some authors argue that the Schlegel reddening may be too high by $30$ to $50$\% for regions of $A_V > 0.5$ \citep{ArceGoodman1999, Cambresy2005}. Although our fields were selected to have relatively low reddening,  the regions of highest reddening in our survey which are closest to the Galactic plane also correspond to the locations of the most metal-rich stars. On the other hand, (1) our stars all lie at Galactic latitude $|b| > 4^\circ$; in the inner Galaxy. These stars are at least $300$ pc above the Galactic plane, so the Schlegel reddening should be appropriate; (2) when selecting fields, we found that the Schlegel reddening was in excellent agreement with the reddening inferred from the morphology of the near-IR colour-magnitude diagram; (3) our most highly reddened field at $(l, b) = (0^\circ,-5^\circ)$ has a mean $A_{V} \sim 2.2$. If the Schlegel reddening is systematically too high, then we would expect to see a corresponding shift relative to the isochrones across all metallicity bins in this field. This was not seen.  In summary, we believe that systematic overestimation of the reddening is not responsible for the offset seen in Figure \ref{fig:stellarparams3} at high metallicities. However, although a systematically high reddening is unlikely to be responsible for the offset seen in Figure \ref{fig:stellarparams3} at high metallicities, an overestimation of the reddening for the highest reddening fields only may explain the offset. Figure \ref{fig:offset1} (a) shows the position of the stars on the isochrone for the most metal-rich stars only ([Fe/H] $>$ 0.25), where the temperature has been calculated for a reddening of 0.9 $\times$ E(J-K) only for E(J-K) $>$ 0.3, where E(J-K) denotes the Schlegel reddening. The metal-rich stars now lie closer to the isochrone and the other metallicity bins are not affected. The effect on the derived MDF of using these lower temperatures is that the number of metal rich stars with [Fe/H] $>$ 0 would decrease by about 15\% and the mean [Fe/H] of the MDF would decrease by 0.06 dex.

We performed our analysis prior to the availability of the \citet{Gonzalez2011} reddening map, which provides only partial coverage of our fields. The \citet{Gonzalez2011} reddening values in our fields are lower than the Schlegel values. For our field, $ (l, b) = (0^\circ ,-5^\circ)$, the Gonzalez maps are about E(J-K) = 0.1 on average lower in estimated extinction compared to the Schlegel maps.  This reduces the determined temperature by about 300 K (which corresponds to a lower mean [Fe/H] by about 0.2 dex).  For the highest metallicity bin (Fe/H $>$ 0), this lower photometric temperature gives a better fit to the stars on the isochrone, but at all other metallicities (Fe/H $<$ 0), the stars are shifted too low in temperature and off the isochrone.

The temperature calibration is the other possible source of the offset. Our adopted temperatures have been used for the calibration of our parameters against a number of well-measured stars and clusters (see ARGOS II) and this temperature calibration is able to return the \logg\ and [Fe/H] of our comparison stars within $\sigma_{log g}$ = 0.3 and $\sigma_{[Fe/H]} = 0.09$, respectively. The temperature relationships selected were for alpha-enhanced models for [Fe/H] $\le$ --0.25 and for solar scaled models for [Fe/H] $>$ -0.25. Over the \teff\ range $4000$ to $6000$ K, a change of $50$ K in \teff\ corresponds to a change of $0.05$ in [Fe/H] and $0.15$ in \logg. A systematic temperature error as large as the offset seen in Figure \ref{fig:stellarparams3} seems to be excluded (unless there is a systematic problem with the grid of model spectra such that the model spectra give the right [Fe/H] and \logg\ values for stars whose temperatures have been overestimated from the photometry).   We note however that, over our temperature range, the \citet{Alonso1999} temperatures derived from $(J-K)$ are typically up to $180$ K cooler than the scale which we have adopted; the difference is largest for the more metal-poor stars.  Figure \ref{fig:offset1} (b) shows the improved fit to the isochrone by adopting the alpha-enhanced temperature relationship from \citet{Bessell1998} also for the metal-rich stars ([Fe/H] $>$ -0.25).  The cooler temperatures move the metal-rich stars closer to the isochrone. The effect on the derived MDF using these lower temperatures is that the total number of stars with [Fe/H] $>$ 0 decreases by 15\% and the mean [Fe/H] of the MDF decreases by 0.03 dex. In summary, although the adopted temperatures give accurate [Fe/H] and \logg\ values, the temperature scale remains a possible reason for the offsets relative to the isochrones. Note that for the \teff\ -- (J-K) relation, the J-K changes relatively slowly with \teff, so it is at more risk of errors than an alternative colour index such as V-K. However, our stars are sourced from the 2MASS catalogue for which J-K is available and this colour relation is also less sensitive to differential interstellar reddening which may be a problem for a few of our $b=-5^\circ$ fields.  

Following this discussion, we formulate the following procedure for estimating stellar distances (see also paper II). For stars that are located near the clump (\logg\ $ = 1.8$ to $3.2$ and \teff $= 4500$ to $5300$ K), we adopt an absolute magnitude $M_K = -1.61$ \citep{Alves2000} which is appropriate for stars with [Fe/H] $> -1.0$ ($95$\% of our sample). See the Appendix for further discussion of the absolute magnitude of the red clump adopted for distance calculations. As can be seen from Figure \ref{fig:stellarparams}, the majority of stars are taken to be clump stars.  For stars outside this range of \teff\ and \logg, we derive $M_K$ values using the BaSTI isochrone of age $10$ Gyr and appropriate [Fe/H], adjusting only the temperature of the star to place it on the closest branch of the isochrone. 

For the red clump stars there are three main sources of distance uncertainty: (i) the Schlegel reddening uncertainty, (ii) the 2MASS uncertainty in the K magnitude and (iii) the red clump intrinsic magnitude dispersion. The reddening uncertainty of the Schlegel maps is estimated to be 16\% \citep{Schlegel}. The mean reddening in our fields is $E(B-V) = 0.2$. A 15\% reddening error would typically contribute a $\sigma_{K_{S}}$ = 0.02 to the K magnitude. In our stellar selection, we selected only those stars with 2MASS photometric errors $\le$ 0.06 in each band. The mean error in each magnitude for the stars in our sample is much smaller, about $0.03$ mag.

The red clump magnitude dispersion estimated by  \citet{Alves2000} is $\sigma_{K_{A}}$ = 0.22. Although this value includes the intrinsic error of the parallax measurement which has a $\sigma_{M_{K_\pi}} = 0.1$ \citep{Alves2000}, we adopt the intrinsic dispersion of the clump to be $\sigma_{K}\approx 0.22$ to include the small contribution from reddening and magnitude errors described above. 
As discussed in paper II, in the (\logg\, \teff) plane, the red clump giants lie on a background of first ascent giants.  The first giants with \logg\ values close to the red clump have $M_K$ values
close to that of the clump.  However, because our measuring errors for \logg\ are relatively large
(0.30 dex), we have taken clump stars to be those with \logg\ between 1.8 and 3.2, and our sample of 
clump giants therefore includes some brighter and fainter stars from the first giant branch (see 
paper II, Figure 7). For an unbiased population, most of these first giant branch stars are fainter 
than the clump stars.  Our sample is not unbiased, because the 2MASS catalog is becoming incomplete 
for our fainter stars, and the numbers of first giant branch stars that we have included as clump stars is therefore reduced relative to the unbiased sample.

For stars not on the clump, the distances are determined from the temperature and gravity via the the BaSTI \citep{Cassisi2006} isochrones. To determine its absolute magnitude, the star is shifted across in temperature to fit the nearest branch of isochrone. The error on \logg\ ($\pm$ 0.30) dominates the error of the absolute magnitude. For the 10 Gyr BaSTI isochrones, the absolute K magnitude increases almost linearly with \logg, with d$K$/d(\logg) $= 2.35$. Therefore, $\sigma_{M_K} = 2.35 \times 0.30 = 0.70$ for stars not on the clump. This corresponds to a distance error of about $38$\%. 

\begin{figure}
\centering
\subfloat{\label{fig:onesmall}\includegraphics[scale=0.25]{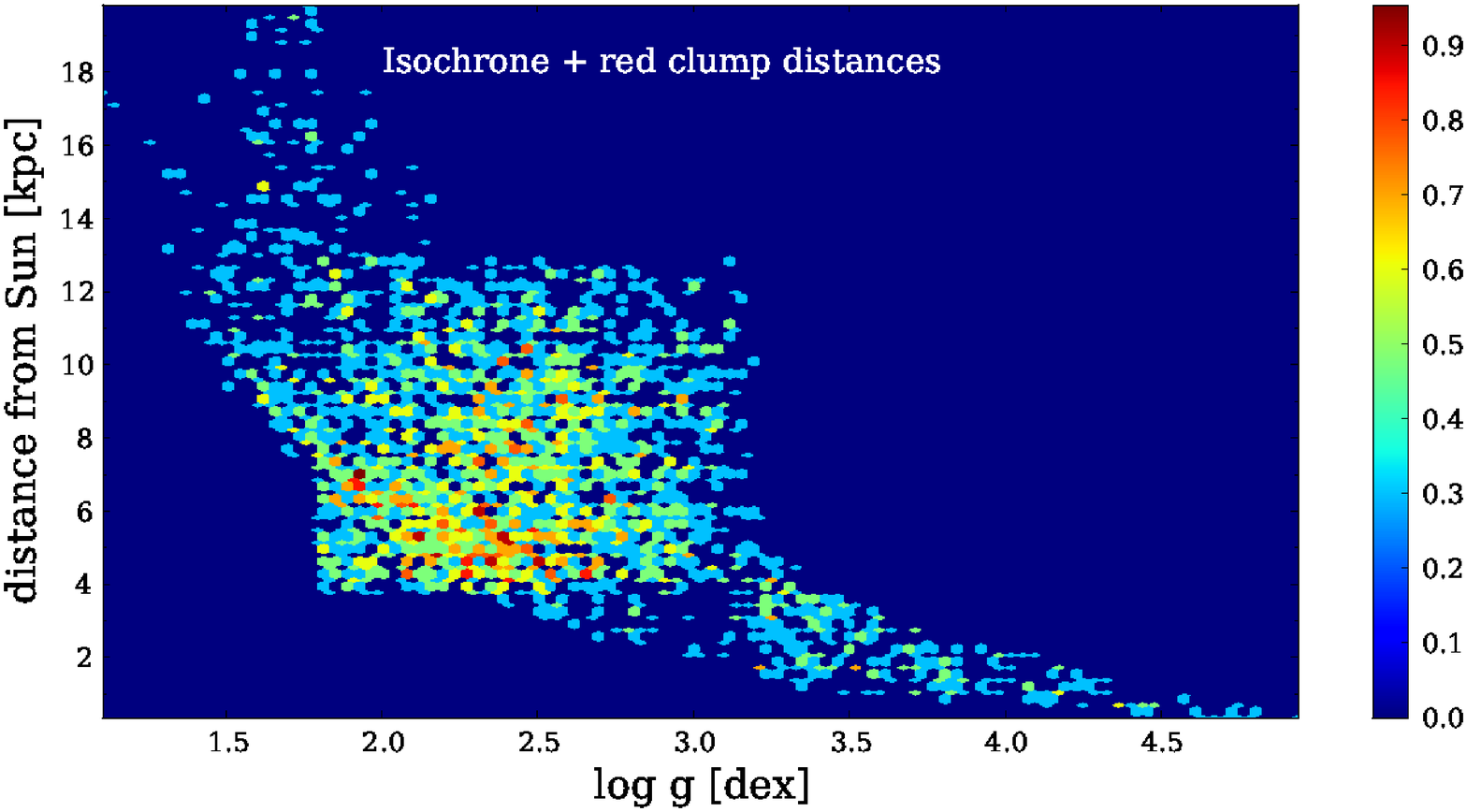}}  \\ 
\vspace{-10pt}
\subfloat{\label{fig:twosmall}\includegraphics[scale=0.25]{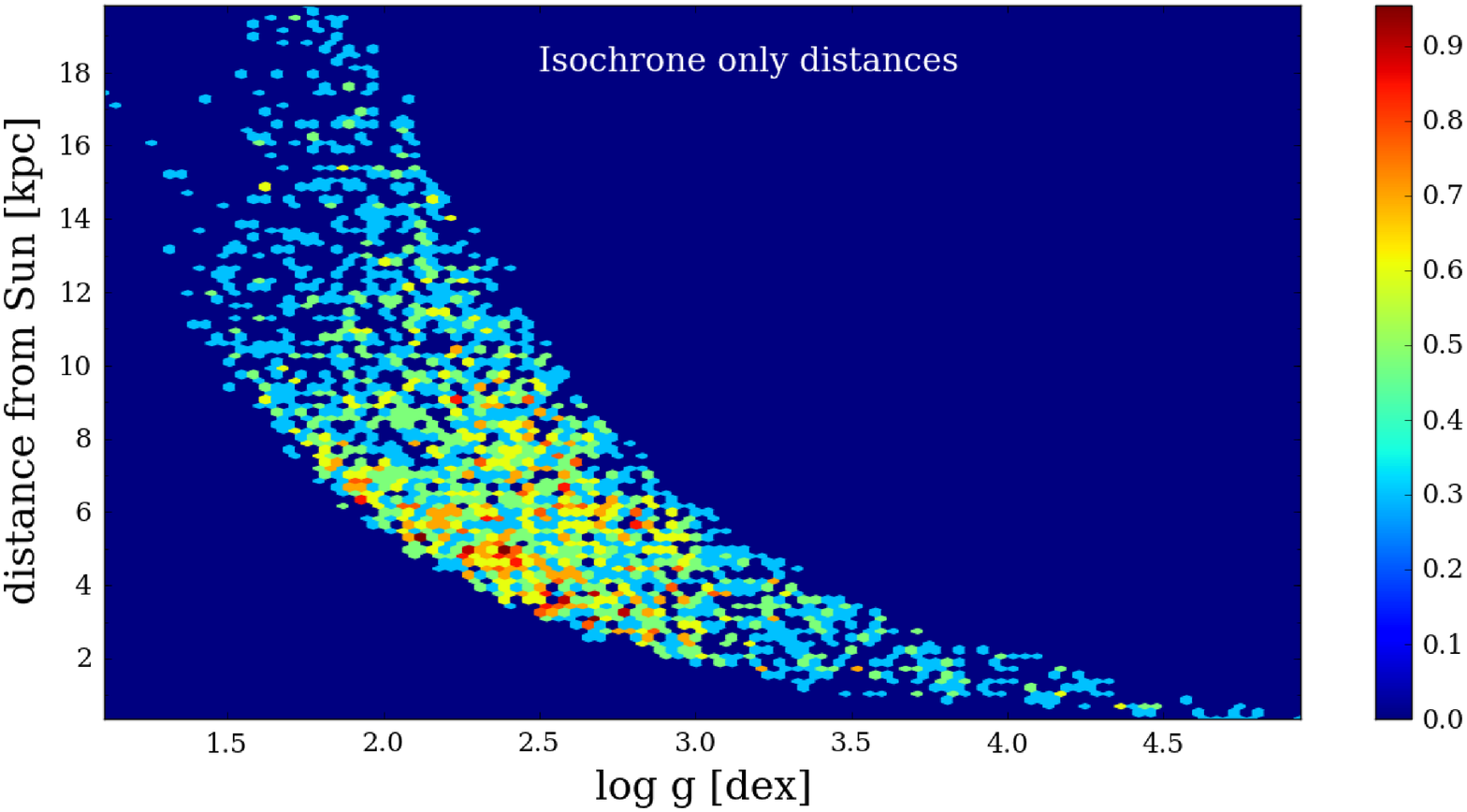}}  \\ 
\caption{The distance, \logg\ distributions for the ARGOS stars using distances calculated for distances calculated from the isochrone + red clump in the interval $1.8 <$  \logg\  $< 3.2$ (top) and from the isochrone only (bottom). Stars are binned into 100 equal width bins for each figure. }
\label{fig:select}
\end{figure}

\begin{figure}
\centering
\includegraphics[scale=0.22]{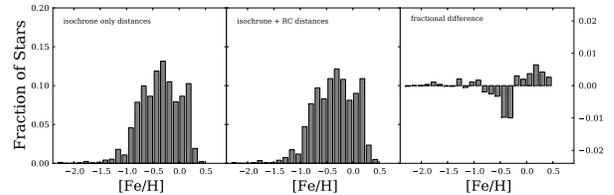}
\caption{The MDFs for all stars on the minor axis within \rgc $<$ 3.5 kpc generated with the isochrone distances only (at left) and with isochrone + red clump distances (centre). The panel on the far right shows the fractional difference in the MDF due to the assumption of red clump magnitudes over this range in 1.8 $<$ \logg\ $<$ 3.2}
\label{fig:difference}
\end{figure}

By using the red clump membership for as many of our potential clump stars as possible, across 1.8 $<$ \logg\ $<$ 3.2 we are able to take advantage of the precision of the red clump absolute magnitude in calculating distances. Selecting stars over this interval has the advantage of including 99\% of our red clump stars given our error of 0.3 dex in \logg. The compromise in doing this is the contamination of non-clump stars that will be intrinsically brighter or fainter which will be bought into  the inner region. For example, stars with a true \logg\ at the upper limit of this selection will in reality lie closer to us than the bulge region which we define in our analysis to be \rgc $<$ 3.5 kpc. These stars will be moved further in distance from the Sun into the bulge. Similarly stars with a real \logg\ at the low end of this selection will in reality be at larger distances than the bulge region.  This kind of contamination by first ascent giants will be a problem for any distance estimates using the red clump magnitude \citep[e.g.][]{Rattenbury2007}, unless gravities are accurately know \citep{Valentini2010}. Figure \ref{fig:select} shows the distribution of the stellar distance to the Sun as a function of \logg\ for stars taken with distances from the isochrone only (at bottom) and using the red clump magnitude across 1.8 $<$ \logg\ $<$ 3.2 (at top) for our minor axis fields. These figures demonstrate how taking a range in \logg\ as assigned clump stars to allow for our observational errors, moves a fraction of closer and farther stars into the bulge region.  The boxy structure seen in Figure \ref{fig:select} across $xy$ reflects the limits across our \logg\ range that we have assumed for the clump and the K-magnitude range selected for our observations of 11.5 to 14. The extent of the spread introduced in taking a range for the clump along the $y$-axis is set by this magnitude range. We can reduce only our interval in \logg, along the $x$-axis. We could have used a narrower interval in \logg\ of say 2.0 -- 3.0 but then we would have lost about 13\% of the clump stars.
Figure \ref{fig:difference} evaluates the effect of our selection on the MDF for stars within \rgc\ $<$ 3.5 kpc on the minor axis using this interval in \logg\ for red clump membership. The MDF at left is for the population  where all distances have been taken directly from the isochrone. There are about 10,000 stars in the inner region using isochrone distances only. The centre panel shows the MDF generated for stars calculated with isochrone + red clump distances and there are about 14,150 stars in the inner region using isochrone + red clump distances. The MDFs are only weakly affected. The far right panel shows the fractional difference of these distributions of MDF (isochrone + red clump) -- MDF (isochrone only).  The selection difference is small in the inner 3.5 kpc, and taking selections of smaller radii increases the relative number of metal rich stars and decreases the number of metal poor stars. For a selection cut of \rgc\ $<$ 1.5 kpc there are about 10\% fewer stars in total with $-0.2 >$  [Fe/H] $> -1.0$  and 10\% more stars with $-0.2 >$ [Fe/H] $> +0.5$.

To summarise, we take all of our potential clump stars with \logg\ between 1.8 and 3.2 to have $M_K =
-1.61 \pm 0.22$  (about 900 pc at the distance of the Galactic center), recognising that for some of the underlying first ascent giants the uncertainty in $M_K$ will be somewhat larger. For the non-clump
giants the distance error is $0.70$ mag, dominated by the uncertainty in the \logg\ estimates. In our data analysis we verified that similar results were obtained for the smaller sample selection obtained using isochrone-only distances. In Section 3 we evaluate Gaussian component fits to our MDFs.  The relative fractions of the components are only slightly changed if we use isochrone-only distances.

To evaluate the properties of the stars which lie in the inner Galaxy, we take stars with cylindrical Galactocentric $|$\rgc$|$ $\le 3.5$ kpc to be part of the inner Galaxy. This sample includes about $14,150$ of the original $25,500$ stars.  The metallicity distribution of these inner stars is summarised in Table \ref{table:fehs}. They are a mixture of all of the populations present in the inner Galaxy: the metal-poor halo, the thick disk, the thin disk around the bulge, and the bulge itself.

\begin{table}
 \centering
     \caption{Metallicity distribution of the 14,150 stars with $|$\rgc$|$ $\le 3.5$ kpc.}
  \begin{tabular}{| l | l |}
 \hline
Number of stars & [Fe/H] range\\
 \hline
16 stars & [Fe/H] $\leq -2.0$ \\
84 stars & $-2.0 <$ [Fe/H] $\leq -1.5$ \\ 
522 stars & $-1.5 \leq$ [Fe/H] $< -1.0$ \\
4219 stars & $-1.0 <$ [Fe/H] $\leq - 0.5$ \\
6914 stars & $-0.5 <$ [Fe/H] $\leq 0$ \\
2392 stars & [Fe/H] $>$ 0 \\
\hline
\end{tabular}
\label{table:fehs}
\end{table}

\subsection{The line of sight distribution of giants in the inner Galaxy } 

\begin{figure}
  \centering
       \includegraphics[scale=0.26]{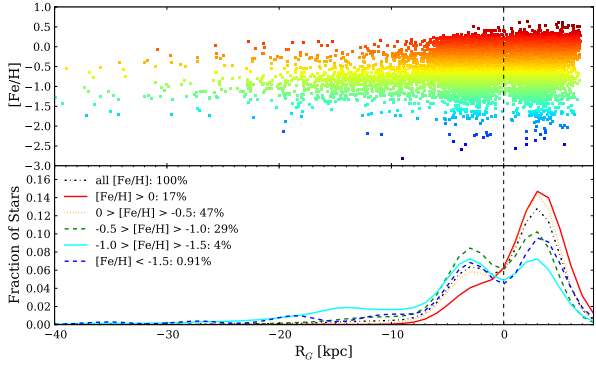} \\
   \caption{Radial distribution of ARGOS stars: In each panel,  the x axis is the cylindrical distance \rgc\ from the Galactic centre. Negative radii are for stars on the far side of the bulge and the vertical line represents the location of the Galactic center. Each star is plotted in the upper panel and the generalised and normalised histogram of the radial distribution is shown in the lower panel. The number of stars as a percentage in each [Fe/H] bin is indicated in the legend. The histograms are shown for all stars and for stars in bins of metallicity. The colours in the upper panels correspond to the colour coding in the lower panels. The [Fe/H] values are colour-coded from red (metal-rich) to blue (metal-poor). Note the changing line of sight distribution with metallicity and the dip near \rgc\ = 0 for some of the metallicity bins. The dip near \rgc\ = 0 comes from the distribution of survey field centres; only the three fields on the minor axis contribute to the distribution near \rgc\ = 0. The kernel size for the generalised histograms is $\sigma$ = 1.5 kpc, the largest estimated distance error.}
   \label{fig:diststars}
         \end{figure}        

Our sample includes giants all along the lines of sight with distances up to $\sim$ 55 kpc from the Sun on the far side of the bulge. The observed distribution of stars over distance is affected by our selection criteria, observational strategy and the stellar metallicities.  Figure \ref{fig:diststars} shows the distribution of radii \rgc\ for our entire sample of 25,500 stars. This figure does not reflect the underlying density distribution. Negative radii are for stars on the far side of the bulge, and the Sun is located at \rgc\ = 8 kpc. The observed stellar density distribution is shown as the smooth curves (generalised histograms with kernel $\sigma = 1.5$ kpc), for (i) all stars and (ii) stars in intervals of [Fe/H] to compare the relative density distribution of different metallicity groups to the total distribution of stars. In total, $60$\% more stars are observed per kpc along the line of sight on the near side of the bulge than on the far side. However, the line-of-sight distribution is metallicity dependent. The majority of observed giants are located about 4 to 10 kpc from the Sun, with the peak about  3 kpc in front of the bulge. \citet{Robin2012} estimates a peak in the thin disk number density at $\sim 2.3$ kpc from the centre of the Galaxy which may contribute to the larger number of metal rich stars seen at low Galactic latitudes on the near side of the bulge. The more metal-poor stars are less biased towards the near side of the Galactic center, with only 15\% fewer stars observed in this range on the far side of the bulge. We preferentially sample the more metal-poor stars at larger distances because they are bluer, brighter and located at higher galactic latitudes.

\begin{table}
\centering
\begin{tabular} {| p{2cm} | p{3cm} }
\hline \hline
$R_{\rm G} $ (kpc) & Number of stars \\
\hline
 0 -- 2 & 5100 \\
2 -- 4 &  12200  \\
4 -- 6 & 5150  \\
6 -- 8 & 2100  \\
$> $ 10 & 900  \\
$> $ 15 & 450  \\
$> $ 25 & 100    \\
$> $ 35 & 30   \\
$\le $ 3.5& 14,150  (55\% of total) \\
$\le $ 4.5& 19,650   (75\% of total) \\
$|y| \le 3.5$ & 17,350 (65\% of total)  \\
\hline
\end{tabular}
 \caption{ The number of stars in different intervals of radius \rgc\ (about $25,500$ stars). In the last line, $y$ is a Cartesian coordinate along the Sun-center line, with origin at the Galactic center.}
\label{table:distnum}
\end{table}

Figure \ref{fig:diststars} shows a minimum or inflection in the stellar number density around \rgc\ = 0. This is due to the location of our fields; stars with $|$\rgc$|$ $< 1.0$ kpc come only from our three fields at longitude $l = 0^\circ$ and at latitudes $b$ between $4$ and $11^\circ$.  This central feature is much less prominent for the more metal-rich stars, indicating that the more metal-rich stars are more concentrated to the inner regions of the bulge. The upper panel of Figure \ref{fig:diststars} shows that most stars with $|$\rgc$|$ $< 0.75$ kpc have [Fe/H] $> -1.0$: only two stars from our sample in this region have [Fe/H] $< -1.0$.  We will return to these properties of the inner region when we examine the metallicity distribution function (MDF) as a function of radius \rgc\ (Figures 12 and 13).

Although the lower limit of metallicity for our stars depends on radius (higher for $|$\rgc$|$ $< 0.75$ kpc), the upper limit of the [Fe/H] distribution is similar at about [Fe/H] $= 0.4$ for all radii up to about $|$\rgc$|$ = 5 kpc, with only a few stars above this level. Table \ref{table:distnum} shows the distribution of the number of stars (giants) with \rgc\ for stars seen out in the halo on the far side of the bulge. 

The most metal-poor stars in our sample, with [Fe/H] $< -2.0$, are located in the inner Galaxy but not in the innermost region. Of the 27 stars with [Fe/H] $< -2.0$,  $18$ lie between $|$\rgc$|$ $= 0.9$ and $4.5$ kpc.  In the inner region, our most metal-poor star has [Fe/H] $= -2.6$ and lies at \rgc $= -3.8$ kpc on the far side of the bulge. Our most metal rich star with [Fe/H] $= 0.68$ is at \rgc\ $= 2.0$ kpc on the near side of the bulge. Although we find a few stars at metallicities [Fe/H] $\sim 0.6$, which are among the highest [Fe/H] values for bulge stars reported in the literature, we note that our $\chi^2$ abundance scale is externally calibrated only for $0.1 >$ [Fe/H] $> -2.1$.  Outside the range, the abundance scale comes directly from the grid of model spectra. 

We have not included the solar neighbourhood dwarfs in Figure \ref{fig:diststars}. In total, our sample has about $2000$ dwarfs located between about $1$ and $3$ kpc from the Sun. They have a median [Fe/H] $= -0.1$.  These stars are not relevant to the present discussion of the inner Galaxy, and are not included in the discussion here.

\subsection{The [Fe/H] distribution in the inner Galaxy}

\begin{figure}
 \centering
 \includegraphics[scale=0.3]{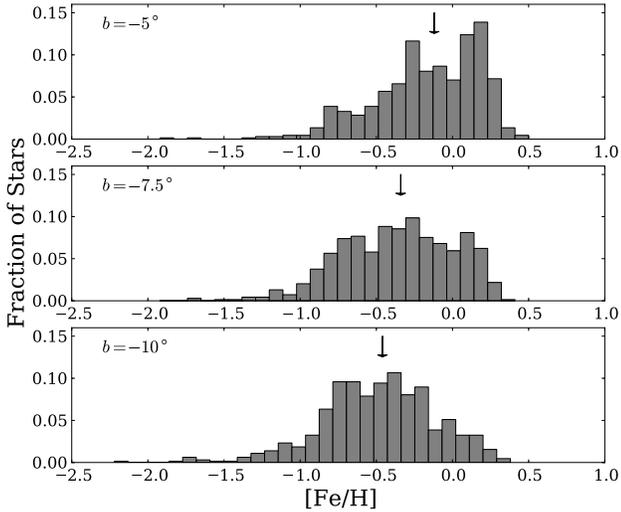} \\
    \caption{Histograms of metallicity for our three fields along the minor axis of the bulge at $b = -5^\circ$, $-7.5^\circ$ and $-10^\circ$ and with $R_{\rm G} \le$ 3.5 kpc. For $(l,b) = (0^\circ,-5^\circ)$ 670 stars (73\% of the total in the field) lie in this \rgc\ range with median [Fe/H] = --0.12. For $(l,b) = (0^\circ,-7.5^\circ)$ there are 690 stars (73\% of the total) and the median [Fe/H] = --0.34. For $(l,b) = (0^\circ,-10^\circ)$ there are 650 stars (67\% of the total) and the median [Fe/H] = --0.46. The arrows indicate the median of the total distribution.}  
  \label{fig:histmine}
\end{figure}

\begin{figure}
 \centering
 \includegraphics[scale=0.3]{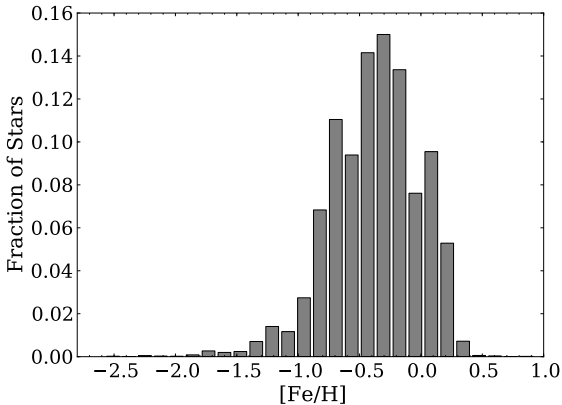} \\
    \caption{Metallicity Distribution Function for all stars within $R_{\rm G} \le 3.5$ kpc. The vertical axis is the fraction of stars in each [Fe/H] bin.  }
\label{fig:mdf}
\end{figure}

From the mean of the abundance distributions in our three latitude zones along the minor axis, we confirm the metallicity gradient seen in the bulge by \citet{Zoccali2008}. Our Metallicity Distribution Functions (MDFs) in the three minor-axis  fields include a total of about $2000$ stars with $|$\rgc$|$ $\le$ 3.5 kpc (see Figure \ref{fig:histmine}), which is about $70$\% of the whole sample of stars observed on the minor axis. Figure \ref{fig:mdf} shows the MDF for all of our survey stars with $|$\rgc$|$ $\le 3.5$ kpc. This distribution shows the relative proportion of stars in each [Fe/H] bin; it includes about 14,150 stars with contributions from fields across longitudes between $\pm 20^\circ$ and all four latitudes of $b=-5^\circ,-7.5^\circ$, $-10^\circ $ and $b=+10^\circ$. 

Our inferred MDFs shown in Figure \ref{fig:histmine} are qualitatively similar to those obtained by \citet{Zoccali2008} from their survey of about 800 K--giants observed at high spectral resolving power ($R \approx 45,000$) in their minor axis fields at $ b = -4^\circ , -6^\circ$ and $-12^\circ$.  However, our selection includes a larger fraction of metal poor stars due to our significantly bluer color cut in our stellar selection compared to that of \citet{Zoccali2008}. Quantitatively, our mean [Fe/H] values are about 0.15 dex lower on average, with our mean [Fe/H] decreasing from --0.16 at $b = -5^\circ$ to [Fe/H] = --0.5 at $b = -10^\circ$. We find a somewhat lower vertical gradient of the mean abundance along the minor axis: $-0.45$ dex/kpc compared to their value of $-0.6$ dex/kpc.

\section{The components identified in the [F\lowercase{e}/H] distribution}

\begin{figure}
  \centering
        \includegraphics[scale=0.25]{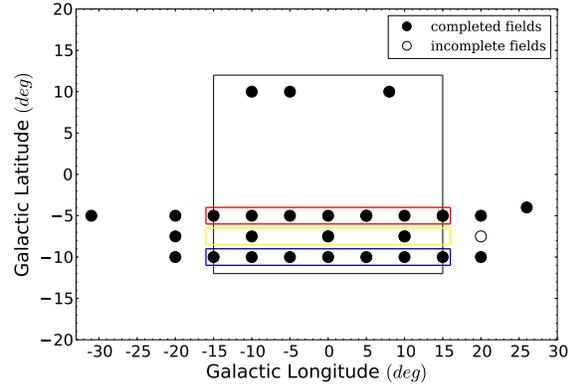} \\
   \caption{Fields with common latitude binned together for generating generalised histograms of the metallicity distributions }
       \label{fig:binning}
    \end{figure}

We now examine the structure of the MDFs in our three latitude zones. To eliminate the effect of binning we examine the generalised histograms for our data in our fields, where each star is represented by a Gaussian distribution centered on its [Fe/H] value and with a standard deviation of $0.13$ (we adopted a kernel that is marginally wider than our estimated measuring error). To increase the number of stars, we have binned our data across all of the bulge fields, out to $l = \pm 15^\circ$, for each latitude; $b= -5^\circ$, $b = -7.5^\circ$ and $b = -10^\circ$. The fields in the three latitudes which are grouped together are shown in Figure \ref{fig:binning}. For a cut in radius of $|$\rgc\ $|$  $\le$ 3.5 kpc, there are 4200 stars in the lowest latitude fields ($b=-5^\circ$), 2000 stars in the intermediate latitude  fields ($b = -7.5^\circ$) and 4000 stars in the highest latitude fields $(b=-10^\circ)$.  It was immediately obvious that the MDFs showed some structure which was qualitatively similar over our three latitude intervals. 

\begin{figure*}
  \centering
 \subfloat[$l \pm 15^\circ$, $b=-5^\circ$]{\label{fig:gull}\includegraphics[width=0.3\textwidth]{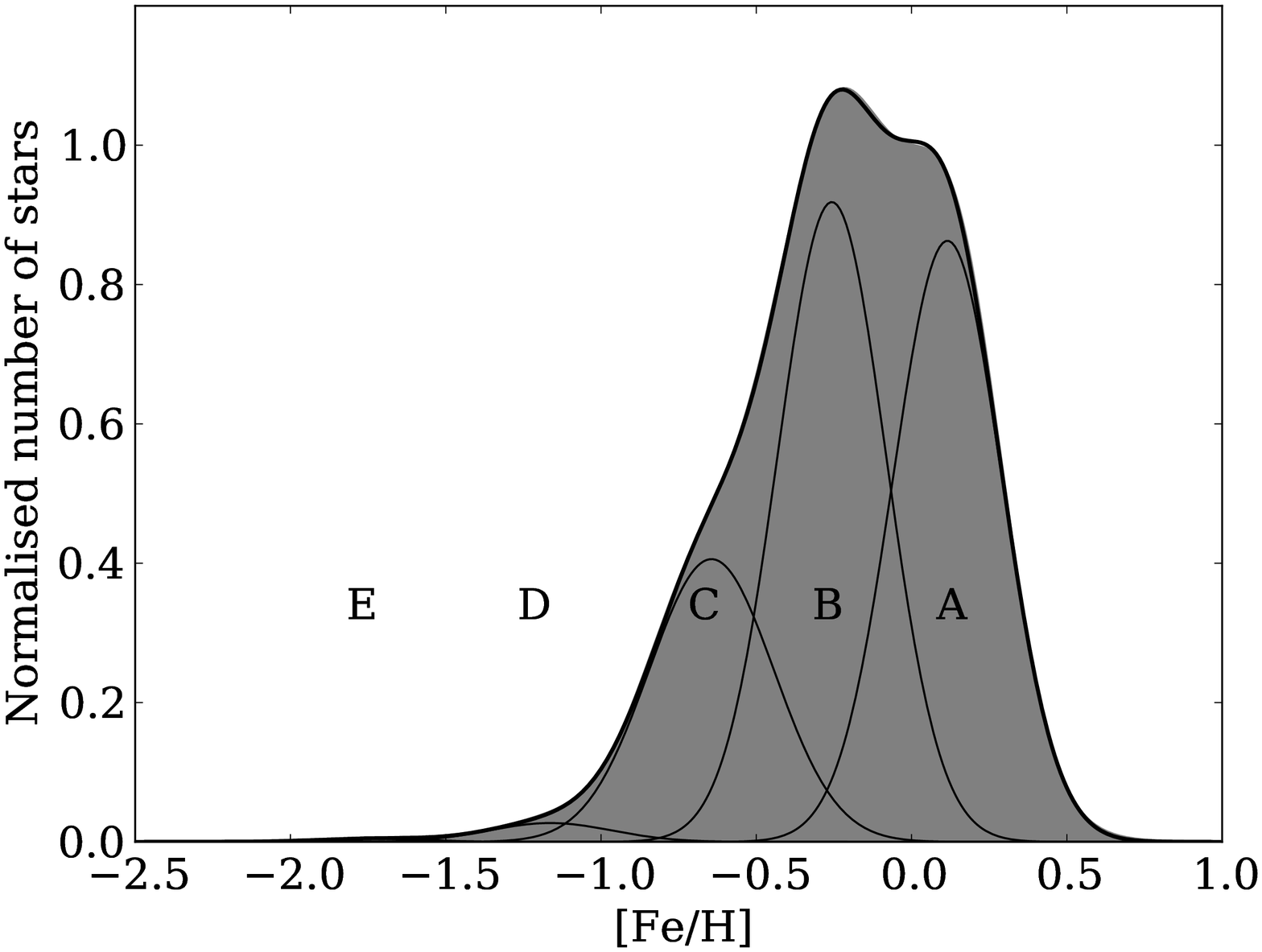}}  
    \subfloat[$l \pm 15^\circ$, $b=-7.5^\circ$]{\label{fig:gull}\includegraphics[width=0.3\textwidth]{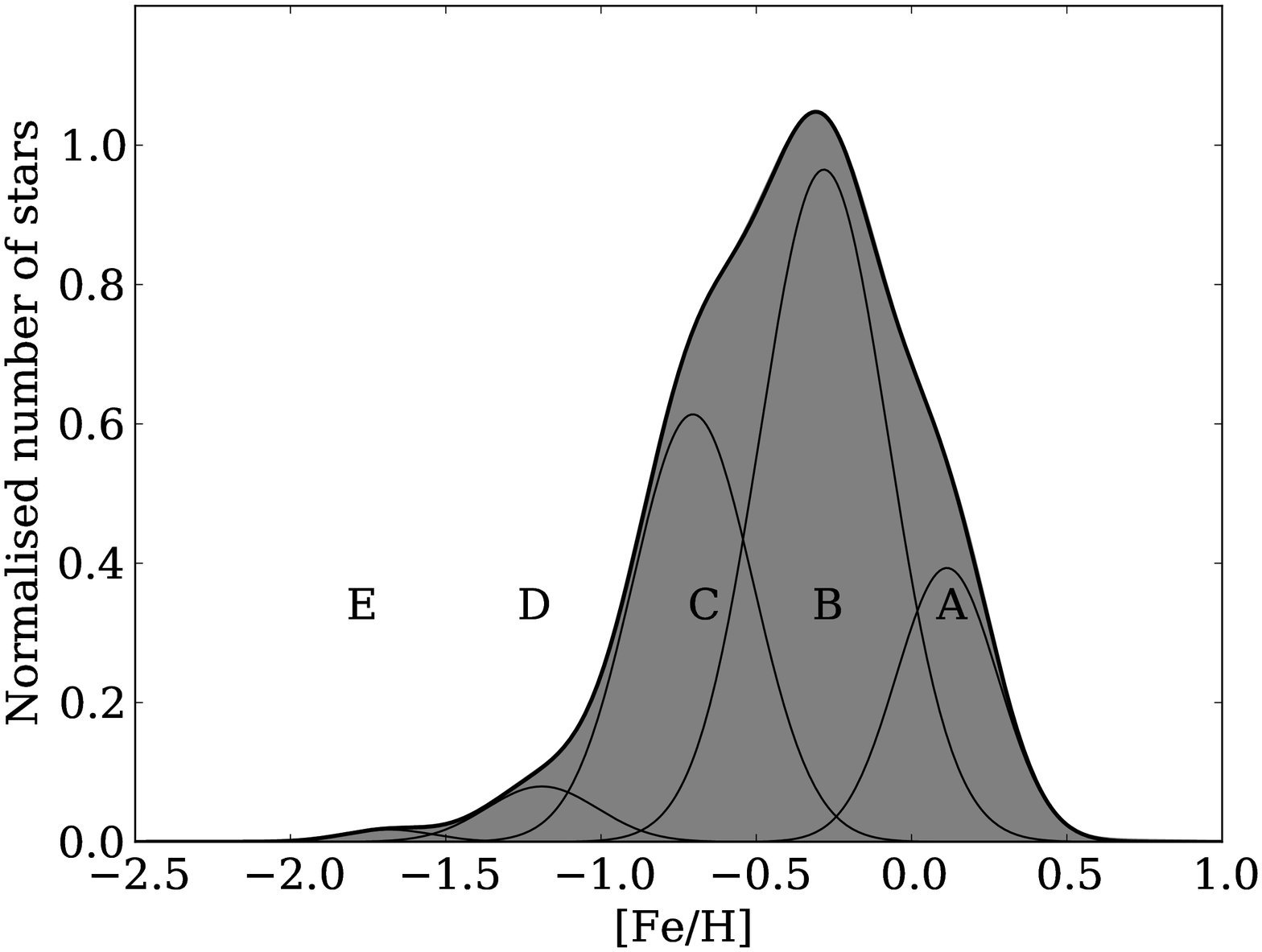}}  
\subfloat[$l \pm 15^\circ$, $b=-10^\circ$]{\label{fig:gull}\includegraphics[width=0.3\textwidth]{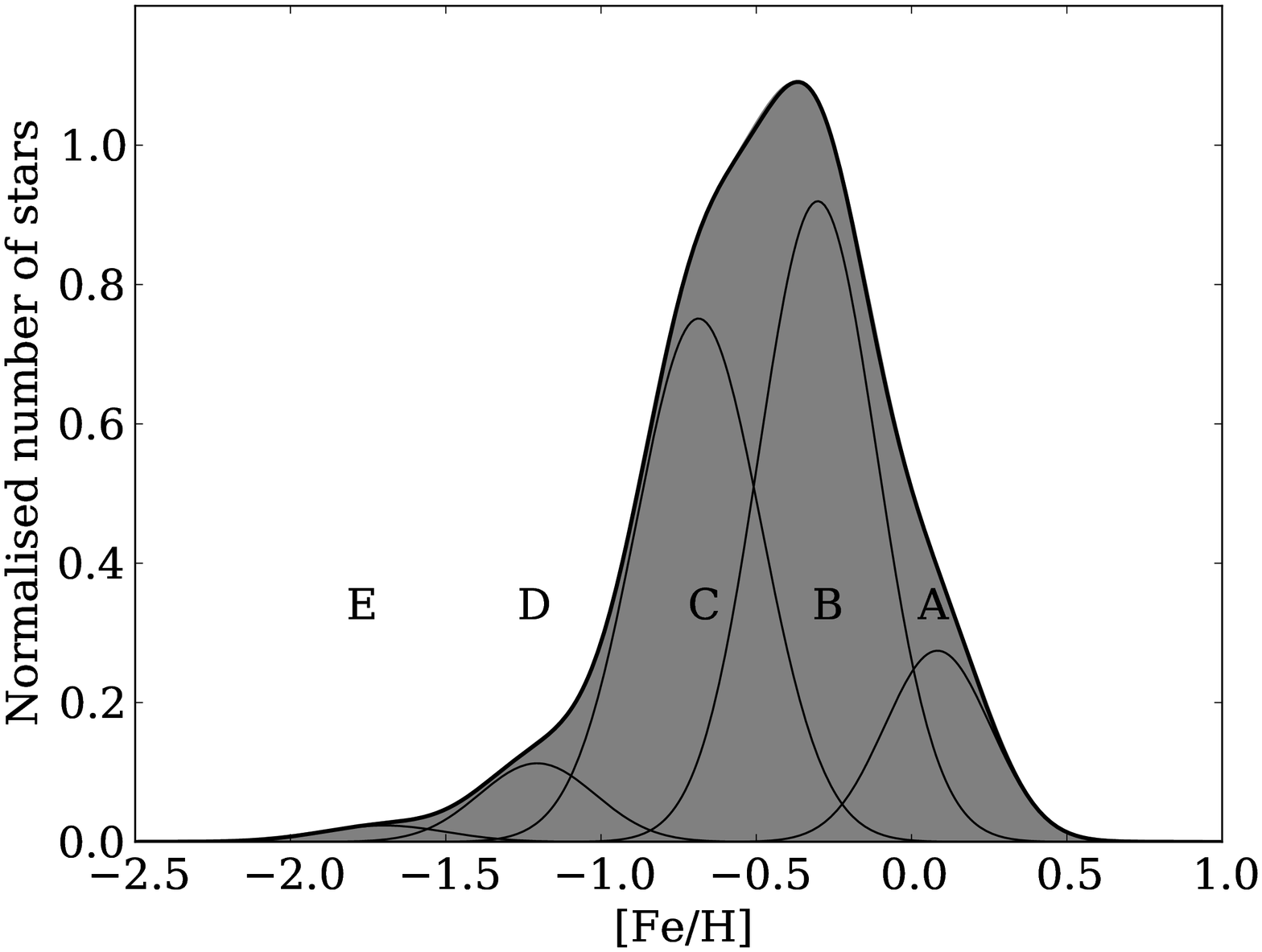}}  \\ 
           \caption{ Panels (a),(b),(c) show the generalised histograms for stars in the inner Galaxy with \rgc\ $\le$ 3.5 kpc  at  latitudes $b = -5^\circ$ , $b = -7.5^\circ$, $b = -10^\circ$. Our smoothing kernel is $\sigma = 0.13$. The number of stars in each panel is 4200, 2000 and 4000 stars  respectively.  Our Gaussian decomposition of the MDFs is shown; A -- E.}
   \label{fig:components}
  \end{figure*}

\begin{figure*}
      \subfloat[$l \pm 15^\circ$, $b=-5^\circ$]{\label{fig:gull}\includegraphics[width=0.3\textwidth]{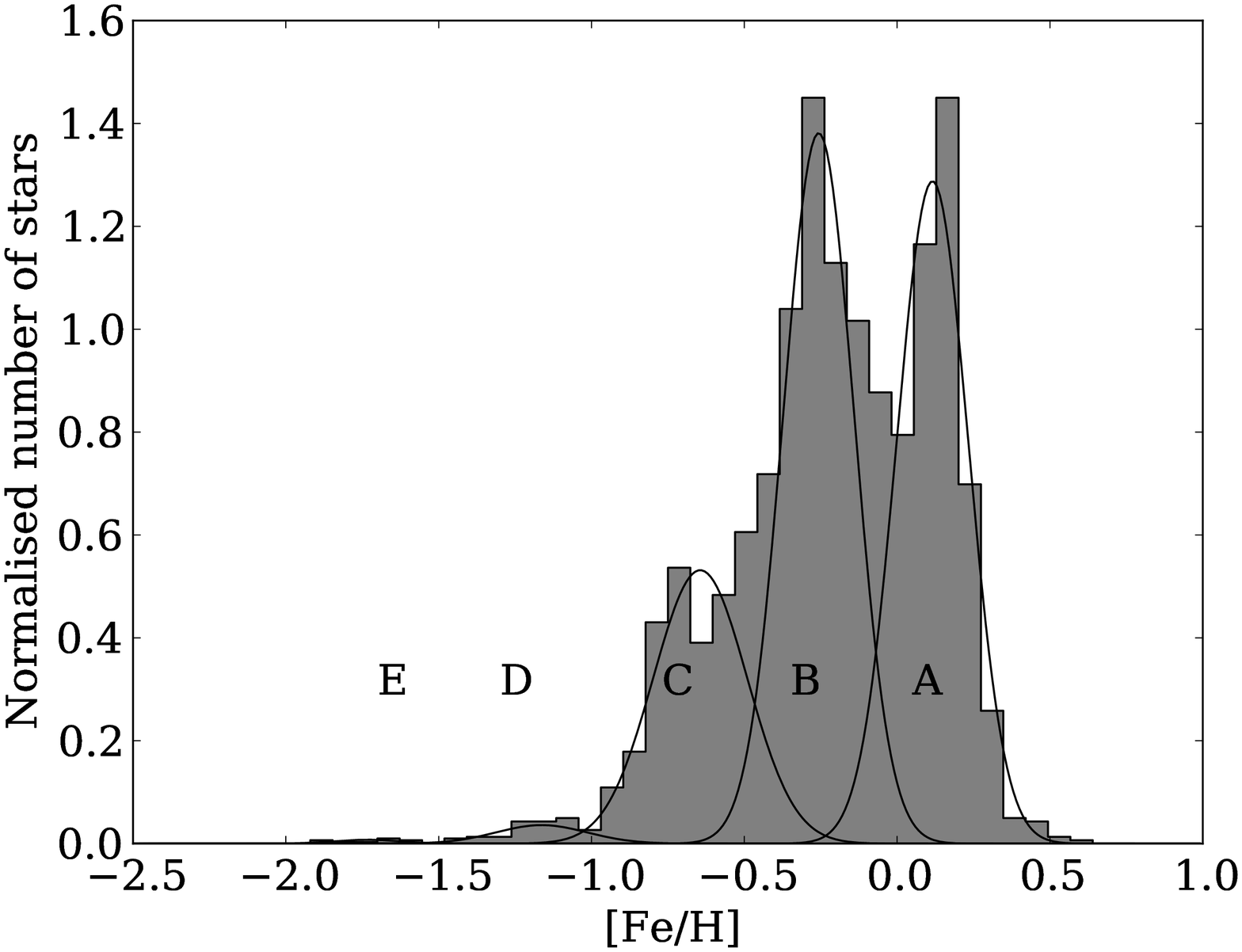}}  
  \subfloat[$l \pm 15^\circ$, $b=-7.5^\circ$]{\label{fig:gull}\includegraphics[width=0.3\textwidth]{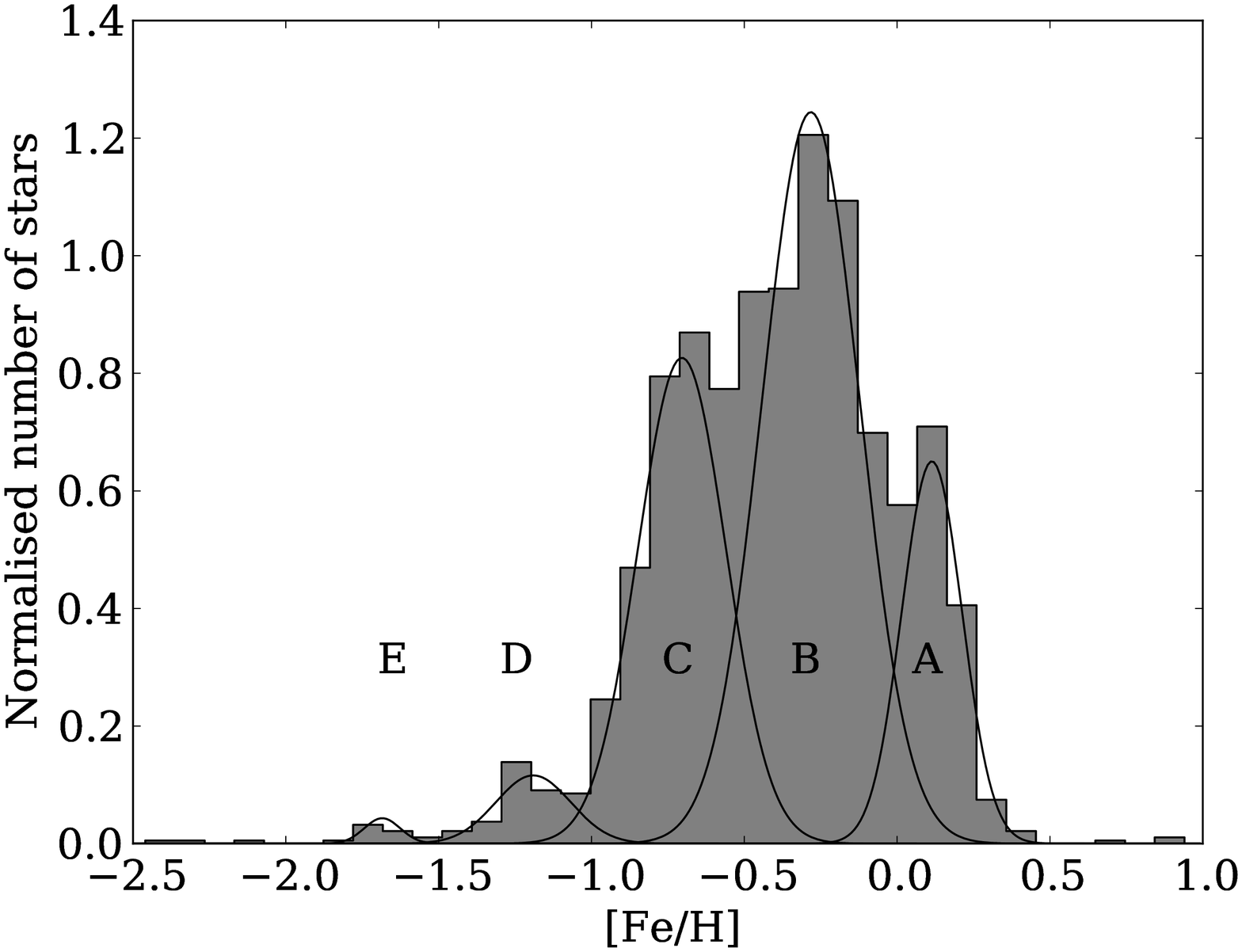}}  
  \subfloat[$l \pm 15^\circ$, $b=-10^\circ$]{\label{fig:gull}\includegraphics[width=0.3\textwidth]{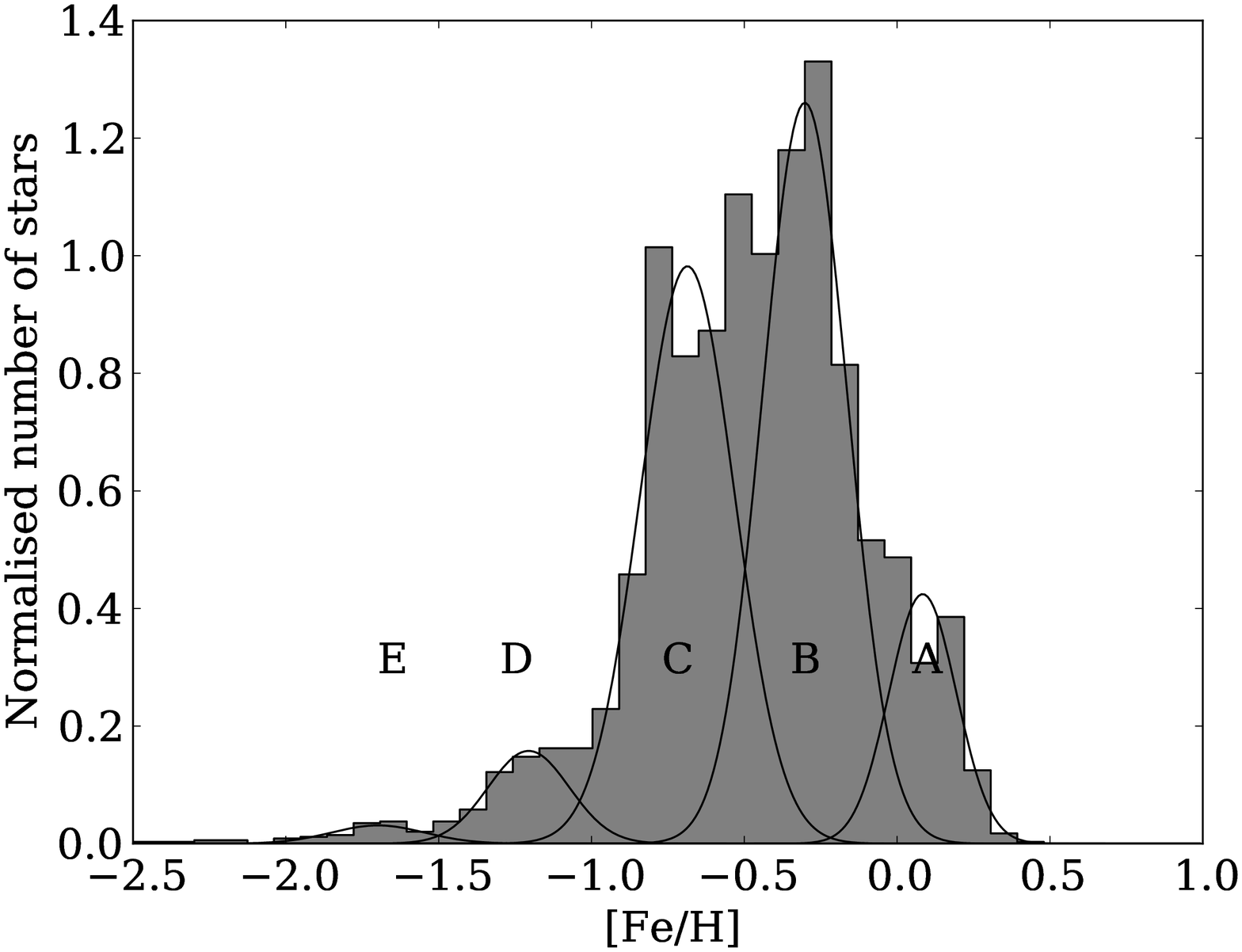}}  \\ 
  \caption{ Panels (a),(b),(c) show the MDFs for stars in the inner Galaxy with \rgc\ $\le$ 3.5 kpc  at  latitudes $b = -5^\circ$ , $b = -7.5^\circ$ , $b = -10^\circ$. Our Gaussian decomposition of the MDFs is shown: see text for more details}
  \label{fig:components2}
\end{figure*}

Figures \ref{fig:components} (a), (b) and (c) show the generalised histograms at these three latitudes for stars within the bulge region $|$\rgc$|$ $ \le 3.5$ kpc. The histograms show some structure and we can decompose them into a few Gaussian components. For all of the three latitudes, we can fit almost the same four Gaussian components, which we denote A,B,C,D\footnote{see Section 3.1 for justification of number of components} plus an additional more metal-poor component E in the higher latitude fields (E is almost negligible at $b=-5^\circ$ and comprises only a very few stars at all latitudes). 
Figures \ref{fig:components2} (a), (b) and (c) show conventional histograms of the metallicity distribution in the three latitude zones.  The individual MDF components can be identified. The smooth curves are gaussians with the same mean [Fe/H] as for the components in the generalised histograms of Figure \ref{fig:components}, but with standard deviations $\sigma_{real}$ reduced to account for the broadening effect of the  generalised histogram kernel ($\sigma_k = 0.13$):  the $\sigma_{real}$ values for the gaussians shown in Figure \ref{fig:components2} are given by $\sigma^2_{real} = \sigma^2_{GH} -  \sigma^2_k$, where $\sigma_{GH}$ are the standard deviations derived in the multi-gaussian fit to the generalised histograms.  

Table \ref{table:mycomponents} summarises the parameters for the MDF fits, corrected for the standard deviation of the generalised histogram kernel. We find slightly decreasing mean abundances of components A, B and C with increasing height below the plane $|z|$ \footnote{By fitting all Gaussian parameters in the more general method used here, we find slightly decreased gradients relative to those reported in \citet{proceedings}}. 

Gaussian fits to the components have been adopted here for simplicity. It may be more correct to use skewed distributions for the components \citep[e.g.][]{Gilmore1995}. This may remove the need for the extra Gaussian components at the low metallicity tail of the distribution, for which there are only a few stars. Components D and particularly E in the metal-poor tail are fit here only to demonstrate the changing fraction of metal-poor stars. Where stars with Fe/H $< $--1.5 are present, component E is included to highlight their relative fraction, although this component is generally not statistically significant. It is instructive however to see how this metal-poor contribution changes with latitude and longitude. Table \ref{table:mycomponents} demonstrates how the fraction of metal-poor stars increases with latitude from an almost negligible value in our lowest latitude zone.

A Monte Carlo analysis was performed to determine the error in the MDF component parameters due to the fitting and sampling. We took the observed sample of [Fe/H] values and generated a new value of each [Fe/H], drawn from a Gaussian distribution around the mean of the data point and its standard deviation (we adopted a $\sigma$ = 0.13, fractionally larger than the measuring error). Revised Gaussian fits were then made to the generalised histogram generated from this new distribution and the process was repeated $10^3 $ times to determine the variance in the mean, amplitude, and standard deviation of the 5 components. Table \ref{table:mycomponents} summarises the outcome:  for the mean [Fe/H] of the three main components A,B,C shown in Figure \ref{fig:components} (a)--(c), the errors are $\approx 0.01$. For the amplitudes they are typically $0.01$, and for the standard deviations of the Gaussian components, they are about $0.01$ to $0.05$. The corresponding Gaussian components along the minor axis fields only are shown in Table \ref{table:mycomponentsminor}.

\begin{table}
\centering
\begin{tabular}{ | l | l | l | l | }
\hline
 & \multicolumn{3}{|c|} { Gaussian Components Fit to Data} \\ [0.5ex]
 \hline\hline
  & \multicolumn{3}{|c|} {$ -5^\circ$}   \\ [0.5ex]
\hline
&Mean [Fe/H]& Weighting& $\sigma_{real}$ \\
A &   +0.12 $\pm 0.01$ & 0.38  $\pm 0.01$  & 0.12 $\pm 0.01$\\
B &  --0.26 $\pm 0.01$ & 0.40  $\pm 0.01$  & 0.12 $\pm 0.01$ \\
C &  --0.66 $\pm 0.02$ & 0.20  $\pm 0.01$  & 0.15 $\pm 0.05$\\
D &  --1.16 $\pm 0.02$ & 0.01  $\pm 0.01$  & 0.15 $\pm 0.02$\\
E & --1.73 $\pm 0.08$ & 0.002 $\pm 0.01$  & 0.10 $\pm 0.04$ \\
\hline
  & \multicolumn{3}{|c|} {$ -7.5^\circ$}   \\ [0.5ex]
\hline
&Mean [Fe/H]& Weighting& $\sigma_{real}$ \\
A &   +0.11 $\pm 0.01$ & 0.16  $\pm 0.02$  & 0.10 $\pm 0.02$\\
B &  --0.28 $\pm 0.01$ & 0.50  $\pm 0.02$  & 0.16 $\pm 0.02$ \\
C &  --0.70 $\pm 0.02$ & 0.30  $\pm 0.02$  & 0.14 $\pm 0.02$\\
D &  --1.19 $\pm 0.09$ & 0.04  $\pm 0.01$  & 0.12 $\pm 0.05$\\
E & --1.68 $\pm 0.3$ & 0.006 $\pm 0.01$  & 0.06 $\pm 0.10$ \\
\hline
  & \multicolumn{3}{|c|} {$ -10^\circ$}   \\ [0.5ex]
\hline
&Mean [Fe/H]& Weighting& $\sigma_{real}$ \\
A &   +0.08 $\pm 0.02$ & 0.12 $\pm 0.02$  & 0.11 $\pm 0.02$\\
B &  --0.30 $\pm 0.01$ & 0.44  $\pm 0.02$  & 0.14 $\pm 0.01$ \\
C &  --0.69 $\pm 0.01$ & 0.38  $\pm 0.02$  & 0.15 $\pm 0.02$\\
D &  --1.21 $\pm 0.03$ & 0.05  $\pm 0.01$  & 0.13 $\pm 0.02$\\
E & --1.70  $\pm 0.10$  & 0.01   $\pm 0.01$  & 0.15 $\pm 0.04$ \\
\hline
\hline

\end{tabular}
\caption{The mean, $\sigma$ and weights of the five metallicity components identified in the Galactic bulge in the three latitudes out to $l=\pm15^\circ$, for $|$\rgc$|$ $\le 3.5$ kpc}
\label{table:mycomponents}
\end{table}

\begin{table}
\centering
\begin{tabular}{ | l | l | l | l | }
\hline
 & \multicolumn{3}{c|} { Gaussian Components Fit to Data} \\ [0.5ex]
 \hline\hline
  & \multicolumn{3}{c|} {$ -5^\circ$}   \\ [0.5ex]
\hline
&Mean [Fe/H]& Weighting& $\sigma_{real}$ \\
A &   +0.15 $\pm 0.02$ & 0.42  $\pm 0.02$  & 0.11 $\pm 0.02$\\
B &  --0.25 $\pm0.02$ & 0.42  $\pm 0.02$  & 0.15 $\pm 0.02$ \\
C &  --0.71 $\pm 0.02$ & 0.14  $\pm 0.02$  & 0.13 $\pm 0.02$\\
D &  --1.18 $\pm 0.1$ & 0.01  $\pm 0.005$  & 0.08 $\pm 0.05$\\
E & --1.80 $\pm 0.1$ & 0.003 $\pm 0.002$  & 0.11 $\pm 0.07$ \\
\hline
  & \multicolumn{3}{c|} {$ -7.5^\circ$}   \\ [0.5ex]
\hline
&Mean [Fe/H]& Weighting& $\sigma_{real}$ \\
A &   +0.13 $\pm 0.02$ & 0.19  $\pm 0.02$  & 0.08 $\pm 0.02$\\
B &  --0.26 $\pm0.03$ & 0.46  $\pm 0.02$  & 0.18 $\pm 0.02$ \\
C &  --0.70 $\pm 0.02$ & 0.31  $\pm 0.02$  & 0.16 $\pm 0.02$\\
D &  --1.19 $\pm 0.1$ & 0.04  $\pm 0.005$  & 0.16 $\pm 0.05$\\
E & --1.68 $\pm 0.1$ & 0.003 $\pm 0.002$  & 0.08 $\pm 0.07$ \\
\hline
  & \multicolumn{3}{c|} {$ -10^\circ$}   \\ [0.5ex]
\hline
&Mean [Fe/H]& Weighting& $\sigma_{real}$ \\
A &   +0.09 $\pm 0.03$ & 0.12 $\pm 0.03$  & 0.12 $\pm 0.01$\\
B &  --0.31 $\pm 0.02 $ & 0.41  $\pm 0.02$  & 0.15 $\pm 0.02$ \\
C &  --0.70 $\pm 0.02$ & 0.40  $\pm 0.02$  & 0.15 $\pm 0.02$\\
D &  --1.17 $\pm 0.03$ & 0.06  $\pm 0.01$  & 0.11 $\pm 0.02$\\
E & --1.68 $\pm 0.2$  & 0.01 $\pm 0.01$  & 0.08 $\pm 0.06$ \\
\hline
\end{tabular}
\caption[The mean, $\sigma$ and weights of the five metallicity components identified along the minor axis]{The mean, $\sigma$ and weights of the five metallicity components identified along the minor axis inside $R_{\rm G} \le 3.5$ kpc}
\label{table:mycomponentsminor}
\end{table}

\subsection{Justification of components}

The maximum likelihood method was employed using the Expectation 
Maximisation (EM) algorithm of \citet{Meng1993} for multi Gaussian fits 
within the statistical package R\footnote{http://www.r-project.org/}, 
to determine the log likelihood for N=2,3,4,5,6 components for our MDF 
distributions given our probability distribution function.\\

\noindent{The probability density $f(x)$ that is most likely to have generated the data points is represented by a mixture of $k = 1....n$ Gaussian functions $G_{k}$; with parameters $\theta_k$ = $(\mu_k, \sigma_k) $ and mixture weights $p_{k}$ so the equation takes the form }

\begin{equation} f(x; \theta) = \sum\limits_{k=1}^n (p_{k}G_{k})
\end{equation}
where each Gaussian function is:\begin{equation} G_{k} = G(x; \mu_{k}, \sigma_{k}) =  \frac{1}{\sqrt{2\pi}\sigma_k} e^{-\frac{1}{2}\left({\frac{x - \mu_{k}}{\sigma_{k}}}\right)^2}
\end{equation}
where the probability distribution integrates to 1 and the weights of the components sum to 1.  \begin{equation}  \int_x f(x; \theta) \,dx = \sum\limits_{k=1}^n p_{k} = 1  \end{equation} 
$x$ is the data of length $N$ (number of [Fe/H] data points) \\

\noindent{The maximum log likelihood is then defined as }

\begin{equation}
$L$(x; \theta) = \prod_{i=1}^N f(x_{i}; \theta)	
\end{equation}
and in our case where we have a Gaussian mixture model
\begin{equation}
L(x; \theta) = \prod_{i=1}^N \sum_{k=1}^{n}p_{k}G(x_{i}; \mu_{k}, \sigma_{k})
\end{equation}
The log likelihood becomes 
\begin{equation}
ln \: L(x;\theta) = \sum_{i=1}^N ln \sum_{k=1}^n p_{k} G(x_{i}; \mu_{k}, \sigma_{k})
\end{equation}

\noindent{The EM algorithm maximises the likelihood function (minimises the negative log likelihood) and so returns a single value given a Gaussian mixture distribution. To assess the relative fits of the models, the Bayesian Information Criterion (BIC) was then applied which compares the relative fit of the models by incorporating a penalty on the maximum likelihood:}

\begin{equation} BIC = -2\, \ln L + n\, \ln(N) 
\end{equation}

\noindent{where $N$ is the number of data points and $n$ is the number of free parameters which are being estimated.}

\noindent{The Bayesian Information Criterion (BIC) method of assessing goodness of fit, which uses the maximum likelihood (and associated Akaike Information Criterion (AIC)) are simplified metrics from a fully Bayesian analysis (Heavens 2011). These methods find only a maximum and not average likelihood, however. The complete Bayesian approach calculates the likelihood over the full parameter space, thus assessing the predictive power of the model, given a prior. However, for our purposes, the information theory approach is sufficient to find the best model. This is because, although we aim to demonstrate that our fits are valid, our decomposition is also motivated by the peaks we can see in the smoothed distribution, the behaviour we find for the red clump stars along the minor axis \citep{Ness2012a} and the clear dependency of kinematics on [Fe/H] \citep[][in preparation]{Ness2012c}

The BIC test returns a minimum (optimal) value for a 4-component fit (A,B,C,D) at $b=-5^\circ$ and $b=-7.5^\circ$ and a minimum for a 5-component fit (A,B,C,D,E) at $b=-10^\circ$. In the Figures 
\ref{fig:components}, \ref{fig:bfive}, \ref{fig:dist10} and Tables 
\ref{table:mycomponents} and  \ref{table:mycomponentsminor}, the 
component E at the lowest [Fe/H] group is included to demonstrate the changing number of the most metal-poor stars seen in the sample as a function of latitude, even though this is not a statistically significant fit.  The maximum likelihood fitting routine returns the parameters of the $\chi^2$ minimisation fitting of Table \ref{table:mycomponents} (except at the metal-poor tail) for the MDFs at $b=-7.5$ and $b=-10^\circ$. For these mixture distributions, component D is returned with weighting $\approx$ 0.01 and with a wider standard deviation than for the $\chi^2$ fit  ($\sigma \approx$ 0.5). This discrepancy does not affect the 
other Gaussians fits nor does it alter the conclusions from the 
decomposition. The agreement of the two approaches indicates that the majority of stars are in components A,B,and C and these components are securely fitted.  The metal-poor components D and E involve relatively few stars and their contributions for the different fields can be seen in Table \ref{table:mycomponents}.  They are not the main target of our discussion here.

\subsection{Consequence of mixture model}

The component fitting described above identifies the three major Gaussian components A,B,C which are directly visible in the MDFs, plus the two minor components D and E. We are motivated here not only by these fits but also by the split clump \citep[see][]{Ness2012a} and kinematics which are clearly different for our identified components \citep[see][in preparation]{Ness2012c}. Independent of our MDF component fits, when we examine the kinematics and distribution of the stars partitioned into [Fe/H] intervals of $0.5$ dex from $+0.5$ to $-1.5$ (these intervals correspond approximately to our components A-D), these subgroups show different kinematics and different split clump characteristics.

If we were to adopt two components for our data rather than our best fit representation of 4 or 5 components, the optimal two components comprise a narrow metal rich distribution and a wider metal poor distribution, with $(\mu_1, \sigma_1)$ = (0.15,0.1), and  $(\mu_2, \sigma_2)$ = (--0.3, 0.35).  This is similar to the components identified by \citet{Hill2011} and by \citet{Babusiaux2010} in their $b=-4^\circ$ field. Both argue for a two-component bulge population, comprising a metal-rich population and an old spheroid. The number of components one considers clearly plays a important role in the interpretation.

\section{Properties of the components}

The decomposition of the MDF into the A,B,C (D and E) components allows us to quantify the varying contributions of the metallicity components over $l$ and $b$ and also over distance along the line of sight. 

Figure \ref{fig:components} shows how the relative contributions of the components change with latitude.  At $b = -5^\circ$,  the more metal--rich components A and B dominate, with a smaller contribution from component C and only a few stars from the metal-poor components D and E. At $b = -7.5^\circ$, components B and C are the strongest; component A is clearly weaker than at
$b = -5^\circ$.   At $b = -10^\circ$,  components B and C still dominate and component A is now quite weak. The metal-poor components increase in significance with increasing $|\,b\,|$, and more metal-poor stars ([Fe/H] $< -1$) appear at higher latitudes. 

In preparation for the discussion in Section 7 about the nature of the metallicity components seen in the inner Galaxy, we show MDFs for intervals of $R_{\rm G}$ which demonstrate the changing contribution fraction of the different components with distance from the Galactic center. Figure \ref{fig:bfive} shows the MDF components in various distance intervals as indicated below the individual panels for $b=-5^\circ$, and Figure \ref{fig:dist10} is similar for $b=-10^\circ$. The data for each latitude zone are combined as before for $l = \pm1 5^\circ$.

 \begin{figure}
  \centering    
        \vspace{-5pt}
   \subfloat[\tiny{$|R_{\rm G}|$ $<$0.75 kpc }]{\label{fig:gull}\includegraphics[width=0.25\textwidth]{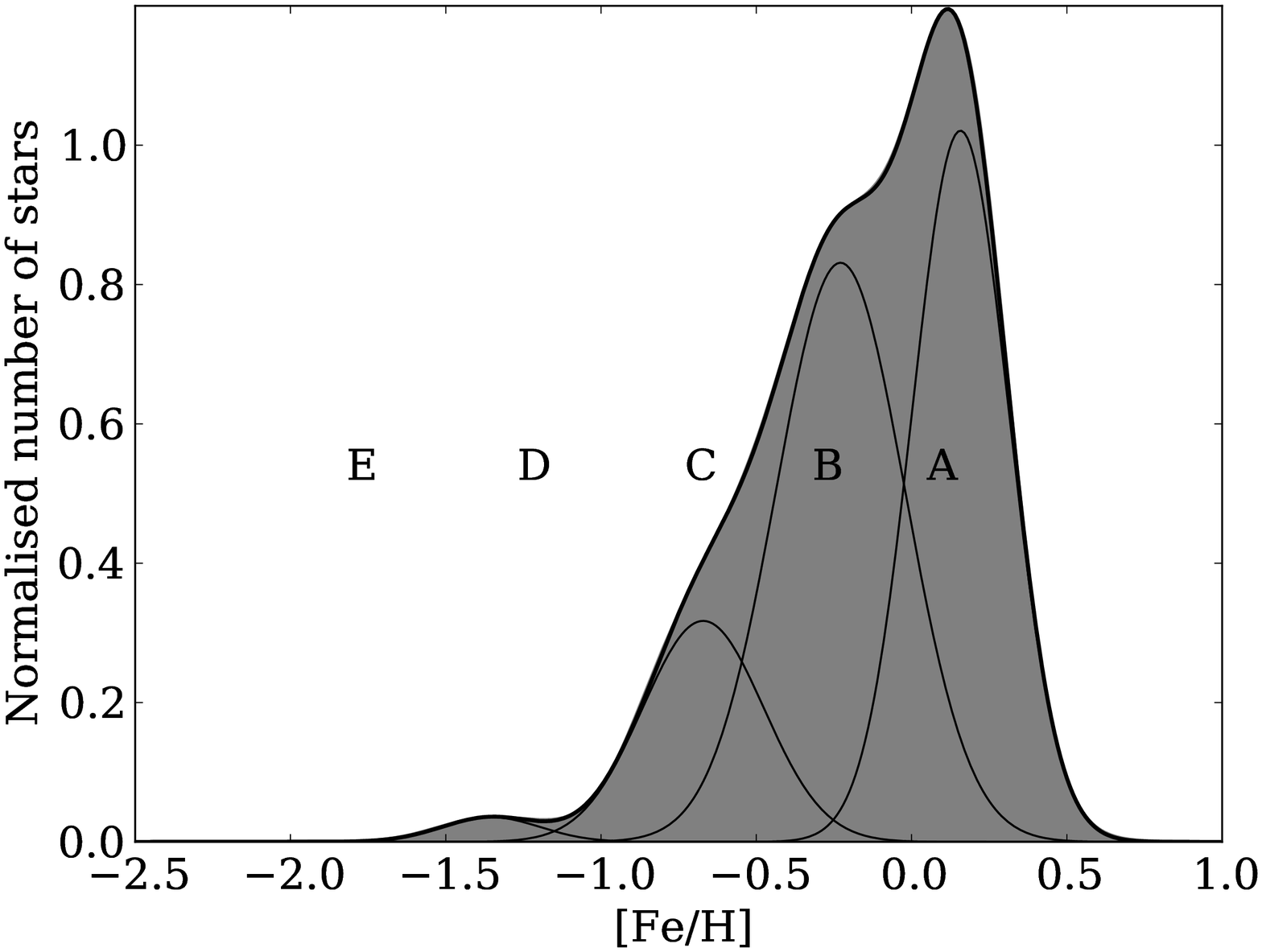}}   
 \subfloat[\tiny{$|R_{\rm G}|$ $<$ 1.5 kpc}]{\label{fig:gull}\includegraphics[width=0.25\textwidth]{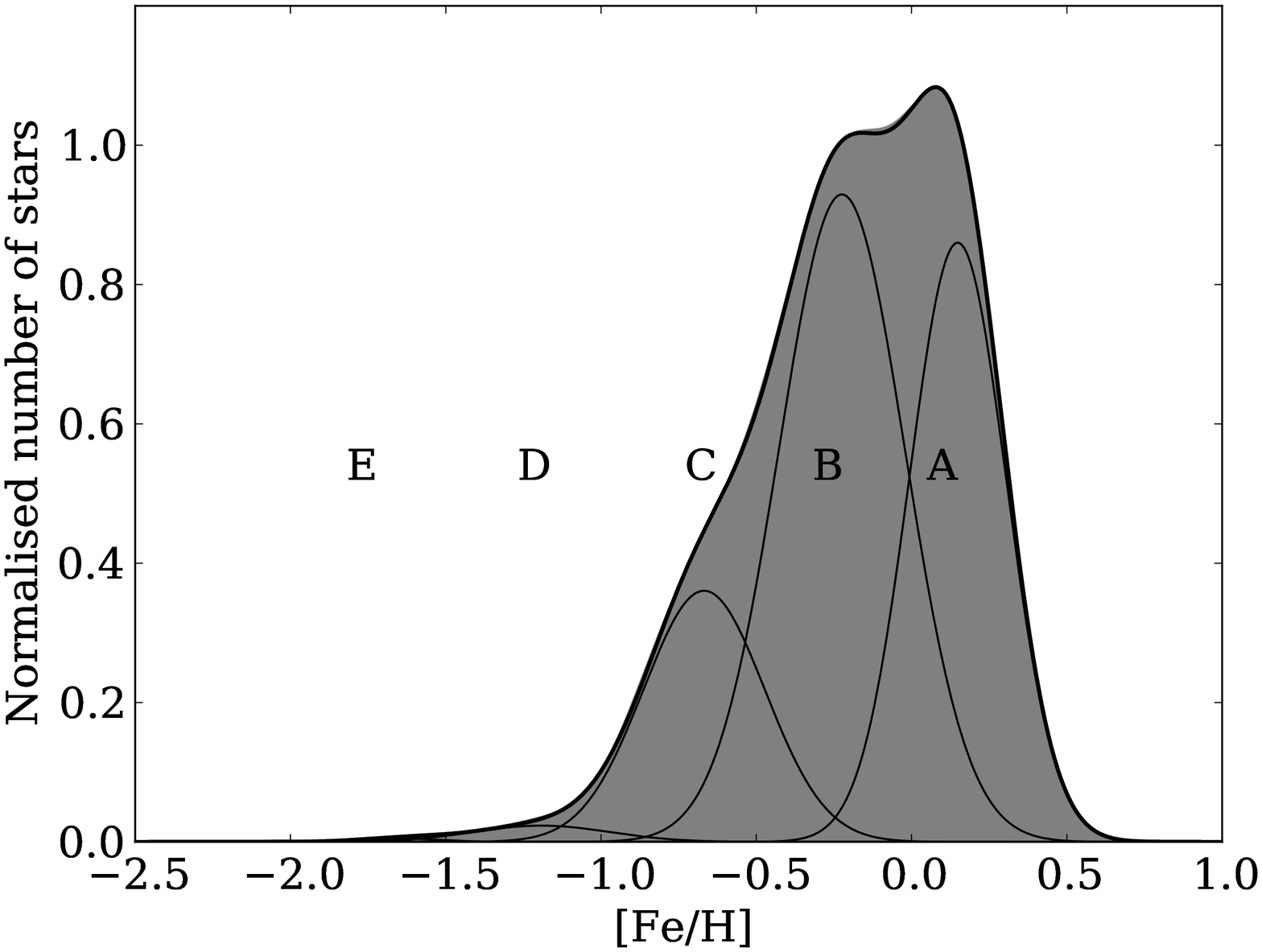}}     \\    
 \vspace{-5pt}
                         \subfloat[\tiny{1.5 kpc $<$ $|R_{\rm G}| <$ 3 kpc}]{\label{fig:gull}\includegraphics[width=0.25\textwidth]{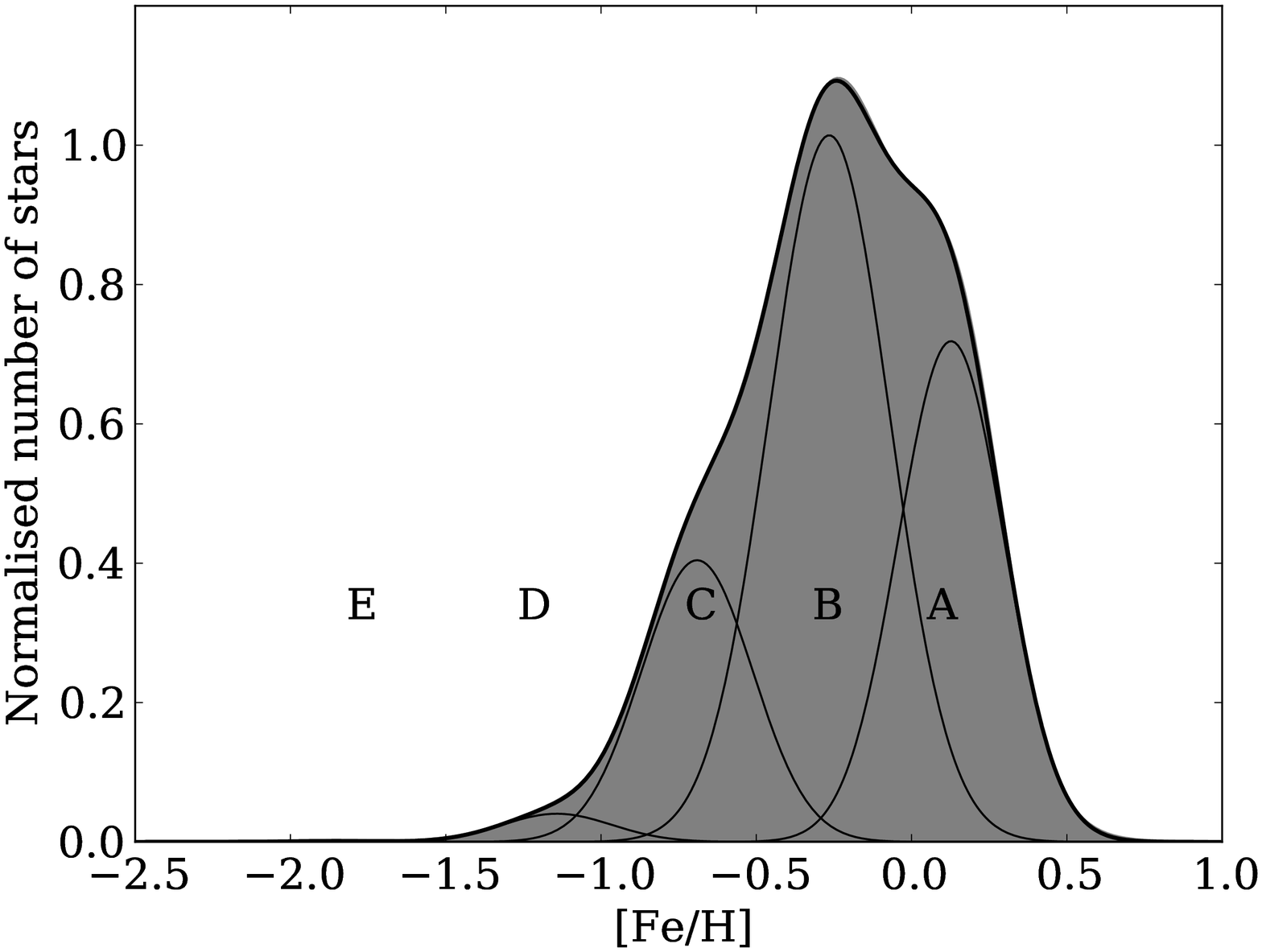}}    
               \subfloat[\tiny{3 kpc $<$ $|R_{\rm G}|$ $<$ 4.5 kpc}]{\label{fig:gull}\includegraphics[width=0.25\textwidth]{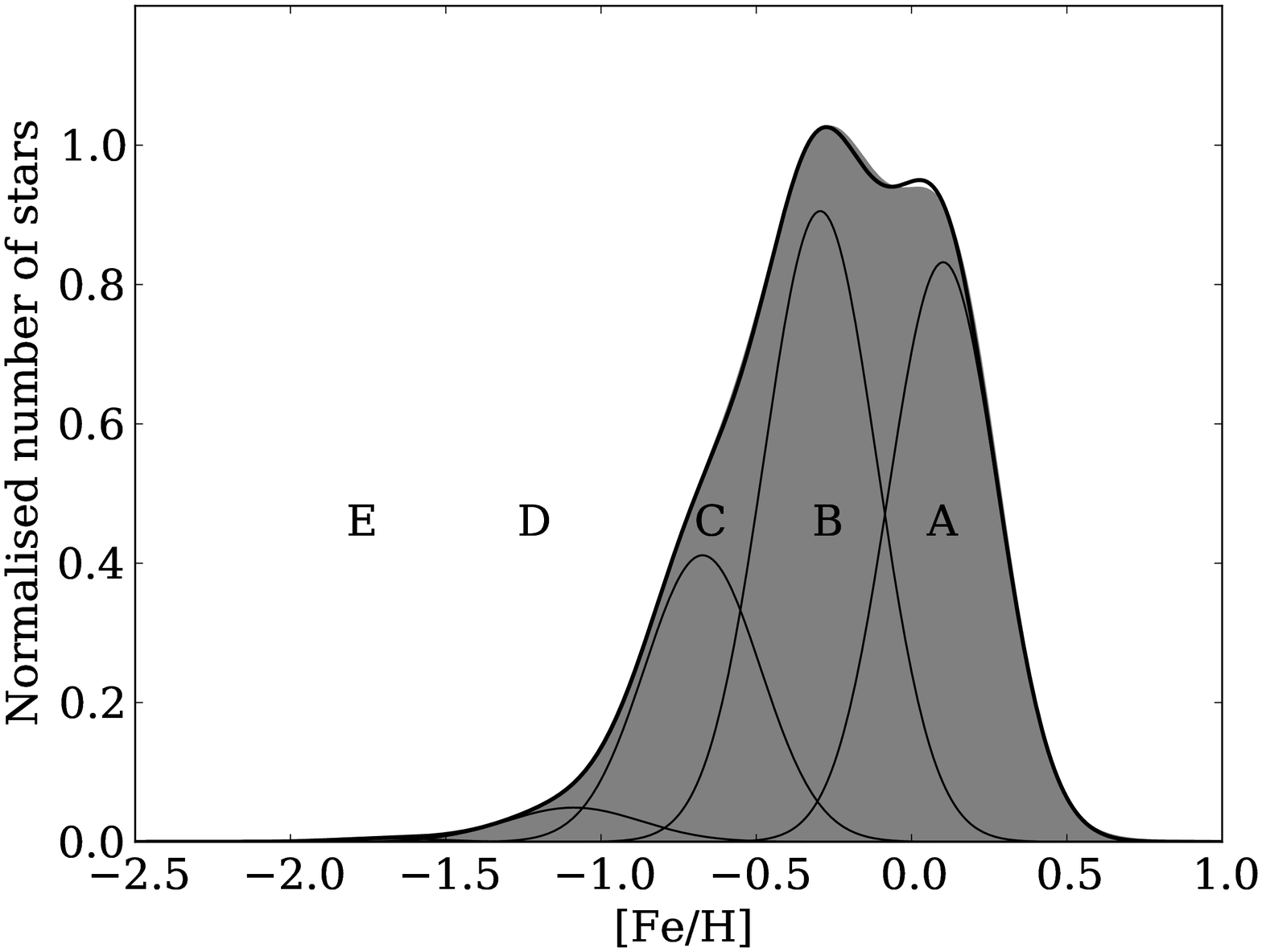}}  \\
                          \vspace{-5pt}    
                      \subfloat[\tiny{4.5 kpc $<$ $|R_{\rm G_{near}}|$ $<$ 6 kpc}]{\label{fig:gull}\includegraphics[width=0.25\textwidth]{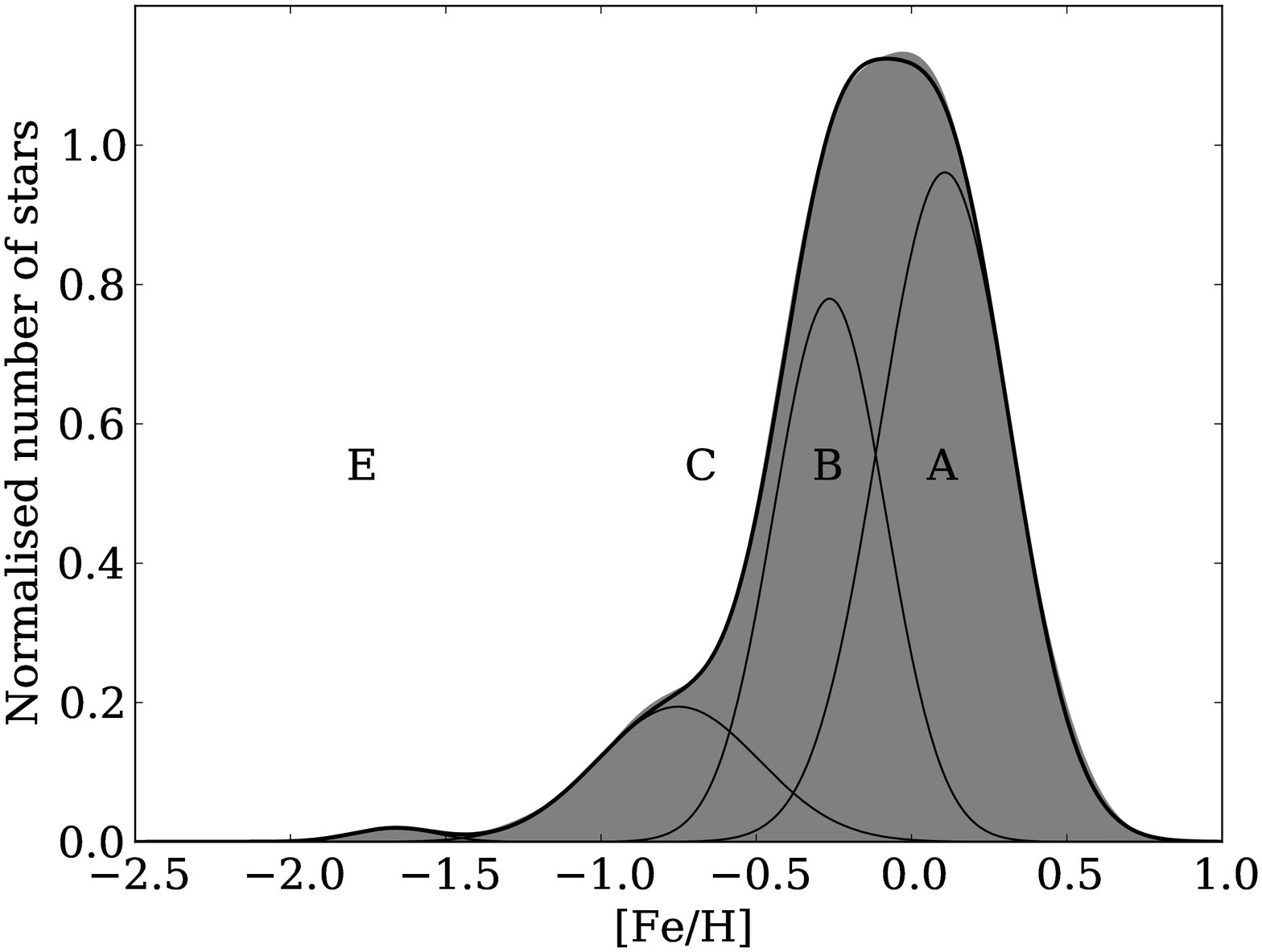}}   
                             \subfloat[\tiny{5 kpc $<$ $|R_{\rm G_{near}}|$ $<$ 7 kpc}]{\label{fig:gull}\includegraphics[width=0.25\textwidth]{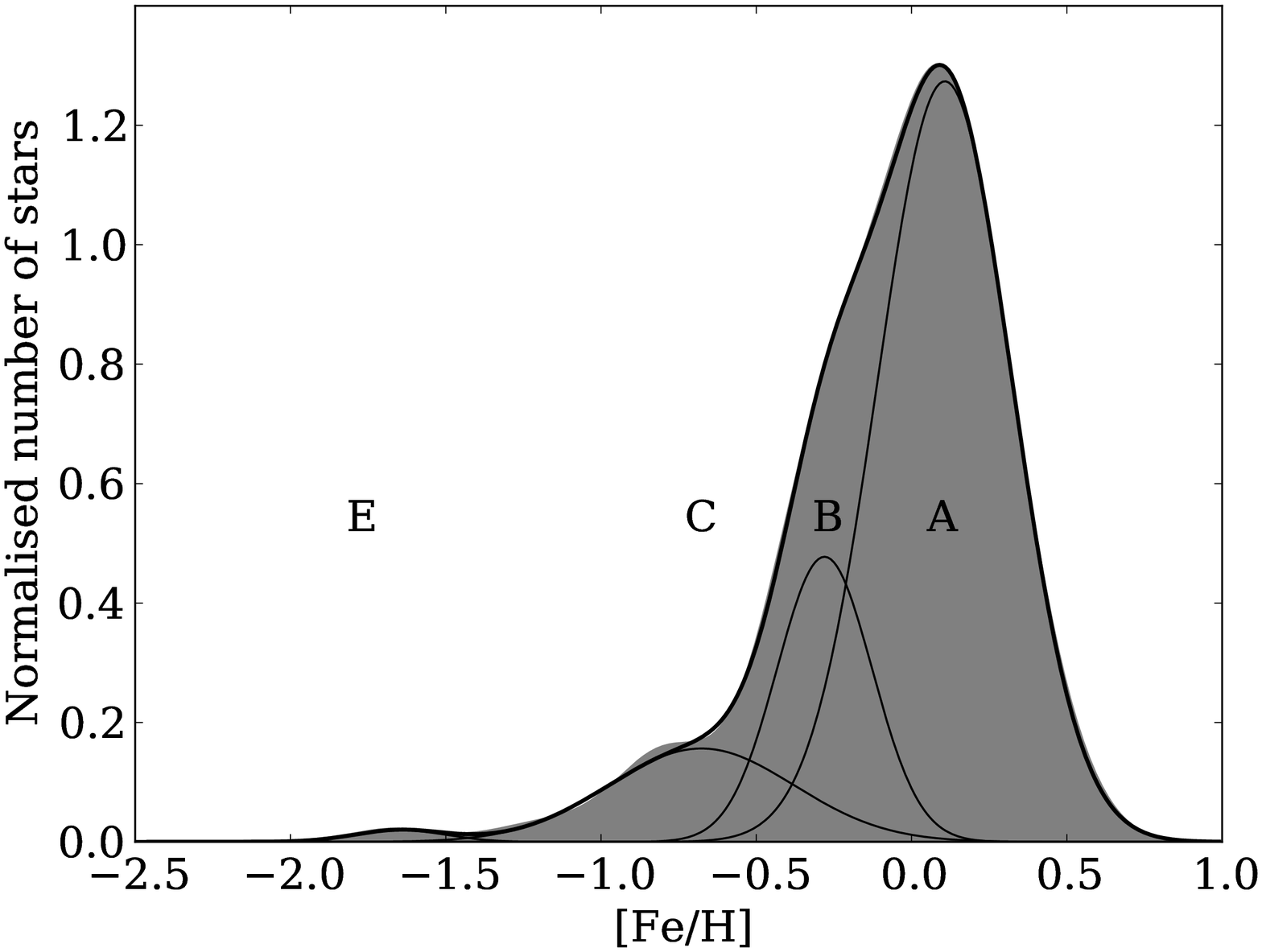}} \\
   \vspace{-5pt}   
\subfloat[\tiny{4.5 kpc $<$ $|R_{\rm G_{far}}|$ $<$ 6 kpc}]{\label{fig:gull}\includegraphics[width=0.25\textwidth]{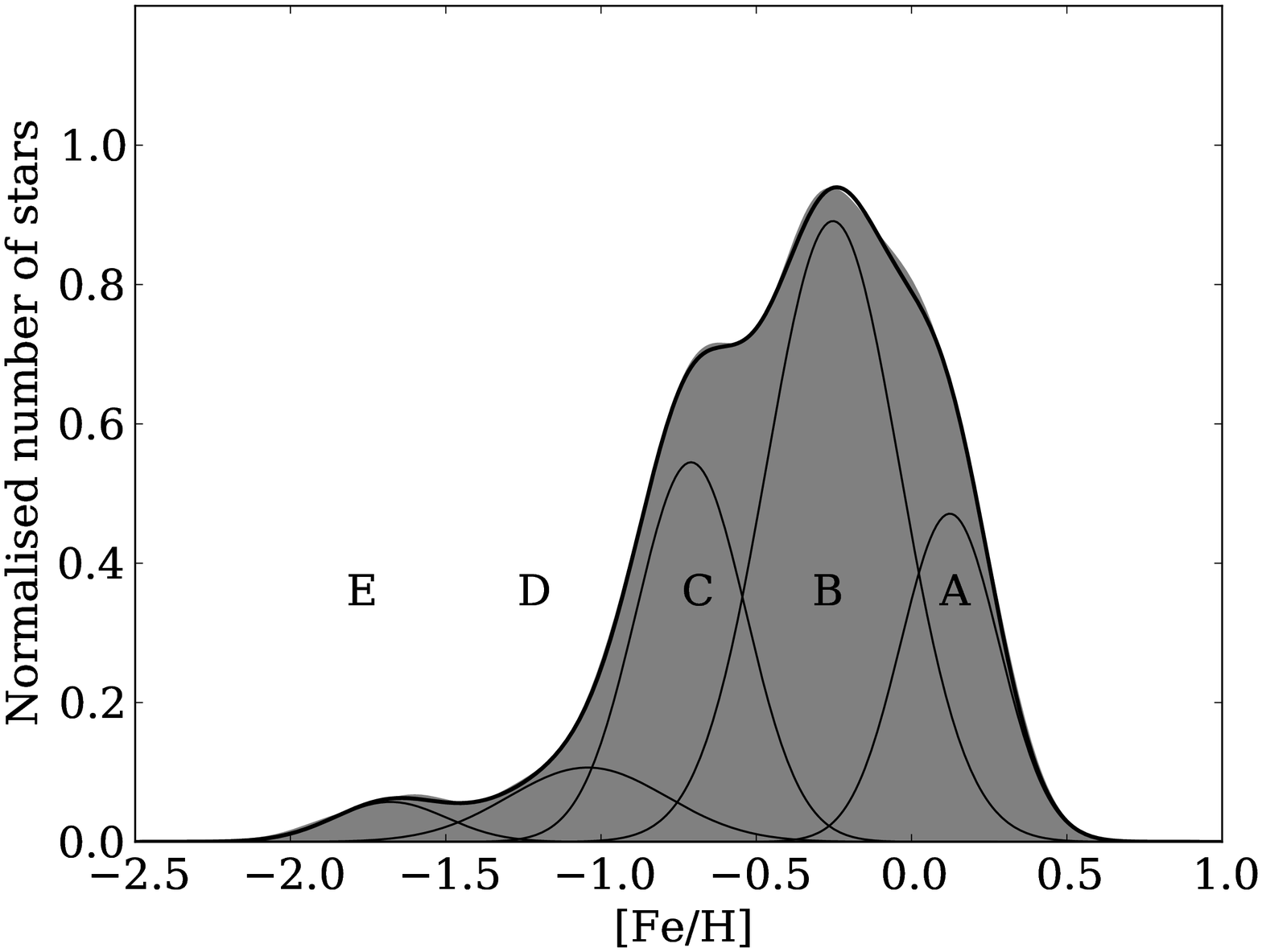}}      
  \subfloat[\tiny{5 kpc $<$ $|R_{\rm G_{far}}|$ $<$ 7 kpc}]{\label{fig:gull}\includegraphics[width=0.25\textwidth]{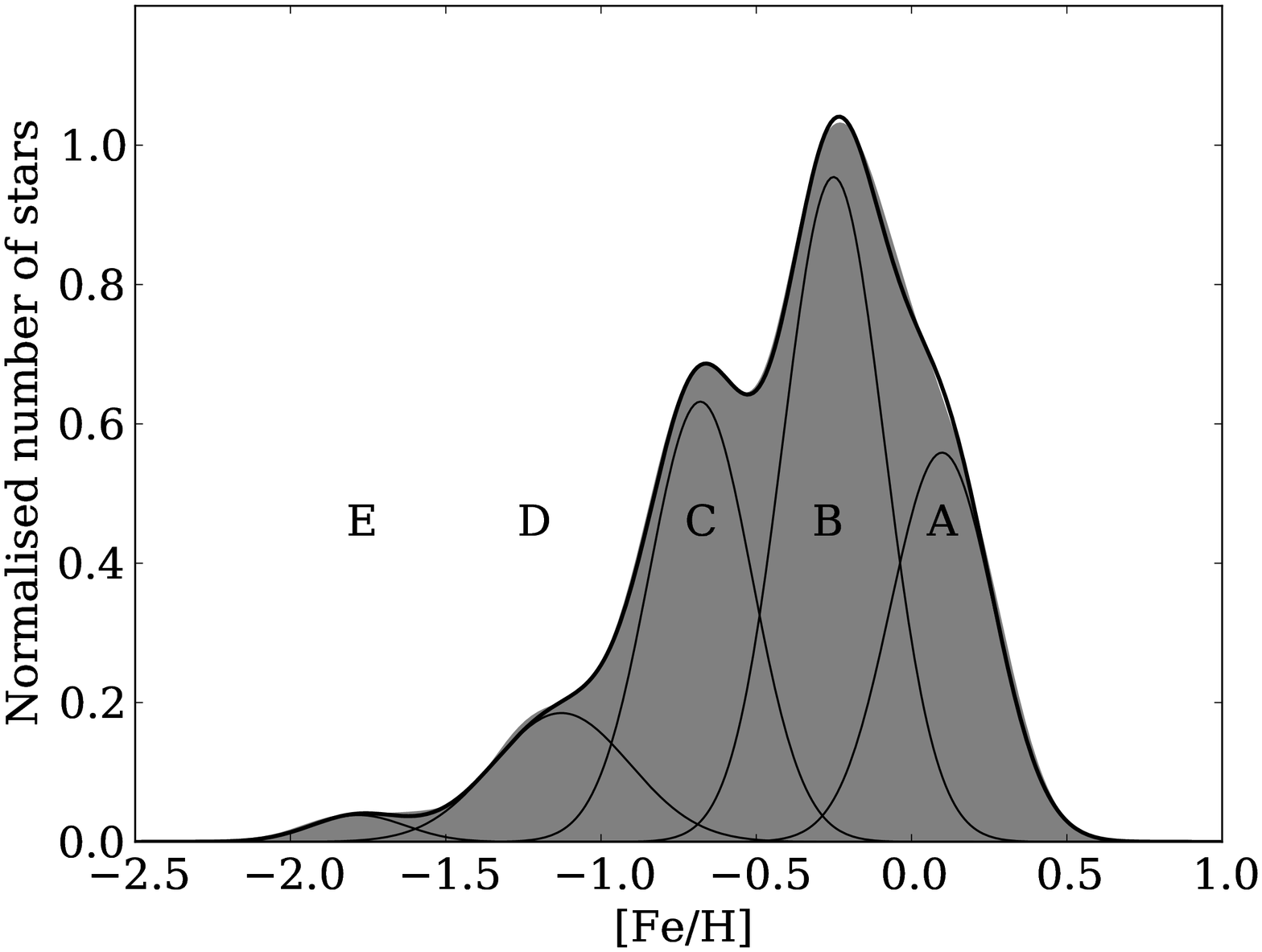}} \\
\subfloat[\tiny{8 kpc $<$ $|R_{\rm G_{far}}|$ $<$ 10 kpc}]{\label{fig:gull}\includegraphics[width=0.25\textwidth]{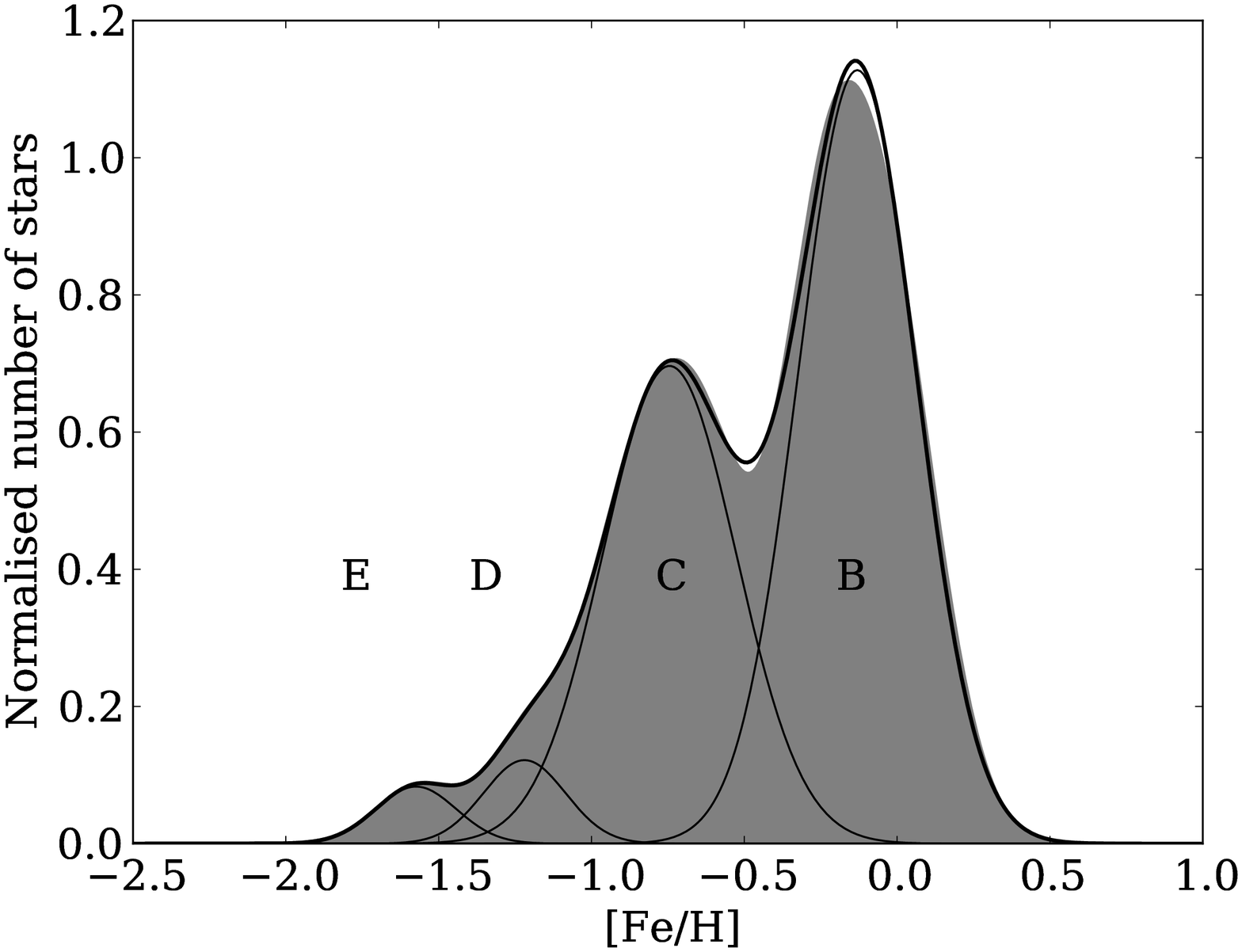}}       
\caption[MDF decompositions for different radial intervals at $b = -5^\circ$]{MDF decompositions for different radial intervals at $b = -5^\circ$.  The number of stars in panels a--i are 300, 1100,  2300, 2000,  600, 200, 120 and 100, 70, respectively.}  
\label{fig:bfive}\label{fig:bfive}
     \end{figure}
     
 \begin{figure} 
  \centering 
          \vspace{-5pt}
                 \subfloat[\tiny{$|R_{\rm G}|$ $<$ 0.75 kpc}]{\label{fig:gull}\includegraphics[width=0.25\textwidth]{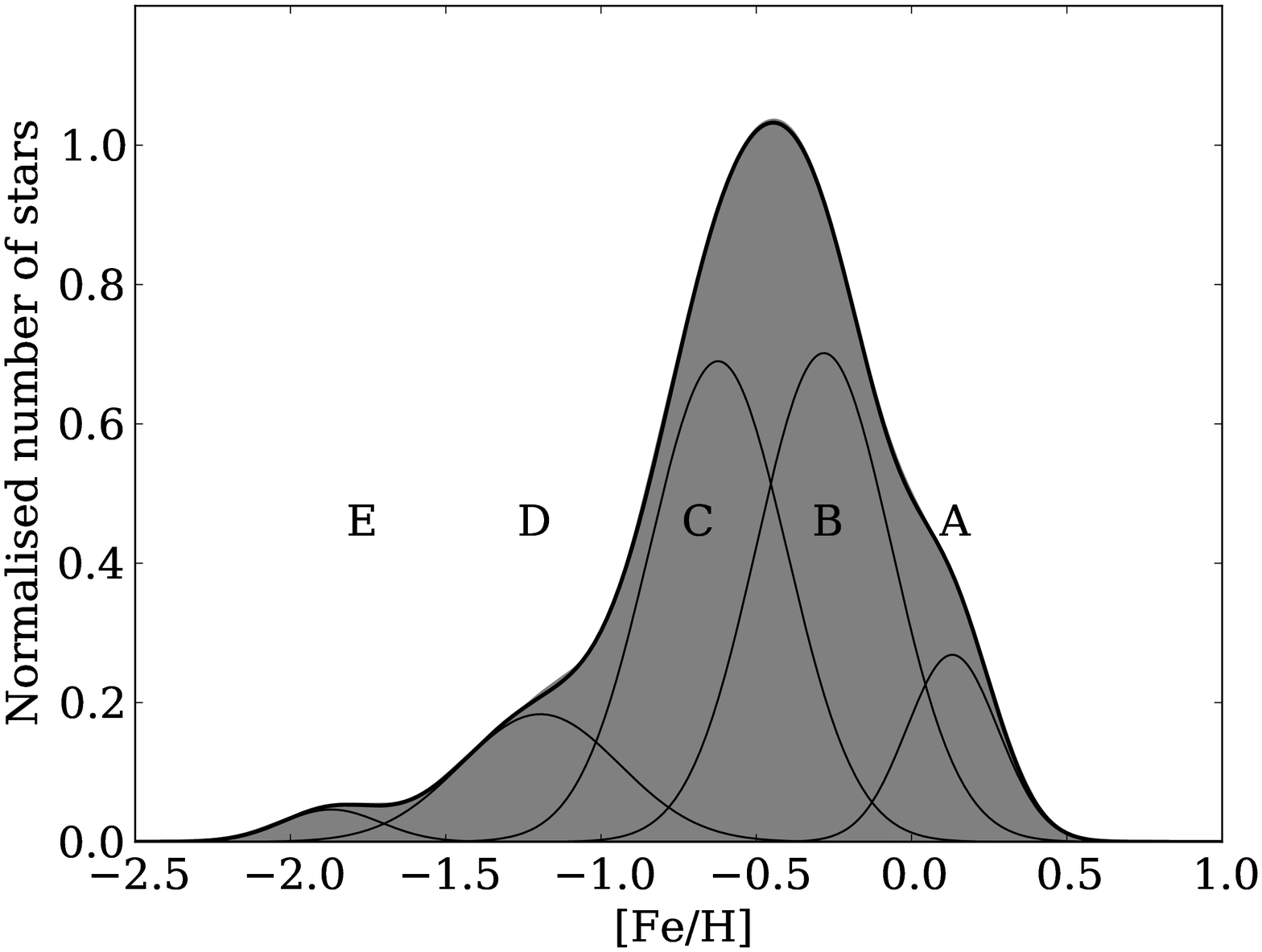}}  
                       \subfloat[\tiny{$|R_{\rm G}|$ $<$ 1.5 kpc}]{\label{fig:gull}\includegraphics[width=0.25\textwidth]{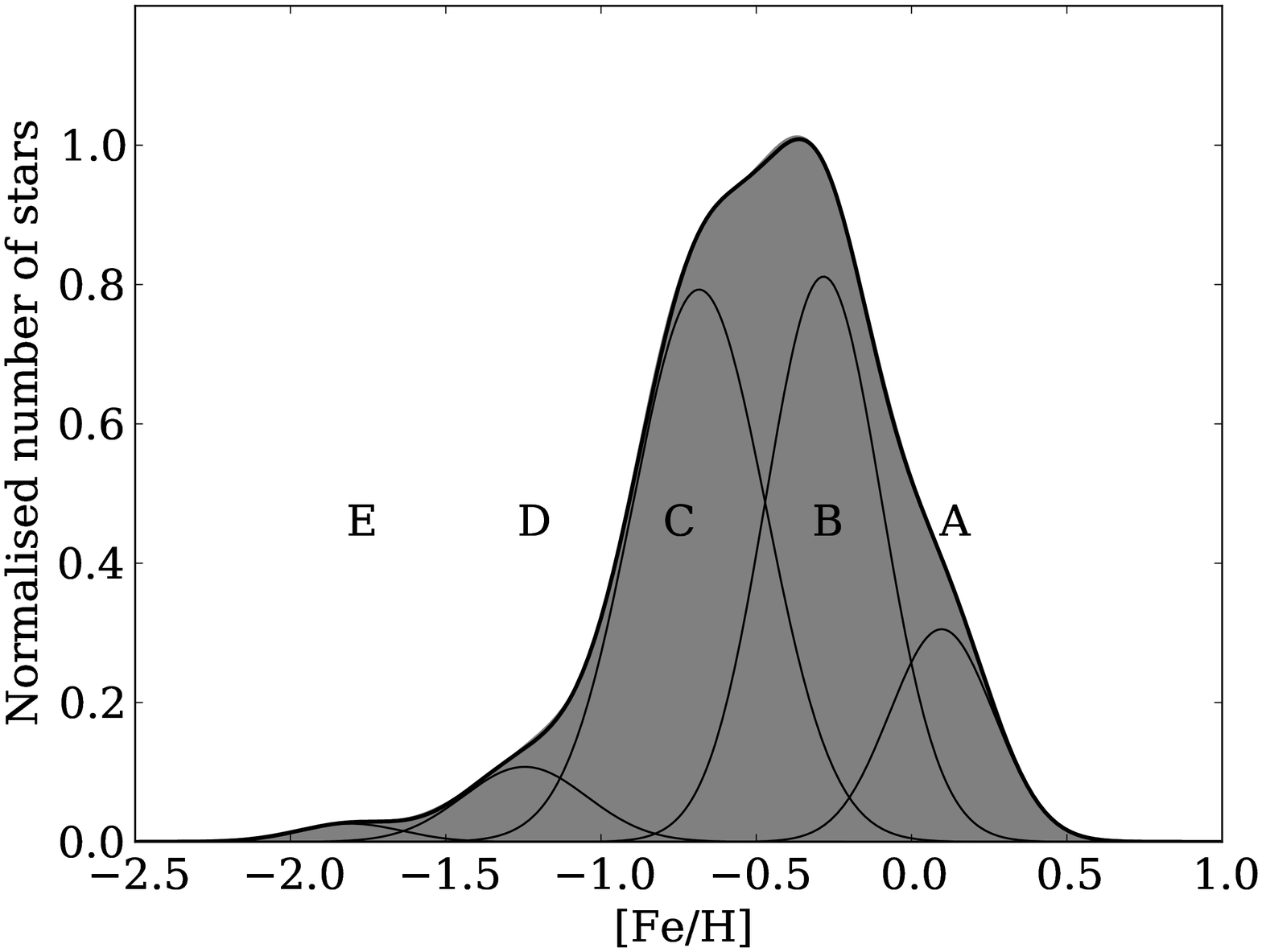}}         \\ 
      \vspace{-5pt}
                         \subfloat[\tiny{1.5 kpc $<$ $|R_{\rm G}|$ $<$ 3 kpc}]{\label{fig:gull}\includegraphics[width=0.25\textwidth]{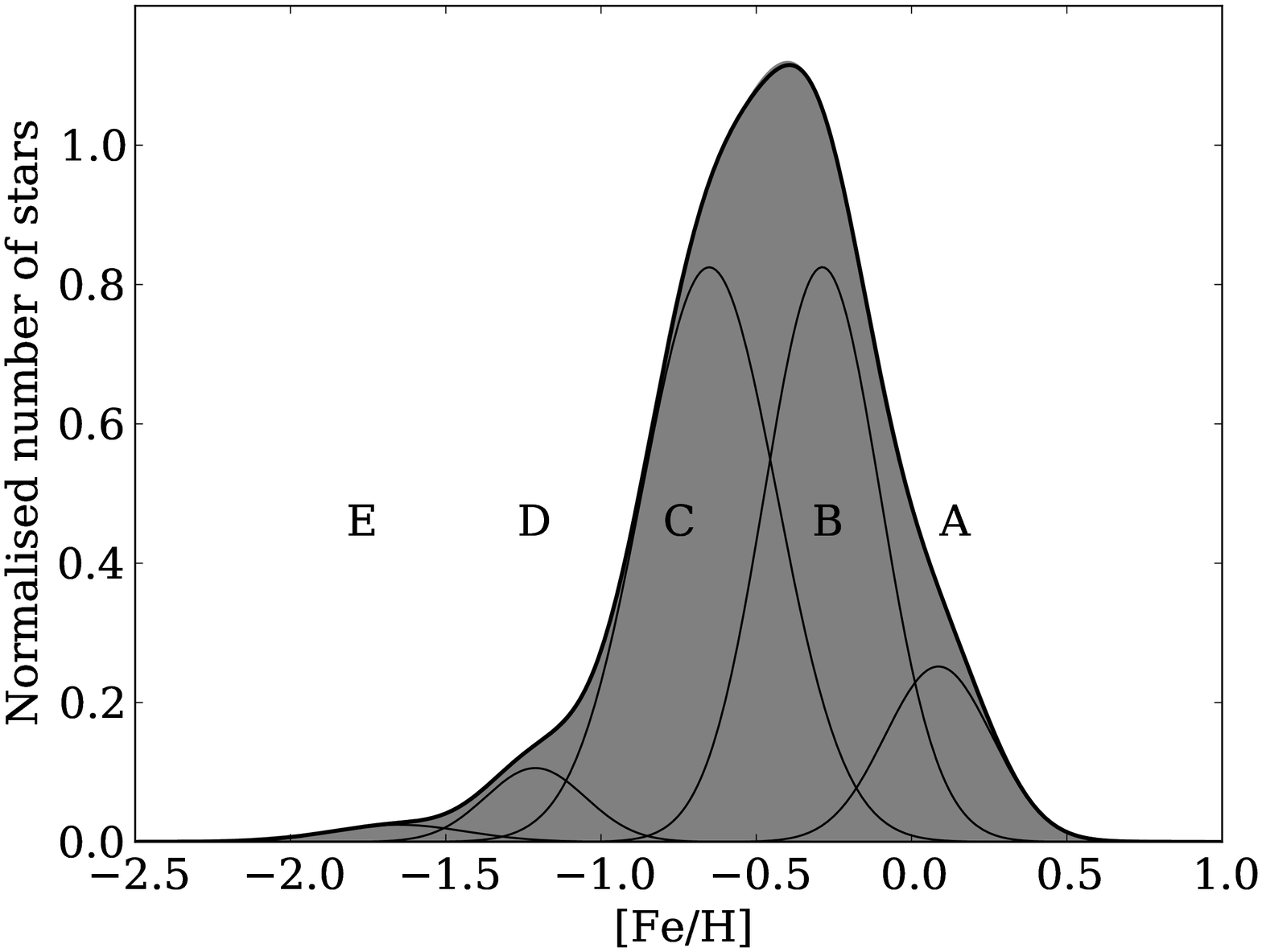}}   
        \subfloat[\tiny{3 kpc $<$ $|R_{\rm G}|$ $<$ 4.5 kpc}]{\label{fig:gull}\includegraphics[width=0.25\textwidth]{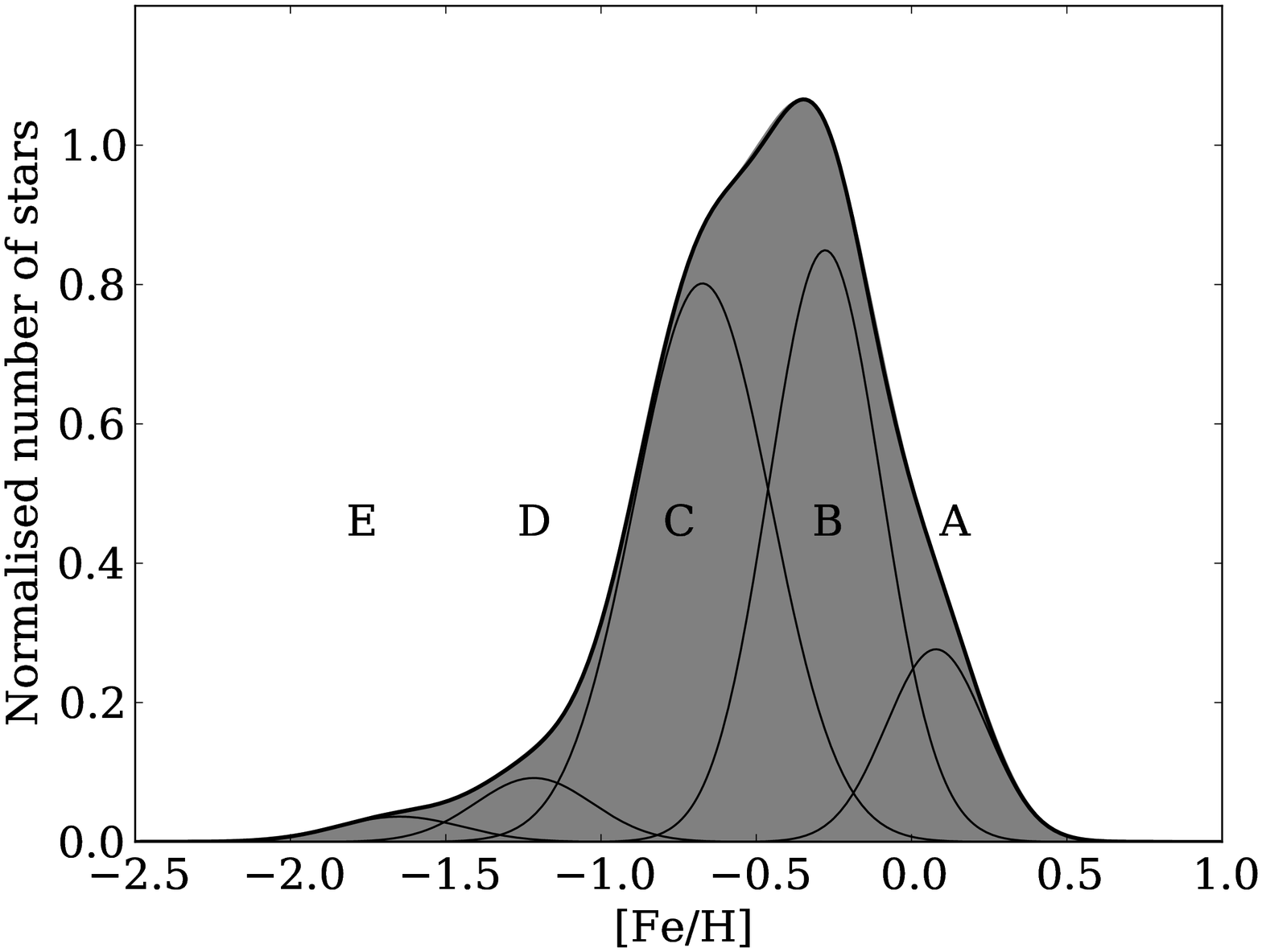}}  \\
     \vspace{-5pt} 
    \subfloat[\tiny{4.5 kpc $<$ $|R_{\rm G_{near}}|$ $<$ 6 kpc}]{\label{fig:gull}\includegraphics[width=0.25\textwidth]{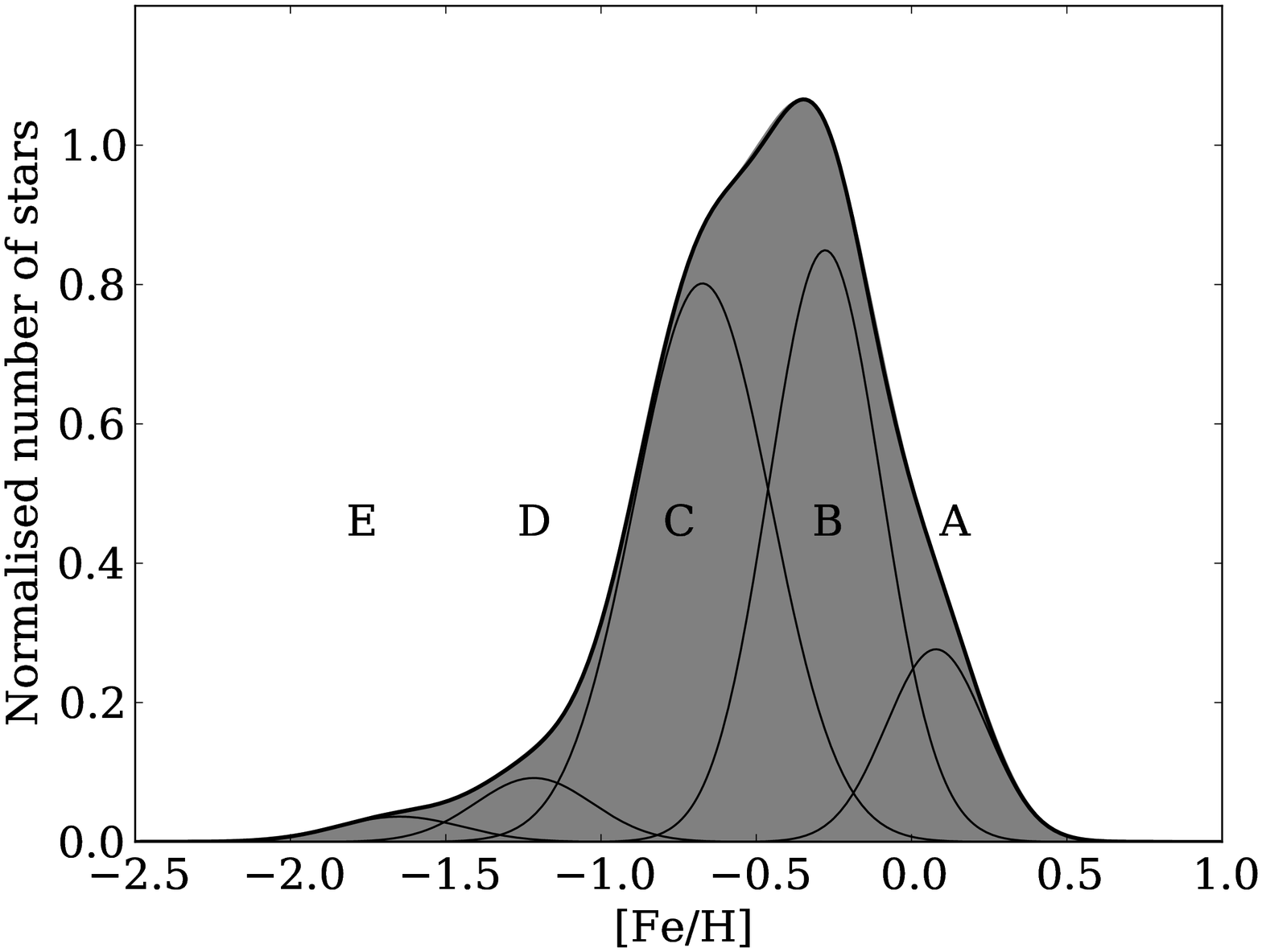}}   
       \subfloat[\tiny{5 kpc $<$ $|R_{\rm G_{near}}|$ $<$ 7 kpc}]{\label{fig:gull}\includegraphics[width=0.25\textwidth]{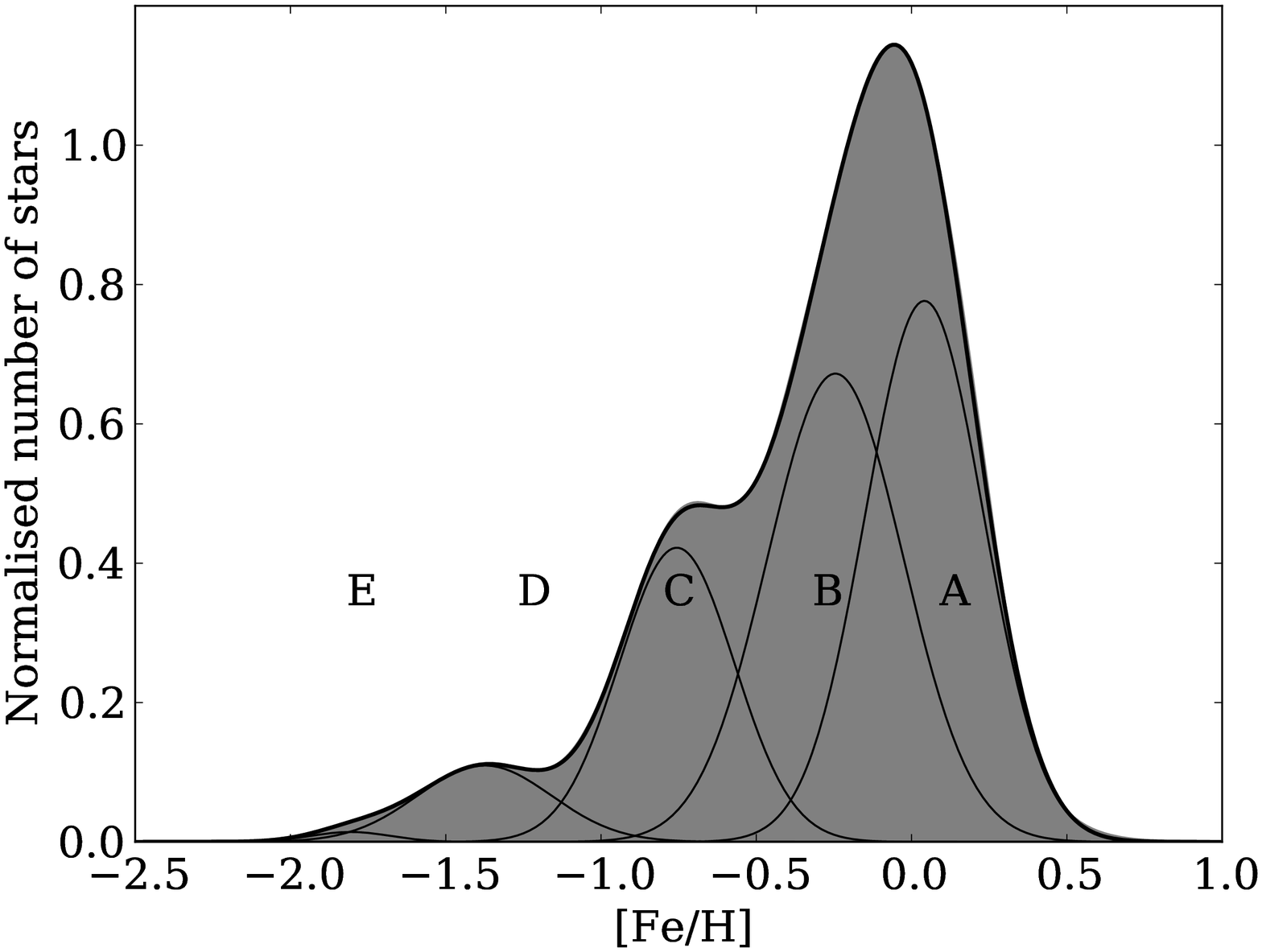}}  
    \vspace{-5pt}
        \subfloat[\tiny{4.5 kpc $<$ $|R_{\rm G_{far}}|$ $<$ 6 kpc}]{\label{fig:gull}\includegraphics[width=0.25\textwidth]{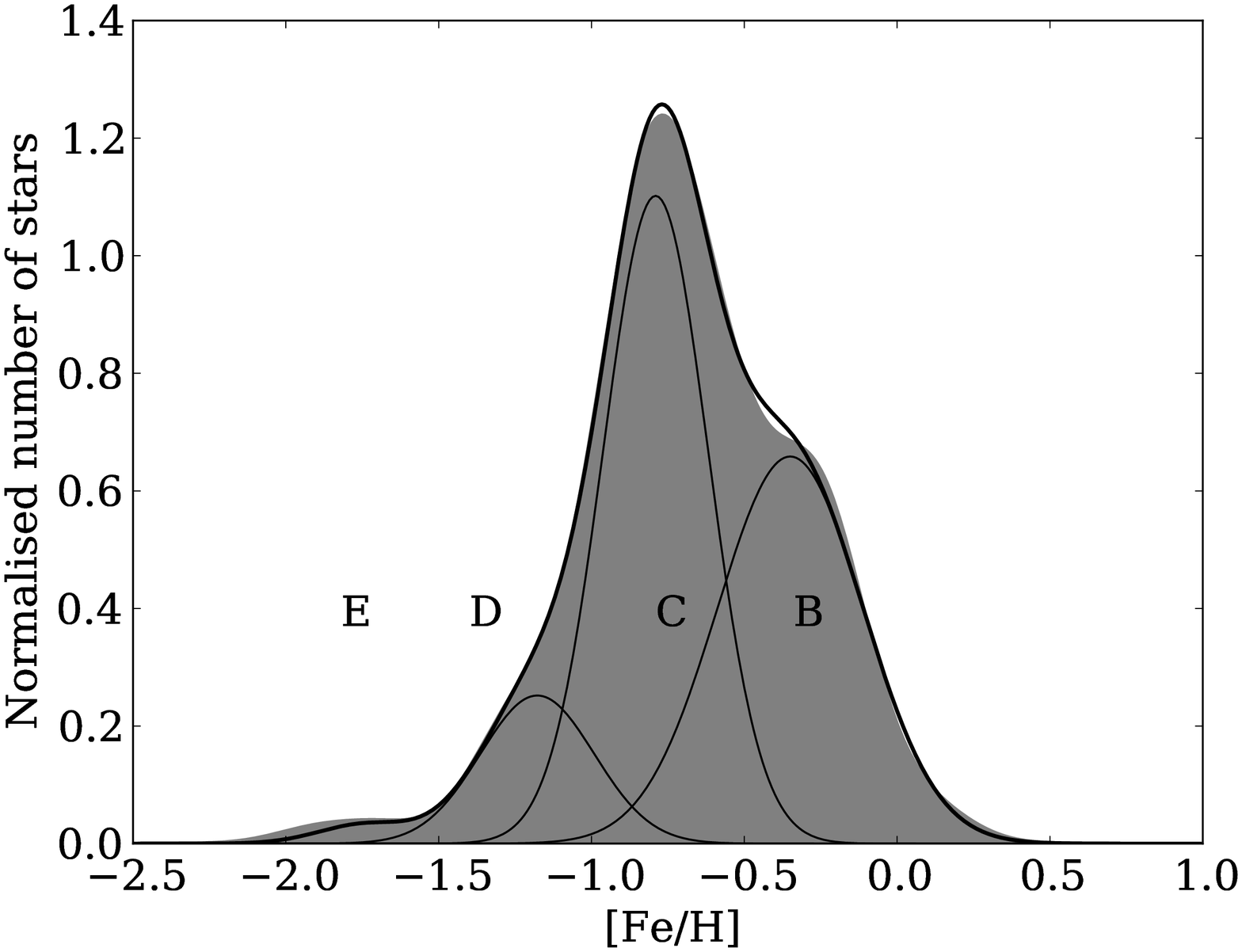}}  
 \subfloat[\tiny{5 kpc $<$ $|R_{\rm G_{far}}|$ $<$ 7 kpc}]{\label{fig:gull}\includegraphics[width=0.25\textwidth]{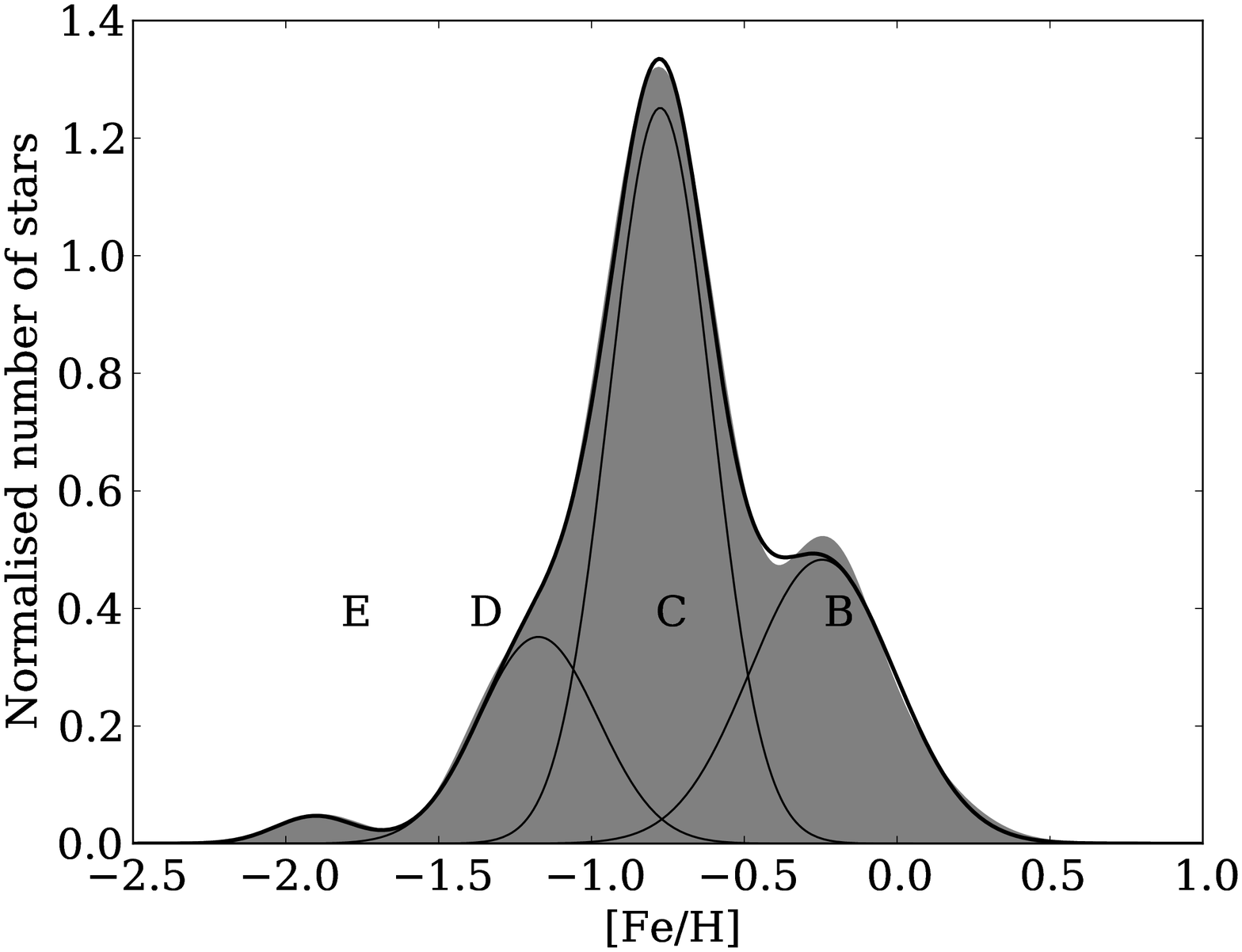}}  \\
  \subfloat[\tiny{8 kpc $<$ $|R_{\rm G_{far}}|$ $<$ 10 kpc}]{\label{fig:gull}\includegraphics[width=0.25\textwidth]{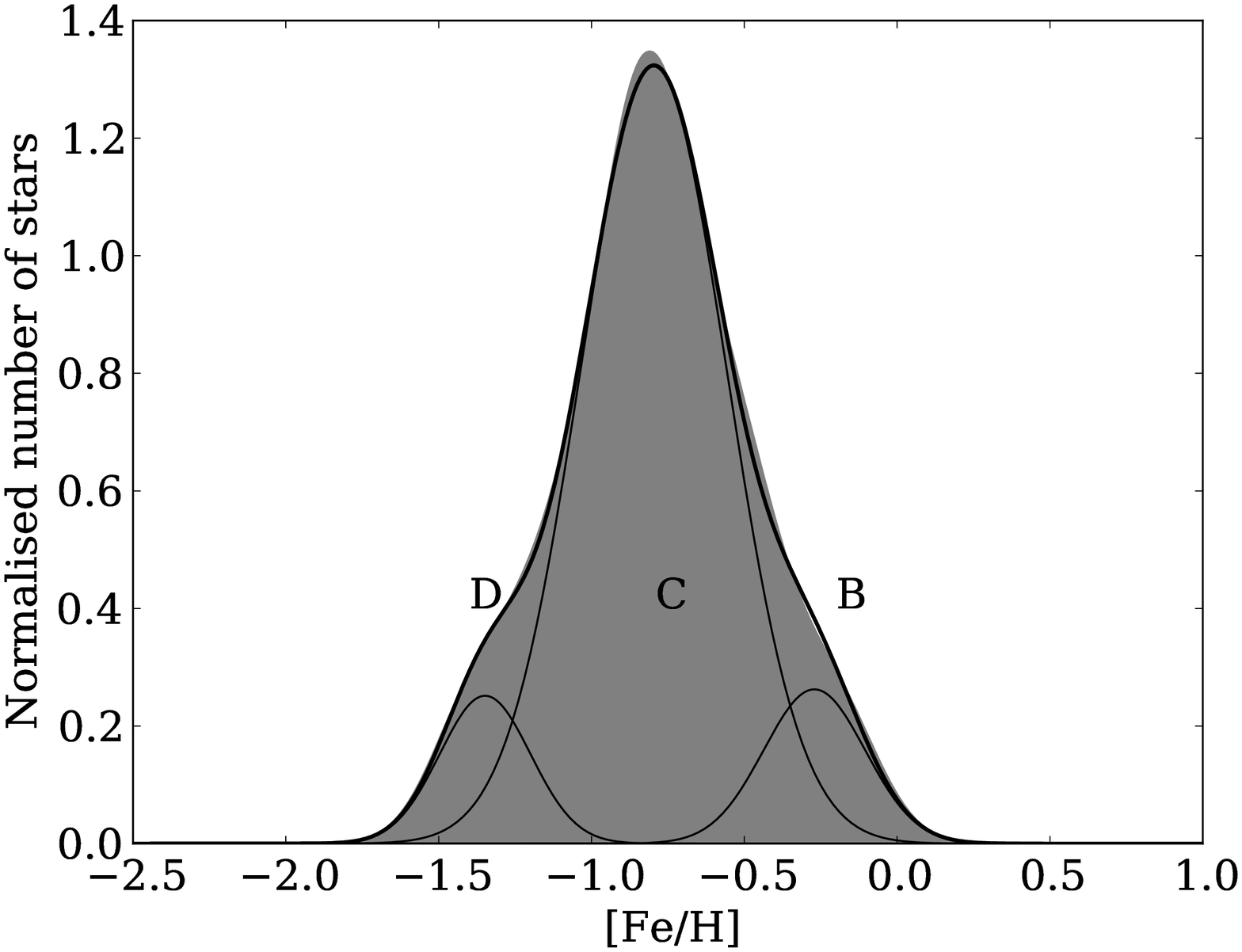}}  
                    \caption[MDF decompositions for different radial intervals at $b = -10^\circ$]{MDF decompositions for different radial intervals at $b = -10^\circ$.  The number of stars in panels a--i are 180, 850, 2450, 1700, 250, 150, 200 and 120 and 50, respectively.}
          \label{fig:dist10}
     \end{figure}

Figures \ref{fig:bfive} and \ref{fig:dist10} show that the components A -- E are present at both latitudes out to $|R_{\rm G}| \approx$ 5 kpc. The widths of the individual components are approximately unchanged, but their contribution fraction changes with latitude and $R_{\rm G}$. Examined separately, the near and far sides along the line of sight show comparable MDFs out $|$\rgc$|$ $<$ 4.5 kpc. There are fewer metal rich stars observed on the near side however and a radial change in the morphology of the MDF occurs for $|R_{\rm G}|$ $>$ 4.5 kpc: at these larger radii, components A and B merge (see for example components A and B in panels $e$ and $f$ of Figure \ref{fig:bfive}). Note that this conclusion depends on the adopted metallicity gradient for component A as discussed below. The distance cuts show that within \rgc $<$ 4.5 kpc the two components A and B are necessary to fit to the stars with $\approx$ [Fe/H]  $>$ --0.5. However, outside the bulge, (for a radius of \rgc\ $>$ 8kpc at $b=-5^\circ$, and  \rgc\ $>$ 4.5 kpc at $b=-10^\circ$), a single component is sufficient to fit to the  stars with [Fe/H] $>$ --0.5 (see Figures \ref{fig:bfive} (i) and \ref{fig:dist10} (i)).

At $b=-5^\circ$, component B dominates between 0.75 to 3.0 kpc but inside of this region the most metal-rich component A is marginally strongest. Out to 4.5 kpc, component C is no more than about 50\% of that of A. At $b =-10^\circ$, component B is the strongest out to $|R_{\rm G}| = 6$ kpc  and component C is a larger fraction of component B out to $4.5$ kpc compared with the lower latitude. Within \rgc\ $<$ 4.5 kpc, component A is relatively weak, at no more than about $30$\% of the strength of C.

At $b=-10^\circ$, component A which appears as a small fraction in the inner region is no longer present on the far side of the bulge (see Figure \ref{fig:dist10} (g) and (h). At both latitudes, component B appears in the largest fraction on the far side of the bulge and component A is the largest fraction on the near side of the bulge outside of $|$\rgc$|$ $>$ 4.5 kpc. 

Components D and E include only a small number of stars. At $b=-5^\circ$, their fraction increases with Galactic radius. The fraction of very metal-poor stars (in D and E) is higher at $b=-10^\circ$, particularly near the Galactic centre ($|R_{\rm G}| < 0.75$ kpc). \\

\subsection{Comparison with disk fields}

One of our fields lies out in the disk at $(l, b) = (-31^\circ , -5^\circ)$, for which the minimum value of $|$\rgc$|$ is about $4$ kpc. The same components A,B,C are seen in this disk field (Figure \ref{fig:components_one}). From Figures  \ref{fig:bfive} and \ref{fig:dist10}, these three components are also seen in the radial intervals extending out into the disk at smaller $|\,l\,|$.  The presence of the same three components A--C in the bulge and the inner disk makes sense.  In the inner disk, these components are present in their undisturbed form, as disk stars, while in the bulge fields the stars of these components have been incorporated dynamically into the bulge.

With increasing distance from the Galactic center, we note again the change in component A relative to components B and C. For $|R_{\rm G}| >$ 4.5 kpc, components A and B merge and, at $b=-10^\circ$, four components (B, C, D and E) are sufficient to fit the data, assuming a radial gradient in the mean abundance for A. At $b=-5^\circ$, where components A and B merge outside of the inner bulge region, component A is dominant close to the Sun. On the far side at \rgc\ $>$ 4.5 kpc, component B is dominant and components A and C is roughly equal out to \rgc\ = 7.0 kpc. At $b=-10^\circ$, component A is similarly the most significant near the Sun and at this higher latitude on the far side of the bulge, for \rgc\ $>$ 4.5 kpc, component C is the strongest. Table \ref{table:ACcuts} shows the relative heights below the plane for each radial cut.  We note that stars with a distance of about 6 kpc on the far side of the bulge at $b=-10^\circ$ are located at $z-$heights of 2.4 kpc below the plane. At $b=-10^\circ$ and $|R_{\rm G}| > 7$ kpc the stars are on the far side of the bulge at mean $z-$heights $>$ $2.6$ kpc. Therefore it is not surprising that the component C (with its thick-disk-like metallicity range) should dominate on the far side of the bulge. 

   \begin{figure}
  \centering
    \subfloat[$l=-31^\circ,b=-5^\circ$]{\label{fig:gull}\includegraphics[width=0.25\textwidth]{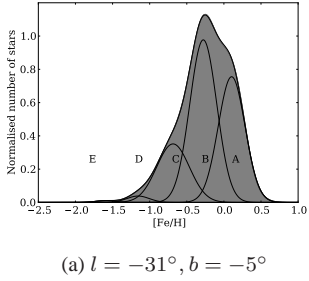}}   \\
  \caption{ MDF for the field $(l, b) = (-31^\circ, -5^\circ)$ from 440 stars with $|R_{\rm G}|$ =  4.0--4.9 kpc. The Gaussian decomposition of the MDFs is shown. We see the same components out in the disk as in the bulge fields}
  \label{fig:components_one}
\end{figure}

\subsection{The metallicity gradient with distance $|z|$ from the plane}

Along the minor axis, the mean [Fe/H] for some of the individual components of the MDF shows a weak decrease with latitude. From $b=-5^\circ$ to $b=-10^\circ$, the total decrease in mean abundance of components A and B is $-0.06 \pm 0.04$ dex and $-0.06 \pm 0.03$ dex respectively. The mean abundance of component C shows no significant decrease with latitude on the minor axis, although the longitude-integrated MDF does show a small decrease in the mean of component C with latitude (see Table \ref{table:mycomponents}) of $-0.03 \pm 0.02$ dex.  The numbers of stars with [Fe/H]$ < -1$ is small (10, 29 and 55 stars at $b=-5^\circ, -7.5^\circ$ and $-10^\circ$ respectively along the minor axis) and we cannot evaluate the abundance trends with height for the metal-poor components D and E.  

A small decrease in the mean abundances of components A and B for the integrated MDFs ($l = \pm 15^\circ$) is also measured for these components. This small change in the mean abundances with latitude, and the large change of the contribution fractions of the components, indicates that the overall observed abundance gradient measured by Zoccali et al. (2008) of about $-0.6$ dex kpc$^{-1}$ from $b = -4^\circ$ to $b = -12^\circ$ comes mainly from the changing proportions of the MDF components with latitude, as shown in Figure \ref{fig:components}, rather than from gradients in the individual components. 

From $b=-5^\circ$ to $b=-10^\circ$, the gradients for A and B with height below the plane calculated from the minor axis MDFs are $-0.08 \pm 0.04$ dex kpc$^{-1}$. No gradient in C is seen along the minor axis but the integrated MDFs show a small gradient of $-0.04 \pm 0.02$ dex kpc$^{-1}$.

\subsection{The radial metallicity gradients}

The mean abundances for components A, B and C as a function of \rgc\ are given in Table \ref{table:ACcuts} with their corresponding errors. A weak radial gradient ($\approx -0.02$ dex kpc$^{-1}$) is evident in the mean abundance of the most metal-rich component A. We note that this gradient may come partly from a radial gradient and partly from a $z-$gradient in these zones of constant latitude. This weak abundance gradient is shallower than the gradient seen in the cepheids \citep{Luck2011} and young (age $\le$ 1 Gyr) open clusters \citep{Yong2012}. However, given an additional negative gradient with height from the plane, this measured gradient will be smaller than the true value. This is because, with increasing distance along the line of sight, the stars sampled are further from the plane and will be relatively more metal poor. 

 At \rgc\ $>$ 8 kpc at low latitudes and \rgc\ $>$ 4.5 kpc at high latitudes, the most metal rich fraction of stars is no longer present and the MDF can be described by two main components: the more metal rich component is component B. This component represents the most metal rich stars that we sample at large heights from the plane; we note that at $b=-5^\circ$ the $|$z$|$-height at 6 kpc on the far side of the bulge  is 1.2 kpc.  From the inner region, component B shows almost no abundance gradient with $|$\rgc$|$. The abundance of component B is higher on the far side of the bulge where components A and B have merged ($b=-5^\circ$) or component A is no longer present ($b=-10^\circ$).

The mean abundance of component C remains relatively flat at both latitudes out to about $|R_{\rm G}|$ = 4.5 kpc. Outside this radius, the errors are too large to evaluate any gradient. At \rgc\ $>$ 8kpc, the $|z|-$height above the plane is about $3$ kpc for $b = -10^\circ$ and there may be increasing contamination of component C by the metal-poor stars which belong to the overlapping lower metallicity component D. 

We have used the above gradient in component A to argue in the previous section that components A and B  merge at a radius $|$\rgc$|$ $ \approx 4.5$ kpc.  Within \rgc\ $<$ 4.5 kpc at both latitudes, the MDF shows three significant peaks; two at [Fe/H] $>$ --0.5 and one at -1.0 $>$ [Fe/H] $>$ --0.5.  Outside of \rgc\ $>$ 4.5 kpc, there is only one clear peak in the MDF present at [Fe/H] $>$ --0.5 and a second significant peak at --1.0 $>$ [Fe/H] $>$ --0.5. At all radii there is the small contribution of stars with [Fe/H] $<$ --1.0 represented by components D and E.

 \begin{table*}
\centering
\caption{Mean metallicity of Gaussian components A, B and C at $b=-5^\circ$ and $b=-10^\circ$, for $|l|$ $<$ $15^\circ$, as a function of Galactocentric radius, $|R_{\rm G}|$. The range of $|z|$-height from the plane for each range is indicated. }
\begin{tabular}{| p{2.0cm} | p{1.5cm} | p{1.5cm} | p{1.5cm} | p{1.5cm} | p{1.5cm} |}
\hline
Distance cut  & Number of stars & \multicolumn{3}{|c|} { Mean [Fe/H]} \\ [0.5ex]
\hline\hline
 $|R_{\rm G}|$ $(kpc)$ &b=--$5^\circ$ &$|z|$ kpc & A & B &  C   \\
\hline
$<$0.75 & 300 &0.6--0.8 & 0.16$\pm 0.02$ & -0.24 $\pm 0.02$ &  --0.67$\pm 0.04$  \\
$<$1.5 & 1100 &0.6--0.8 &0.15$\pm 0.01$ &-0.22$\pm 0.01$ & --0.7 $\pm 0.03$   \\
1.5--3 & 2300 &0.1--0.4  &0.13$\pm 0.01$ & -0.27$\pm 0.01$   &  --0.69$\pm 0.03$   \\
3--4.5 & 2000 &0.1--0.3 &0.10$\pm 0.01$   & -0.28$\pm 0.01$   &  --0.67$\pm 0.02$  \\
4.5--6 (near) &600 &0.2--0.3 & 0.11 $\pm 0.03$ & -0.26 $\pm 0.04$ &  -0.75$\pm 0.07$ \\
5--7 (near) &200  &0.1--0.3 & 0.11 $\pm 0.03$ & -0.28 $\pm 0.04$ &  -0.67$\pm 0.07$ \\
4.5--6 (far) &120 &1.1--1.2 & 0.12 $\pm 0.08$ & -0.25 $\pm 0.06$ &  -0.70$\pm 0.07$ \\
5--7 (far) & 100 &1.1--1.3 & 0.10 $\pm 0.08$ & --0.25 $\pm 0.06$  &  -0.68$\pm 0.07$ \\
8--10 (far) & 70 &1.4--1.6 & -- & -0.13 $\pm 0.04$ & -0.75$\pm 0.08$  \\
10--15 (far) & 130 &1.6--2.0 & -- & -0.26 $\pm 0.08$ &  -0.77 $\pm 0.08$ \\
15--40 (far) & 100 &2.0--4.2 &-- &  -0.21 $\pm 0.08$ & -0.95 $\pm 0.07$\\
\hline
   $|R_{\rm G}|$ $(kpc)$ &b=--$10^\circ$ &$|z|$ kpc   &  A & B &  C   \\
\hline
$<$0.75 &180 & 1.3-1.5 & 0.13$\pm 0.06$& -0.27$\pm 0.03$& --0.67$\pm 0.04$\\
$<$1.5 & 850 & 1.1--1.7 &0.10$\pm 0.03$& -0.28$\pm 0.02$& --0.68$\pm 0.01$\\
1.5--3 & 2450 & 0.9--1.9 &0.09$\pm 0.03$& -0.29$\pm 0.02$& --0.65$\pm 0.01$\\
3--4.5 & 1700 & 0.9--2.2 & 0.08$\pm 0.03$& -0.27$\pm 0.02$& --0.67$\pm 0.01$\\
4.5--6 (near) & 250 &0.3--0.6 & 0.10 $\pm 0.06$ & -0.22 $\pm 0.05$ &  -0.71$\pm 0.05$ \\
5--7 (near) & 150 &0.2--0.5 & 0.04 $\pm 0.05$ & -0.25 $\pm 0.05$ &  -0.76$\pm 0.04$ \\
4.5--6 (far) & 200 &2.2--2.5 & -- &  -0.35 $\pm 0.05$ &   -0.78$\pm 0.06$ \\
5--7 (far) & 120 & 2.3--2.6& -- &  -0.25 $\pm 0.08$ &   -0.77$\pm 0.06$ \\
8--10 (far) & 40 &2.8--3.1 &-- & --0.27$\pm 0.08$ &-0.79$\pm 0.08$\\
10--15 (far) & 150 &3.1--4.0 & -- & --0.13$\pm 0.08$ &  --0.91$\pm 0.06$  \\
15--40 (far) & 170 & 4.0--8.4 & -- & --0.20$\pm 0.08$ &  --0.70$\pm 0.06$\\
\hline
\end{tabular}
\label{table:ACcuts}
\end{table*}

\section{Alpha-enhancement}

The $\alpha- $enhancement reflects the rate of chemical evolution. Stars are $\alpha- $enhanced if the chemical enrichment in their formation environment has been sufficiently rapid that there was not time for the  Fe--enriching SN Ia to contribute fully to their chemical evolution. In the following,  [$\alpha$/Fe] denotes the mean of [Si/Fe],  [Mg/Fe] and [Ti/Fe] (we used 5 Si lines, 3 Mg lines and 9 Ti lines); we were unable to make an accurate estimate of [Ca/Fe]. Figure  \ref{fig:alphadistfarside} compares the radial distribution of $\alpha- $enhancement for stars in  0.5 dex bins of [Fe/H] from --2.0 to 0.6, extending in radius from inside the bulge  to the outer disk and halo at $R_{\rm G}$ = 30 kpc. For this figure, we have not associated the [Fe/H] intervals with the MDF components, because of the  abundance gradient seen in the most metal-rich component A.

\subsection{Alpha-Enhancement trends with Galactic radius}

Stars with [Fe/H] $> -0.5$, and particularly the most metal rich stars, show a decrease in [$\alpha$/Fe] from slightly enhanced values in the inner Galaxy. Stars with [Fe/H] $>$ 0 show  [$\alpha$/Fe] declining to negative values beyond $|R_{\rm G}|$ $>$ 20 kpc (Figure \ref{fig:alphadistfarside}). This decrease in the [$\alpha$/Fe] ratio with radius indicates that the chemical evolution in the inner Galaxy proceeded more rapidly than in the disk at larger radii where the [$\alpha$/Fe] ratio is near solar for the most metal-rich stars. Note that stars at these larger values of $|R_{\rm G}|$ $>$ 7 kpc are mostly on the far side of the bulge, at $z-$heights $> 1.4$ kpc.

For stars with $-2.0 < $ [Fe/H] $< -0.5$,  the $\alpha-$enhancement is fairly flat with galactic radii out to 30 kpc. The most metal poor fraction of stars shows a slight increase in enhancement, although these stars show a larger scatter.

    \begin{figure}
   \centering
    \includegraphics[width=0.52\textwidth]{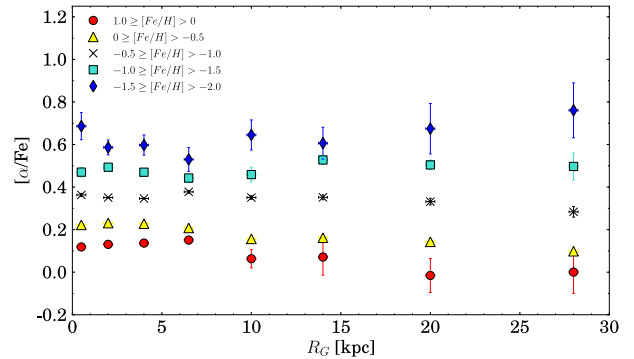} 
   \caption{Alpha-enhancement of stars in $0.5$ dex intervals of [Fe/H] against Galactic radius $|R_{\rm G}|$. The size of the radial bins increase with increasing distance.}
   \label{fig:alphadistfarside}
\end{figure}

        \begin{figure}
   \centering
     \includegraphics[width=0.52\textwidth]{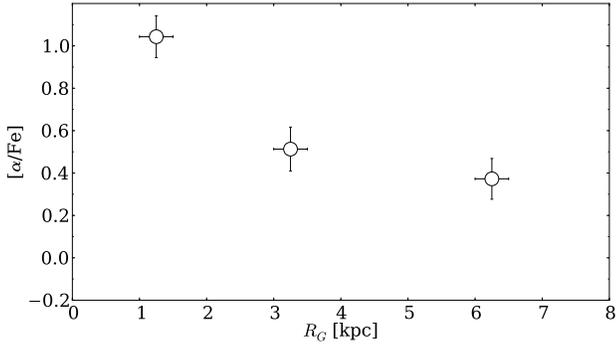} 
   \caption{Alpha-enhancement of our 27 most metal-poor stars with [Fe/H] $< $--2.0 against $|R_{\rm G}|$. For $|R_{\rm G}| > 8$ kpc, we have only three metal-poor stars (at $|R_{\rm G}| \approx$ 9kpc, 11kpc and 12 kpc).}
   \label{fig:alphadistpoor}
\end{figure}    

 Figure \ref{fig:alphadistpoor} shows the radial distribution of mean [$\alpha$/Fe] for the most metal-poor stars with [Fe/H] $<$ --2.0.   (The most metal poor star in the sample is at [Fe/H] $=-2.9$; however it is difficult to measure the gravity from the weak Ca-triplet lines and we have no distance estimate for this star). The distribution of [$\alpha$/Fe] for these metal-poor stars show a steep decrease from the inner region.  In this subsample of metal-poor stars with $|$\rgc$|$ $<$ 8 kpc, $75$\% lie within  $|R_{\rm G}| < 4.0$ kpc. 

\subsection{Alpha-enhancement and components}

We can analyse the $\alpha-$enhancement in the context of the components of the MDF. Figure \ref{fig:bulgealpha} shows contour plots of the stellar density for stars with $R_{\rm G} <$ 3.5 kpc in the [Fe/H] -- [$\alpha$/Fe] plane, within longitude $\pm 15^\circ$, with the generalised MDFs for the components ABC superimposed.  At $b=-5^\circ$ there are two clear peaks in the density distribution, corresponding to components A and B. At $b=-10^\circ$, component A has attenuated relative to component B and the contour plots show only a single density maximum associated with component B. The most metal--rich component A has a mean $[\alpha$/Fe] of about $0.1$ at $b=-10^\circ$ and $0.16$ at $b=-5^\circ$
(in agreement with the weakly enhanced [$\alpha$/Fe] at solar [Fe/H] found by high resolution studies of bulge stars in Baade's window by \citet{AlvesBrito2010}). All of the lower-metallicity components in the bulge are more $\alpha-$enhanced.  The $\alpha-$enhancement increases as expected with decreasing [Fe/H],  from about $0.1$ at solar metallicity to about $0.6$ at 
[Fe/H] $= -2.0$.  The dispersion in [$\alpha$/Fe] increases as [Fe/H] decreases. At latitude $-10^\circ$, the contribution of the more metal-rich component of stars is attenuated, and we see a lower dispersion in the {[$\alpha$/Fe]} for [Fe/H] $> 0$ relative to the lower latitude strip. The $\chi^2$ error on the [$\alpha$/Fe] measurement is $\sigma = 0.1$ and this is consistent with the spread seen in Figure \ref{fig:bulgealpha}.

  \begin{figure}
   \centering
      \includegraphics[width=0.5\textwidth]{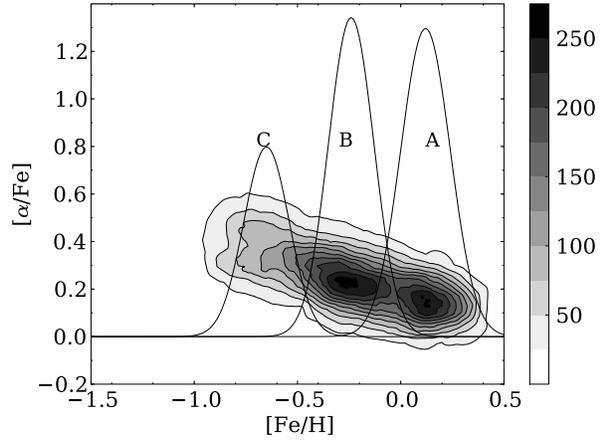} \\
   \vspace{-5pt}
    \includegraphics[width=0.5\textwidth]{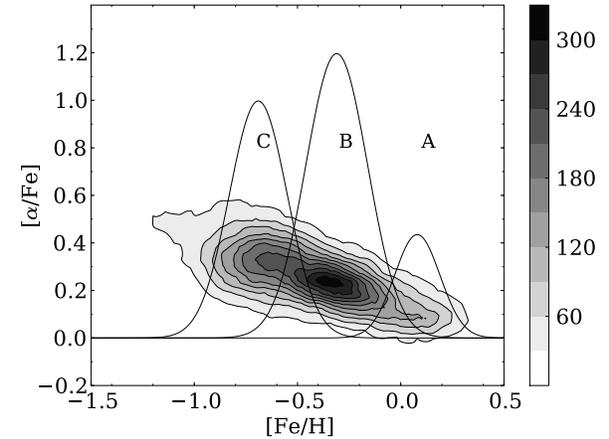} 
         \caption{Alpha-enhancement at $b=-5^\circ$ (top) and $b=-10^\circ$ (bottom) in the bulge region $|R_{\rm G}| <$ 3.5 kpc. The MDF components at each latitude are superimposed for comparison.The grayscale colour legends at right show the number of stars.}
   \label{fig:bulgealpha}
\end{figure}

Figure \ref{fig:alphadistcut} compares the [Fe/H] -- [$\alpha$/Fe] relation for stars in three radial zones: inside the bulge at $|R_{\rm G}| < $ 1.2 kpc and $|R_{\rm G}| < $ 3.5 kpc, and outside the bulge at 5 kpc $< |R_{\rm G}| <$ 10 kpc. For the innermost interval inside the bulge region, the $\alpha-$enhancement increases linearly with decreasing [Fe/H], from about $0.1$ for stars above solar metallicity to about $0.6$ at [Fe/H] $= -2.0$. The dispersion in [$\alpha$/Fe] increases more slowly, from about $0.1$ at the most metal-rich end to about $0.15$ at [Fe/H] $= -2.0$. This dispersion at the more metal-poor end is higher for stars within $|R_{\rm G}| < 3.5$ kpc, at $\sigma \approx 0.35$; other dispersion values in this radial interval follow those of the inner region $R_{\rm G} < 1$ kpc.  

Comparing the $\alpha-$enrichment of stars in the disk and the bulge, we see that the stars  with 5 kpc $< |R_{\rm G}| < 10$ kpc and [Fe/H] $> -0.5$ have similar [$\alpha$/Fe] values to those inside the bulge. The [$\alpha$/Fe] values are about $15$\% lower for stars with [Fe/H] $< -1.0$ and about $10$\% higher for stars with [Fe/H] $>$ --1.0 relative to those with $|R_{\rm G}| < 3.5$ kpc. The dispersion is higher than for $|R_{\rm G}| < $ 3.5 kpc by about $10$\% at the more metal-rich end and $30$\% at the more metal-poor end.

    \begin{figure}
  \centering
   \vspace{-5pt}
\subfloat[\small{$R_{\rm G} <$ 1.2kpc}]{\label{fig:gull}\includegraphics[width=0.52\textwidth]{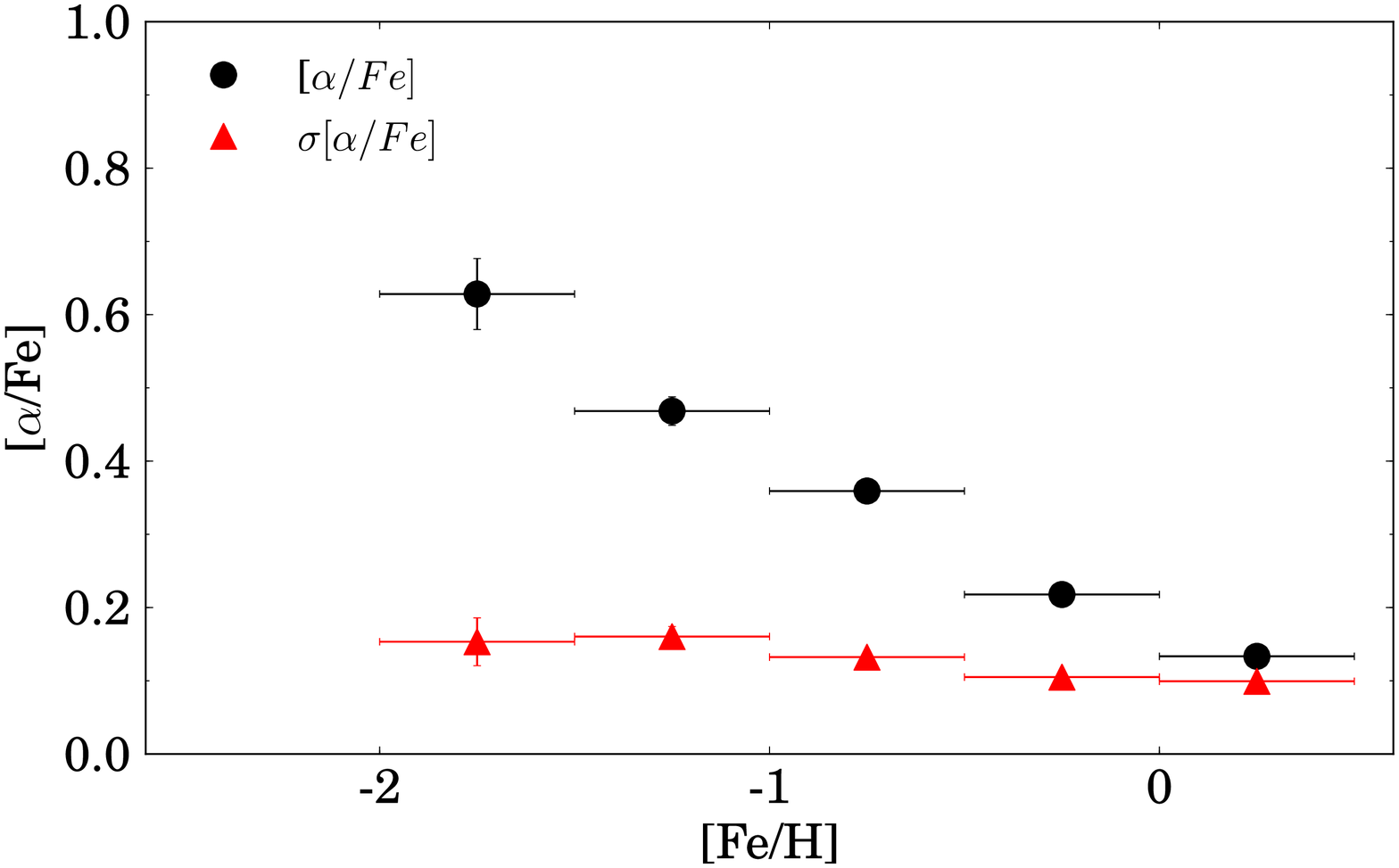}}   \\
\ \subfloat[\small{$R_{\rm G }<$ 3.5kpc}]{\label{fig:gull}\includegraphics[width=0.52\textwidth]{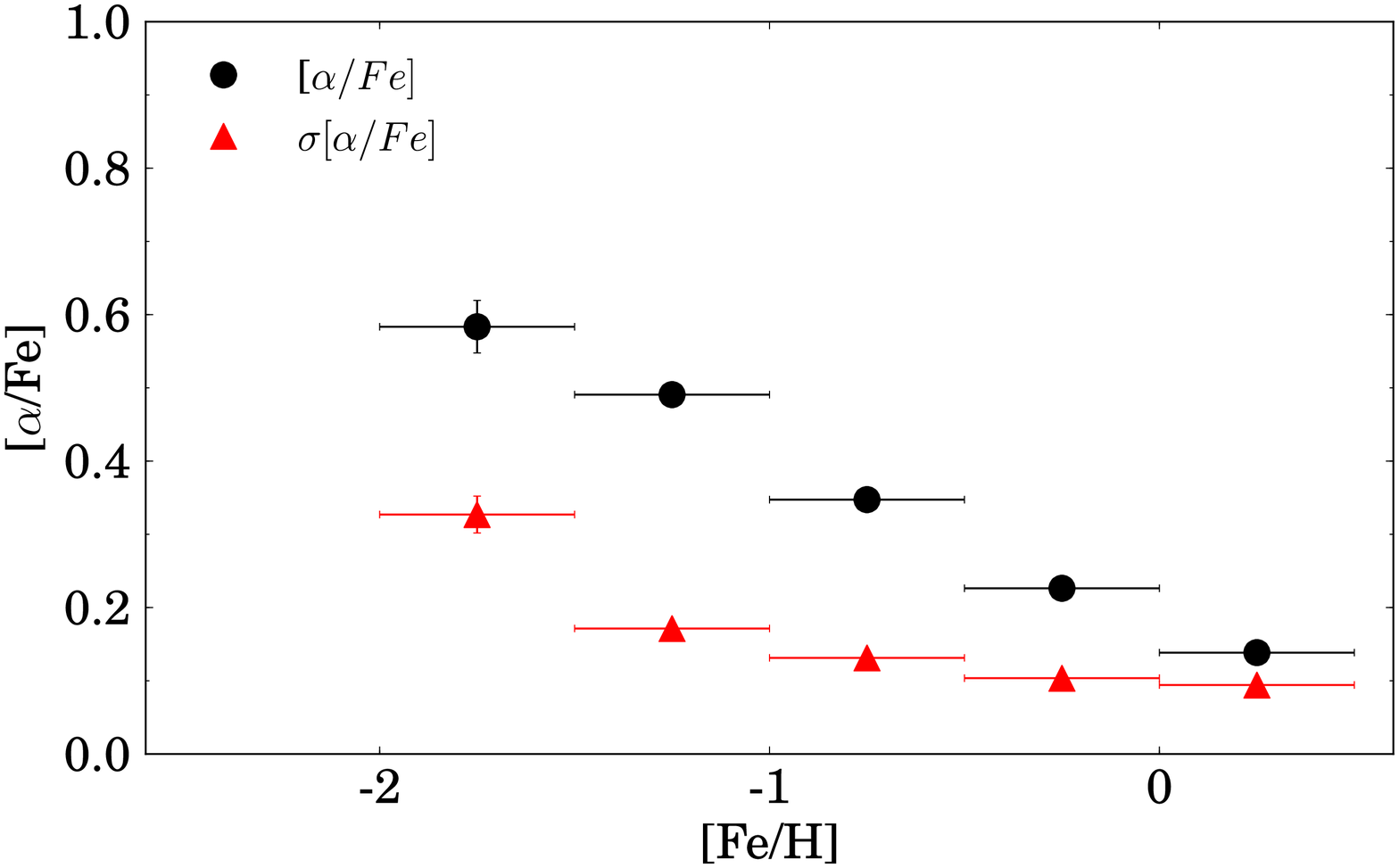}}   \\
 \vspace{-5pt}
\subfloat[\small{5 kpc $< R_{\rm G} <$ 10kpc}]{\label{fig:gull}\includegraphics[width=0.52\textwidth]{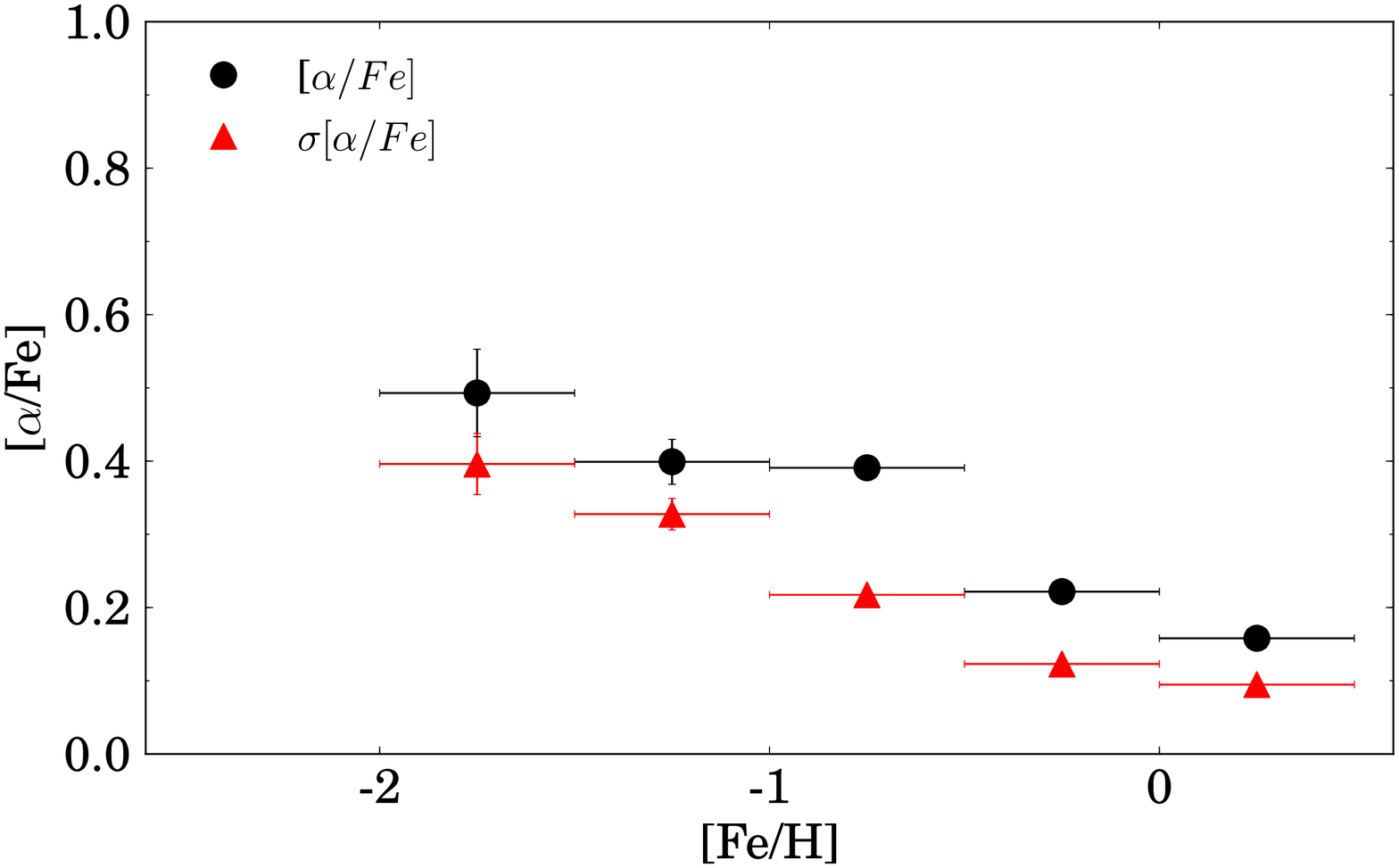}}   \\
   \caption{Mean [$\alpha$/Fe] and dispersion in [$\alpha$/Fe] versus [Fe/H] for three intervals of $R_{\rm G}$: in (a) $R_{\rm G} < 1.2$ kpc, and (b) $R_{\rm G} < 3.5$ kpc,  the stars lie within the bulge region, and in (c) 5 kpc $< R_{\rm G} <$ 10 kpc they lie outside the bulge region. }
        \label{fig:alphadistcut}
     \end{figure}

\section{Interpreting the components}

In this section, we discuss two significant features of the MDF components: (1) their individual vertical abundance gradients and the overall vertical abundance gradients associated with the changing fraction of the major components with height above the plane, and (2) the contribution of the components to the split red clump of the bulge.

\subsection{The vertical metallicity gradient of the bulge}

In the bulge region, components A, B and C individually show weak vertical abundance gradients (section 4.2) of $-0.08 \pm 0.04$ to $-0.04 \pm 0.02$ dex kpc$^{-1}$ over the latitude interval $b=-5^\circ$ to $b=-10^\circ$.  Although our overall vertical metallicity gradient  ($-0.45$ dex kpc$^{-1}$) is comparable to that found by \citet{Zoccali2008},  we interpreted this overall gradient as coming from changing fractions of  components A--D with latitude {\it via} the action of the instability process on the pre-existing early disk. 

It is not clear how to interpret these small vertical abundance gradients in the individual components. Are they the remnants of abundance gradients that were already in place in the thin and thick disks before the instability event? The small gradients which we observe compare well with the fairly flat vertical abundance distribution found by \citet{Bekki2011}, from their simulation of an initial disk system which buckles into a boxy/peanut bulge/bar {\it via} the disk/bar instability. They model both the instability of an initial single thin disk and also a dual (thin $+$ thick) disk population. These disks initially have vertical abundance gradients. Due to the vertical mixing of the stellar populations by the bar, they find that the bulge is unlikely to have a steep vertical metallicity gradient. 
Their model does not include chemical evolution so is appropriate for a system in which the chemical evolution took place in the early disk, before the instability events and the formation of the bar/bulge.

\subsection{The split red clump of the Galactic bulge} 

There is a striking relationship between the spatial distribution of stars in the bulge region and [Fe/H] which was seen in our analysis of the split clump (Ness et al. 2012a). The split clump is a consequence of the peanut structure of the bulge, so the presence of a split clump indicates membership of the peanut bulge. 

Figure \ref{fig:data5to10} shows the magnitude distribution of the red clump stars along the minor axis for our three latitudes, selected with $K_{0}$ = 12.38 to 13.48 and \logg\ = 1.9 to 3.1. The clump stars are split into three [Fe/H] bins; (i) [Fe/H] $>$ 0, (ii) 0 $>$ [Fe/H] $>$ --0.5 and (iii) [Fe/H] $<$ --0.5. In Ness et al. (2012a) we compared the spatial distribution and kinematics of the clump stars with predictions from an evolutionary N-body model of a bulge that grew from a disk {\it via} bar-related instabilities.  The density distribution of the peanut-shaped model is depressed near its minor axis, thus producing a bimodal distribution of stars along the line of sight through the bulge near its minor axis, very much as seen in our observations for stars [Fe/H] $>$ --0.5. We found that only stars with [Fe/H] $>$ --0.5 show the split structure at latitudes $b < -5^\circ$, and concluded that the stars of the boxy/peanut bulge have [Fe/H] $>$ --0.5. Stars with [Fe/H] $<$ --0.5 do not show the split clump at any latitude.  This indicates that the initial disk from which the bulge formed had very few stars with [Fe/H] $<$ --0.5, or that stars with [Fe/H] $<$ --0.5 were not in a region of phase space which is mapped into the boxy structure by the disk instability.

From Figure \ref{fig:data5to10}, stars with [Fe/H] $>$ --0.5 show the split at $b < -5^\circ$. Stars in our component A ([Fe/H] $> 0$) show a deeper split than stars in our component B ($0 >$ [Fe/H] $>-0.5$) at the intermediate latitude of $b= - 7.5^\circ$.  At $b=-10^\circ$ the split is still seen for components A and B but is affected by the asymmetry in the magnitude distribution for the more metal-rich stars: these stars are located preferentially closer to the plane and are therefore found more prominently in the nearer and brighter peak.  The proportion of stars with [Fe/H] $>$ 0 decreases from about $40$\% at $b = -5^\circ$ to about $8$\% at $b = -10^\circ$.  In contrast, for the component B stars with  $-0.5 <$  [Fe/H] $<$ 0, the proportion stays approximately constant from $b = -5^\circ$ to $b = -10^\circ$. 

   \begin{figure*}
   \centering 
\includegraphics[width=0.8\textwidth]{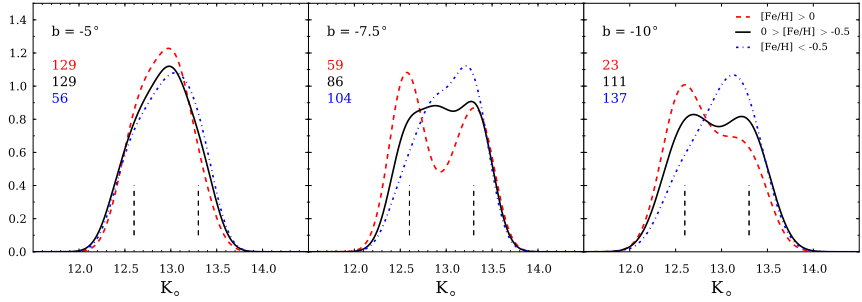} 
 \caption{The magnitude distributions for the red clump stars selected with $K_{0}$ = 12.38 to 13.48 in three [Fe/H] intervals of (i) [Fe/H] $<$ -1.0, (ii) -0.5 $<$ [Fe/H] $<$ 0 , (iii) [Fe/H] $>$ 0. Gaussian smoothing kernel $\sigma$ = 0.15 mag was applied. The y-axis is the normalised number of stars. }
   \label{fig:data5to10}
\end{figure*}

The split distribution of the component A and B stars, and the absence of the split for the more metal-poor component C stars, leads us to infer that the peanut bulge structure is defined primarily by the stars of component A ([Fe/H] $>$ 0) and component B (0 $>$ [Fe/H] $>$ --0.5).

\section{Discussion}

In a survey of the inner Galaxy, we can expect to find stars from the inner regions of all of the major components of the Galaxy. We have identified five metallicity components in total in the stellar MDF of the inner Galaxy, and interpreted the MDF as a composite population made up of these components A--E in order of decreasing [Fe/H]. The two more metal-rich components (A and B) define the boxy/peanut structure of the bulge.  We make the following interpretation of the metallicity components within the disk instability scenario for bulge formation:\\

\noindent $\bullet$ A: The metal-rich ``boxy/peanut-bulge'' with a mean [Fe/H] $\approx +0.15$ that is concentrated towards the plane.  In the inner regions, component A has the chemical properties of a younger chemically evolved thin disk (higher [Fe/H] compared to the thin disk in the solar neighbourhood and slightly higher [$\alpha$/Fe]).  At larger Galactic radii, component A appears to be somewhat more metal-poor (see Table 5) and we associate it with the thin disk at larger Galactic radii. Its MDF at larger radii is similar to that of the thin disk near the Sun \citep[e.g.][]{Haywood2008}. Although chemically similar to the thin disk, it replicates the line-of-sight kinematics and spatial properties of the thicker boxy/peanut-bulge component B in the inner regions of the Galaxy \citep[see][in preparation]{Ness2012c}, but at lower heights above the plane (see Section 4). We note again that stars in the metallicity range of component A are prominently involved in the bimodal peanut structure of the bulge seen at higher Galactic latitudes.  

$\bullet$  B: The vertically thicker ``boxy/peanut-bulge'' contributes a similar fraction of the bulge stars over the latitude range from $b=-5^\circ$ to $b= -10^\circ$. Its mean [Fe/H] value is about $-0.25$ and it is the dominant component in most of our bulge fields inside \rgc\ $<$ 4.5 kpc (see Figures 10 and 11). In the instability scenario, it would represent the early thin disk when there were few stars with [Fe/H] $< -0.5$ dex at the onset of the bar buckling instability. Again, stars in the metallicity range of component B show the split structure of the bulge at higher latitudes. At larger radii, component B is associated with the metal poor stars of the early thin disk that we sample at large heights from the plane. 

$\bullet$  C: We interpret component C as the inner thick disk (mean [Fe/H] about --0.70) which is at most weakly involved in the boxy/peanut structure of the bulge.  Component D is tentatively ascribed to the metal-weak thick disk (mean [Fe/H] about --1.2). These components were likely present in the inner regions of the Galaxy ($|R_{\rm G}| < $ 3.5 kpc) before the instability event, and may have been puffed up by this event.  We note again that components C and D do not appear to be part of the split structure of the peanut bulge (Ness et al., 2012a). At larger $|R_{\rm G}|$, these old components are still present in their relatively undisturbed form.

$\bullet$ E: We interpret this metal-poor component as the stars of the inner Galactic halo.

Components A and B are clearly part of the boxy/peanut structure of the Galactic bulge, as shown by the split red clump discussed in Ness et al., 2012a. Models by \citet{Athanassoula2005} show different levels of boxy/peanut behaviour. For the boxy systems, the isophotes are roughly parallel to the equatorial plane. The isophotes of the peanut systems show a clear dip near \rgc\ = 0. If this dip is relatively deep, the systems give an impression of an X-shaped structure most clear at larger heights above the Galactic plane, particularly after unsharp masking. We have seen in \citet{Ness2012a} that component A shows a more prominent central minimum than component B, but it is not clear whether A is intrinsically more X-shaped or if its deeper minimum reflects a similarly shaped structure but with a smaller scale height than the thicker boxy/peanut-bulge represented by component B.

\subsection{The origin of the components}

In the instability scenario, the boxy/peanut-bulge formed out of the early thin disk: after about 1 Gyr, the instability of the disk forms a bar which then buckles vertically into the observed boxy/peanut morphology.  This scenario gives a guide to interpreting the components of the bulge. We know from the age-metallicity relation in the solar neighbourhood (e.g. Haywood 2008) that the chemical evolution of the solar neighbourhood proceeded rapidly: 10 Gyr ago, the stars of the thin disk in the solar neighbourhood already had a mean [Fe/H] near solar, with a spread from about $+0.5$ to $-0.5$.  In the context of inside-out formation of the disk, it is likely that the chemical evolution of the inner Galaxy was even more rapid. By the time that the bar-forming instability occurred, the chemical evolution of the inner disk was probably already well advanced. 

The instability process would affect all of the inner disk stars, and generates a mapping of the stars of the early disk into the boxy/peanut structure.  In this way, the bulge preserves a dynamical imprint of the chemical distribution of the thin disk at the time that the buckling occurred; the bulge is a chemical snapshot of the MDF of the early thin disk captured in the extended bar. We have seen that two components, A and B, are involved in the boxy/peanut structure of the bulge. Figure \ref{fig:components} shows that the more metal-rich component A (mean [Fe/H] $= +0.15$) is more concentrated to the plane than the thicker dominant bulge component B (mean [Fe/H] $= -0.25$).  This more metal-rich component A is chemically consistent with what we would expect for a younger, colder and more metal-rich thin disk in the inner region at the time of the instability event. 

Examination of the N-body models shows that the mapping from disk to bulge depends on the location and kinematics of the stars at the time of the instability. Kinematically colder stars in the disk can suffer substantial radial and vertical migration, and can therefore be strongly involved in the boxy/peanut structure. We will discuss this mapping of the disk into the bulge in more detail in a later paper.  At this stage, we suggest that component A may simply represent the responsive cold and metal-rich stars of the thin disk in the inner region which were entrained in the potential of the major boxy/peanut component B. The kinematical selectivity of the mapping is likely to explain the presence of the two distinct components A and B in the bulge at the higher latitudes $|\,b\,| > 5^\circ$.

The thick boxy/peanut bulge population B can be interpreted as the product of the bulk of the thin disk (already a few Gyr old at the time of the instability event) which forms a bar and puffs up into a boxy/peanut-bulge \citep{Athanassoula2008}.  The mean abundance and abundance range of component B is similar to that of the intermediate-age disk stars ($3$ to $8$ Gyr old) in the solar neighbourhood at the present time (e.g. Haywood 2008).  
The bulge structure is visible in NIR photometry and star counts out to a radius $R_{\rm G}$ of about 3.5 kpc. Further from the Galactic center, stars of the old thin disk (component B) are still present, but in their relatively undisturbed original thin disk structure. 

Component C has a mean abundance like that of the thick disk in the solar neighbourhood. If it is representative of the thick disk in the inner Galaxy before the instability event, it may have been correspondingly hotter and less dynamically responsive at the time of the instability. That may be the reason why component C does not appear to be involved in the boxy/peanut structure of the bulge. Similarly, component D has a mean [Fe/H] of about $-1.20$,  like that of the metal-weak thick disk near the Sun \citep[e.g.][]{ChibaBeers2000, Carollo2010}.

Component E with mean [Fe/H] $= -1.7$ appears in small fractions at about the same [Fe/H] in most fields, and shows no evidence for an abundance gradient with latitude. It is slowly rotating (see Ness et al., 2012c) and is probably an entirely separate population, not kinematically or chemically  associated with the disk.  We interpret the metal-poor component E stars as inner halo stars on orbits of high eccentricity passing through the bulge region. They may correspond to the high eccentricity stars at an abundance of about $-1.7$ identified near the Sun by \cite {ChibaBeers2000}. Simulations \citep[e.g.][]{Diemand2005,Brook2007,Tumlinson2010} show that the oldest stars in the Galaxy, formed in fragments from high-$\sigma$ density fluctuations, should now inhabit the inner Galaxy.  In this context, the stars of component E in the inner Galaxy would be worth further spectroscopic study at higher resolution. 

We now return to the metal-rich component A, which is more prominent closer to the Galactic plane but at the same time shows most strongly the split red clump which we associate with the boxy/peanut structure of the bulge.  The MDF of the solar neighbourhood may again provide a useful reference.  In the solar neighbourhood, the youngest disk stars (ages $< 2$ Gyr) are relatively metal-rich, have a relatively narrow range of abundance, and are kinematically colder than the intermediate-age disk stars.  The larger width of the MDF for the intermediate-age stars in the solar neighborhood is currently attributed to the secular effects of radial migration, which radially smooths the Galactic abundance gradient and so increases the width of the local MDF. In the same way, the more metal-rich stars of the early inner thin disk were probably kinematically colder and dynamically more responsive, thus preferentially mapped into boxy/bulge-supporting orbits.  

Alternatively it is possible that component A comes, {\it via} a second later and less vertically extended bar-forming instability event, from the more metal-rich thin disk which had continued to undergo star formation and chemical evolution since the first instability episode. Simulations have shown that a second buckling episode can occur \citep{Athanassoula2005b, Inma2006}, well after the first one, and that it leads to a vertically thicker structure than given by the first only buckling event. If we make the plausible assumption that component B forms earlier from the older metal-poor disk, and component A forms later from the metal-rich disk, then component B may well have had sufficient time to undergo a second buckling, while component A has not. This offers a possible explanation for why component B is more vertically extended and more metal poor than A. A secondary instability could also be triggered by a significant accretion event. Models with two disks of different scale heights may also offer more insight.  The current N-body models we are working with do not show a secondary instability event and have a single disk only; we therefore cannot test these effects. 

Component A attenuates in density by a factor of $5$ between $b=-5^\circ$ and $b=-10^\circ$. It is possible that the original bar formed in the early disk may have gradually entrained the chemically evolving thin disk during and after the instability process. In this way, the more metal-rich stars could have experienced only the later fraction of the bar-buckling process, and remained more closely bound to the Galactic plane. 

In these alternative scenarios, we are associating component A with the inner thin disk which has continued to undergo star formation and chemical evolution. We can trace this component out to large Galactic radii, where it becomes gradually more metal-poor (Table $5$), reflecting the abundance gradient of the Galactic disk. In the central region however this component behaves in a clearly bar-like way, reflected in the split clump (see Section 6.2) and the kinematics as a function of longitude (Ness et al., 2012c). 

\subsection{The alpha-enhancement of the components}

Figure \ref{fig:bulgealpha} shows that the [$\alpha$/Fe]-[Fe/H] relation for stars in the inner Galaxy follows the pattern in the solar neighbourhood, except that the most metal-rich stars in the inner regions do show a weak $\alpha$-enhancement. Alpha-enhancement is a measure of the rate of chemical evolution.  The more metal-poor boxy/peanut component B, which we associate with the older pre-bulge disk, is $\alpha$-enhanced above the level of the more metal-rich bulge component A which we associate with the younger pre-bulge thin disk. This is consistent with our argument that the bulge stars of component B were formed earlier as part of the older disk. Their $\alpha$-enrichment would be present not only in the bulge population but also for the other stars of the early disk which are still now in the disk.

The more metal-poor components in our survey, with [Fe/H] $< -0.5$, include the thick disk and metal-weak thick disk. Although there are several possibilities for the formation of the thick disk, the stars of components C and D  may also have originated from the even earlier thin disk, heated  kinematically by early minor merger activity before the instability event \citep[e.g.][]{Freeman2008}. They would be the oldest stars of the original disk, and their level of $\alpha-$enhancement indicates that the duration of their star formation was short. Some of these more metal-poor stars may also have been puffed up into the bar/bulge by the instability event itself. The $\alpha$-enhancement for all stars with [Fe/H] $<$ --0.5 is fairly flat out to $R_{\rm G} = 30$ kpc. As their 
$\alpha-$enhancement is directly related to the duration of their star formation event, this suggests that the more metal-poor stars which contribute to the bulge/bar, the thick disk and metal-weak thick disk may have had a common origin in the oldest stars of the initial Galactic thin disk. In this picture, the different MDF components are fossil remnants from the dynamical redistribution of an initial disk system of the Milky Way. The abundances and $\alpha$-enhancement are proxies for the time at which the stars of each component formed and the rate of their formation.  Chemodynamical models may be able to link the abundances and timescales for this formation and structural dispersion. 

Figure \ref{fig:bulgealpha} shows the clearly bimodal distribution of abundances between the two boxy/peanut components A and B in our $b = -5^\circ$ fields.  Component A has a mean [$\alpha$/Fe] value of 0.16, and the old disk component B has a mean [$\alpha$/Fe] of 0.23. Several authors have argued that the bulge is an $\alpha$-enhanced metal-rich population \citep{Zoccali2008alpha, Fulbright2007, Lecureur2007, Melendez2008}.  We see that the metal-rich component A is only weakly $\alpha$-enhanced, and the thicker $\alpha-$enhanced bulge component B is not so metal-rich. 

From Figure \ref{fig:alphadistcut}, we can compare the $\alpha$-enhancement of stars with similar [Fe/H] abundances in the bulge and in the disk at larger \rgc.  For stars with [Fe/H] $> -1$, which we would identify as originating from the disk, the mean [$\alpha$/Fe] values are very similar in the bulge (Figures \ref{fig:alphadistcut} (a) and (b) ) and the disk (Figure  \ref{fig:alphadistcut} (c) ).  For the more metal-poor stars with [Fe/H] $< -1.0$ the alpha-enhancement at large $|$\rgc$|$ (Figure \ref{fig:alphadistcut} (c)) is similar to stars in the inner region $|$\rgc$|$ $<$ 3.5 kpc. However the dispersion in the alpha enhancement is significantly lower in the inner bulge region (Figure \ref{fig:alphadistcut}) compared to at larger radii. This may reflect that the metal-poor population in the centre is more homogenous than in the outer regions. For the very metal-poor stars with [Fe/H] $< -2$, the $\alpha$-enhancement is significantly higher in the inner bulge region  than in the disk at larger radii (see Figure \ref{fig:alphadistpoor}), indicating a more rapid star formation history for these very metal-poor stars in the inner regions of the Galaxy.

\section{Comparison to other studies}

We have found an overall metallicity gradient in the whole MDF across our latitudes, comparable to that of \citet{Zoccali2008}. Our MDF at $b=-10^\circ$ and our $\alpha$-enhancement results at our higher latitude, also compare well with the results of \citet{Uttenthaler2012} for their high resolution study at $(l,b) = (0^\circ,-10^\circ)$. Note that \citet{Uttenthaler2012}  also confirm the presence of the split clump only for the more metal rich stars. Other spectroscopic studies of bulge stars have found evidence for multiple components in the MDF, and we now compare our results with theirs. Table \ref{table:Argosfeh} summarises the mean abundances for the components found in the ARGOS survey compared with those from the work by \citet{Babusiaux2010}, \citet{Bensby2010} and \citet{Hill2011}. Each of these studies sees some but not all of the components found in the larger ARGOS sample.

\subsection{Vertex deviations and the bulge} 

In the study of about 400 bulge K--giants and clump giants in Baade's window at $b = -4^\circ$ and a field near the minor axis at $b = -6^\circ$ by  \citet{Babusiaux2010}, our components A, B and D can be seen.  They argue that both of the main scenarios of bulge formation have contributed to the bulge.  Using proper motion and radial velocity data, they measure the vertex deviation of the velocity ellipsoid for their identified components. Their more metal-rich component, with [Fe/H] near $0.14$ (our A), shows a significant vertex deviation and they conclude that this component is the bar/bulge that formed via the instability scenario. Their more metal-poor components (our B and D) show a smaller vertex deviation, and they associate these components with a spheroid and thick disk.

\citet{Babusiaux2010} also measured the radial velocity dispersion as a function of [Fe/H] for their fields at $b=-4^\circ, -6^\circ$ and $-12^\circ$ near the minor axis of the bulge. They find that the dispersion increases with [Fe/H] in their $b=-4^\circ$ field (the more metal-rich stars are kinematically hotter) but decreases as a function of [Fe/H] in their higher latitude fields. Similarly, we find that at $b=-5^\circ, -7.5^\circ$ and $-10^\circ$, the velocity dispersion decreases with [Fe/H] along the minor axis \citep[and indeed at all longitudes, see][in preparation]{Ness2012c}. This change with latitude of the sense of the abundance - (velocity dispersion) relationship indicates a structural change of the bulge populations at around $b=-4^\circ$. 

We now discuss the vertex deviation measurements of the bulge near the minor axis and compare these results to our N-body model. We do not have proper motions for our stars so cannot independently measure the vertex deviations. The vertex deviation of the stars in Baade's window has been measured by \citet{Soto2007}  and \citet{Babusiaux2010}. \citet{Soto2007} conclude from their measurements that stars with [Fe/H] $> -0.5$ are supporting a bar. They find no significant vertex deviation for stars with [Fe/H] $< -0.5$. This makes sense in the context of our finding of the metallicity dependence of the split clump and assignment of components to populations. \citet{Soto2007} find a trend of decreasing velocity dispersion with increasing metallicity as do \citet{Babusiaux2010}. \citet{Babusiaux2010} also find the largest vertex deviation for the more metal-rich stars which we would associate with component A. Their vertex deviation in the $lr$ plane (denoted $l_{v}$) for stars with [Fe/H] $< -0.14$ is $-13^\circ \pm 9$. For the more metal-rich stars with [Fe/H] $> -0.1$ the vertex deviation is about $-40^\circ$. They argue that the most metal-rich stars represent the bar/bulge in the central region, and the stars with lower metallicities are associated with an old spheroid. 

We consider first the vertex deviation of stars belonging to our boxy/peanut-bulge component B, with $0 >$ [Fe/H] $> -0.5$. The $l_{v}$ vertex deviation measured by \citet{Babusiaux2010} for stars in this abundance range is about $-30^\circ$.  This agrees well with the estimate from our bar/bulge model, for which we find a vertex deviation $l_{v} = -26^\circ \pm 2^\circ$ for all stars within a galactocentric radius of $0.5$ kpc. 

Now we address the vertex deviation of stars belonging to our boxy/peanut-bulge component A, with [Fe/H] between $0$ and $0.5$. For stars in this abundance range, \citet{Babusiaux2010} measure a vertex deviation in Baade's window of $l_{v} = -43^\circ \pm 5^\circ$ . We examined 6 models with different shapes and scale heights of the central bulge, scaled to match the kinematics of the Milky Way, and found the vertex deviation near the minor axis to be fairly insensitive to the scale heights of the bulge/bar and also to the height above the plane. To reproduce this higher $l_{v}$ in our models, we needed to rotate the bulge/bar in the model from the previously adopted 20$^\circ$ to 45$^\circ$ with respect to the line of sight. The vertex deviation in the model for a bar angle of 45$^\circ$ is $l_{v} = -45^\circ \pm 2^\circ$ similar to that derived by \citet{Babusiaux2010}.  This is an interesting result in light of the work of \citet{CL2008} and others \citep[e.g.][]{Weinberg1992, Hammer1994} who argue for the existence of a long bar in the Milky Way from analysis of 2MASS data  \citep{2MASS}. Using red clump giants as tracers, \citet{CL2008} present evidence for a long bar with a semi-length of 4 kpc, ending around $l=28^\circ$. This long bar inclined at $45^\circ$ coexists with the thicker bulge/bar inclined at $20^\circ$ and $\approx$ 3.5 kpc in length \citep{Gerhard2002} (our component B).  It has been previously suggested  \citep[e.g.][]{CL2008} that the inner Galaxy is dominated at higher latitudes by the bulge and at lower latitudes by the long bar. 

We are not suggesting that the vertex deviation in our model is strong evidence for the coexistence of a boxy/peanut-bulge made of stars with $0 > [Fe/H] > -0.5$ and a long flatter bar that contains the metal-rich stars with [Fe/H] $> 0$ in the bulge. It is however an interpretation, in light of the dependence of vertex deviation of the model on the bar orientation. Changing the angle however may be one of several ways to affect the vertex deviation.  Given the magnitude distribution of the split clump at the higher latitudes as a function of [Fe/H] and the changing relationship between dispersion and [Fe/H] at $b =-5^\circ$, it seems clear that there are two components to the bulge population. It is possible that these results could be explained by a secondary structure which is dominant at low latitudes as per \citet{CL2008}. The metal-rich stars may be associated with a thin disk population associated with a bar orientated at $45^\circ$ to the Sun-centre line.

\subsection{Microlensed dwarfs in the bulge} 

A potential problem with our interpretation of component B as the primary component of the bulge comes from \citet{Bensby2011} who compiled abundances for 26 microlensed dwarfs in the direction of the bulge.  The [Fe/H] values for their stars extend from $-0.72$ to $+0.54$. The MDF of their microlensed dwarfs shows two well--separated components corresponding to our components A and C (which we argue are the thin boxy/peanut-bulge and the old thick disk). As for our sample, they find that the stars of component C are $\alpha$-enhanced, while the metal-rich stars of component A have [$\alpha$/Fe] $\approx$ 0.1, close to solar.  They estimate isochrone ages for these stars, and find that the component C stars are almost all old, around 10--12 Gyr, while the component A stars cover a wide age range from 2 to 14 Gyr.  Their ages are consistent with the identifications of component A  with the thin disk in a boxy/peanut bar structure, which has continued to undergo star formation, and with C being the old thick disk.  Bensby et al, however, find only a small fraction of stars in the [Fe/H] range of what we believe to be the main bar/bulge (component B), around [Fe/H] $= -0.25$. 

We note that the \citet{Bensby2011} stars are mostly closer to the plane than our lowest fields at $b = -5^\circ$. At $b=-5^\circ$, at a radial distance of $\pm$ 0.5 kpc from the centre of the bulge, about $13$\% of our stars have [Fe/H] between $-0.2$ and $0$, but there is only one microlensed dwarf out of 26 in this abundance range.  Along the minor axis at $b=-5^\circ$, $40$\% of our stars have [Fe/H] $> 0$, somewhat lower than the $50$\% of microlensed stars in the same abundance interval which lie closer to the plane. This increase in the fraction of metal-rich stars closer to the plane is consistent with our finding that the fraction of the most metal-rich component A increases towards the plane.   The number of stars in the metallicity range of component B ($0 >$ [Fe/H] $> -0.5$) increases by a factor of only 1.2 from $b=-10^\circ$ to $b=-5^\circ$, whereas the the number of stars in the metallicity range of component A ([Fe/H] $> 0$) increases by a factor of 5.6. Furthermore, if A is associated with the thin disk, then there may be an increasingly mixed-age population of metal-rich stars at low latitudes that is not seen at higher latitudes ($b < -5^\circ$). 

The missing population B at the latitudes measured by \citet{Bensby2011} suggests that the component B stars close to the plane are greatly outnumbered by the metal-rich stars of component A.  The region near the plane may be dominated by the flatter bar and host a younger, thin disk population which may be part of the bar. The low fraction of stars from component B in the \citet{Bensby2011} data suggests that the stars of the early thin disk with [Fe/H] $\approx -0.25$ were mostly displaced to $z-$heights $> 0.7$ kpc during the boxy/peanut-bulge buckling process. As a check on these speculations, it would be interesting to know the abundance distribution of microlensed dwarfs at higher latitudes, closer to ours.  The probability of finding microlensed stars decreases rapidly, however, with latitude. It is also possible that the \citet{Bensby2011} stars
do not lie in the bulge itself: the geometry of microlensing would favor their preferential location on the far side of the peanut/boxy bulge. In this case, one would expect to find relatively few stars of component B (the dominant peanut/boxy bulge population) in the microlensed sample. To check this suggestion, we would need a realistic calculation of the dependence of microlensing optical depth on the position of stars in the inner regions of the Galaxy.  

\begin{table}
\begin{tabular}{| p{0.6cm} | p{0.9cm} | p{0.9cm} | p{0.9cm} | p{0.9cm} | p{0.9cm} | p{0.9cm} |}
\hline
 Component & \multicolumn{3}{| c |}{ARGOS 2012} & Bensby et al. 2010 & Babusiaux et al. 2010 & Hill et al. 2011 \\ 
 \hline
 $b$ &  $-5^\circ$ & $-7.5^\circ$ & $ -10^\circ$ &$< -5^\circ$ & $-6^\circ$ & $-4^\circ$\\
 \hline
A & +0.12 & +0.11 & +0.08 & +0.32 & +0.14 & +0.32  \\
B & --0.26 & --0.28 & --0.30 & -- & --0.27 & --0.30\\
C & --0.66 & --0.68 & --0.70 & --0.60 & -- &  --\\
D & --1.16 & --1.20 & --1.19 & -- & --1.09 & -- \\
E  & --1.73 & --1.67 & --1.68 &  -& -- & --\\
\hline
\end{tabular}
\caption{The components of [Fe/H] identified in ARGOS survey compared to Bensby, Babusiaux and Hill surveys. The ARGOS survey values are taken for the integrated MDF's, across  $l$=$\pm 15^\circ$, for $R_{\rm G} <$ 3.5 kpc. The metallicity of A is 0.15/0.13/0.09 at $b=-5^\circ$,$-7.5^\circ$ and $b=-10^\circ$, taking only minor axis fields.}
\label{table:Argosfeh}
\end{table}

\section{Conclusions}

By measuring stellar parameters for the 25,500 stars in our survey, we are able to identify the nearby dwarfs and subgiants, determine distances to the giants and make a distance cut to separate out approximately $14,150$ bulge candidates with $|R_{\rm G}| \le$ 3.5 kpc. The metallicity distribution function of these $14,150$ stars shows four to five distinct components (denoted A to E). We find that the relative contributions of these components changes with position in the bulge.  Our data extend into the disk beyond the bulge, which helps to identify the nature of the MDF components and interpret how they originated from the early disk of the Milky Way. 

The components show that the stars of the inner Galaxy are a composite population. The inner region of the Milky Way comprises stars with [Fe/H] $> -0.5$, which appear to belong to the boxy/peanut bulge,  and stars with [Fe/H] $< -0.5$ which lie
in the inner regions but are not part of the boxy/peanut structure.  There are two spatially and chemically separated bulge components with [Fe/H] $> -0.5$: (i) component A: a relatively thin boxy/peanut-bulge with mean [Fe/H] $\approx 0.15$ which we associate with the thin disk and (ii) component B: a thicker boxy/peanut-bulge population with mean [Fe/H] $\approx -0.25$.

The duality of the bulge is clear but the mechanism by which this duality has arisen is not so clear. N-body models of disk evolution show that the thick boxy/peanut component B is likely to have formed in a major instability event which occurred many Gyr ago.
The thinner boxy/peanut bulge component A appears also to be associated with such an instability event. It may have originated in the same event from the initial stratification of the thin disk, or it might have formed after the thicker structure. 

Chemically we associate the thin boxy/peanut-bulge (component A) with the younger thin disk, and this component can be followed out to large Galactic radii beyond the bulge at low $|z|-$heights from the plane where it merges with component B. It shows a radial metallicity gradient, becoming more metal-poor in the outer regions. Its gradient on the near side of the bulge from \rgc\ $ = 0$ to $6$ kpc is $-0.02 \pm 0.01$ dex/kpc. Spatially it seems likely that component A comes from the redistribution of the younger thin disk in the inner regions into a boxy/peanut structure, via the bar-forming and bar-buckling instabilities of the disk.  Our MDFs indicate that nearer the Galactic plane this population will be very dominant. It may also be mixed in age, due to ongoing star formation, with younger stars located nearest to the plane. 

Component B is the primary bulge component that is more vertically extended than the metal rich component A. Outside the bulge at high latitudes, where very few metal rich stars are present even inside the bulge, it represents the most metal rich fraction of stars observed in our sample at large $|z|$-heights from the plane. It is associated with the early disk and from the instability scenario represents the primary population which formed the boxy/peanut bulge.

The distinct mean [Fe/H] and [$\alpha$/Fe] values for components A and B suggest that they formed in two distinct star forming episodes. Comparison of vertex deviations observed in the inner Galaxy and derived for the N-body models indicate that the thin boxy/peanut bar component A and the thick boxy/peanut bulge component B may lie at different angles to the Sun-center line: $45^\circ$ and $20^\circ$ respectively. 
The two components A and B show the same vertical abundance gradient of: $-0.08 \pm 0.05$ dex/kpc and $-0.08 \pm 0.04$ dex/kpc, respectively. An even smaller vertical abundance gradient of $-0.04 \pm 0.02$ dex/kpc is seen for the thick disk component C in the inner Galaxy. These small vertical gradients need not be interpreted as favouring a merger scenario for the formation of the bulge, or formation via dissipational collapse. They are likely to be residual gradients of the disk surviving from the instability event. 
The survival of vertical gradients in the initial disk needs to be confirmed by a more detailed study of the way in which the instability process in the N-body models maps the initial phase space into the post-instability phase space. We argue that the much larger overall vertical abundance gradient in the bulge region, of about $-0.45$ to $-0.6$ dex kpc$^{-1}$, comes primarily from changing contributions of components A--C with height above the plane (see Figure \ref{fig:components}). 

We find that the bulge is not a single population. Its metallicity components reflect the Galactic components which were present before the boxy/peanut bulge formation, and the properties of these components reflect the dynamical processes that altered the spatial distribution of stars primarily in the early disk. Stars with [Fe/H] $> -0.5$ are involved in the boxy/peanut bulge structure, and stars with [Fe/H] $< -0.5$ are present in the inner bulge regions and contribute to the total population of the inner Galaxy, particularly at higher latitudes.  Our thin and thick boxy/peanut-bulge/bar structures are both similar to the boxy/peanut bar/bulge seen in our instability driven N-body model, but with different scale heights and correspondingly different kinematics. Our more metal-poor components C, D and E, with [Fe/H] $< -0.5$ can be identified with the pre-instability thick disk, metal-weak thick disk and halo populations. 

The presence of the two kinematically and chemically distinct components A and B in the boxy/peanut bulge is evidence that internal evolution is the dominant process in the formation of the bulge.  It seems unlikely that major mergers have been an important part of the formation of the bulge of the Milky Way. Such disruption to the inner region would probably smooth the two bulge components into a continuous population. The survival of the two components of the bulge constrains the level of any major disturbance that could have occurred in the inner Galaxy after the disk instability event. At this stage, we do not attempt to place limits on the contribution of any merger-generated ``classical'' bulge component. Further N-body modelling is needed to quantify this issue.

\section*{Acknowledgments}

We thank the Australian Astronomical Observatory, who have made this project possible. This publication makes use of data products from the Two Micron All Sky Survey, which is a joint project of the University of Massachusetts and the Infrared Processing and Analysis Center/California Institute of Technology, funded by the National Aeronautics and Space Administration and the National Science Foundation. This work has been supported by the RSAA and Australian Research Council grant DP0988751. E.A. gratefully acknowledges financial support by the CNES and by the European Commission through the DAGAL Network (PITN-GA-2011-289313). J.B-H is supported by an ARC Federation Fellowship. G.F.L thanks the Australian research council for support through his Future Fellowship (FT100100268) and Discovery Project (DP110100678). R.R.L. gratefully acknowledges support from the Chilean {\sl Centro de Astrof\'\i sica} FONDAP No. 15010003. L.L.K is supported by the Lend\"ulet program of the Hungarian Academy of Sciences and the Hungarian OTKA Grants K76816, MB08C 81013 and K83790.

\section{Appendix}

Regarding our red clump magnitude calibration: the Alves (2000) calibration is for stars in the range  0.0 $>$ [Fe/H] $>$ Ð0.5, which is the metallicity range of the majority of our stars in the bulge. Alves investigated the metallicity dependence on the absolute magnitude of the clump and recommended no [Fe/H] correction be made for the stars in this abundance range. \citet{Laney2012} similarly reported no dependence on metallicity and an absolute red clump magnitude of $M_K$ = -1.61, in agreement with \citet{Alves2000}. 

We have adopted his $M_K$  over a broader metallicity range, because there is disagreement in the literature about the existence of a K-band magnitude-metallicity dependence.   It is consistently reported to be weak at most, and there is disagreement about the factor in the equation for the metallicity dependence.  We believe that it is appropriate to adopt a mean value of -1.61 for calculating clump distances for our stars \citep[see][for a review]{Valentini2010}.  Even if a metallicity dependence exists, its effect on our distances is likely to be small. For example, with Udalski's I-band calibration of a 0.13 mag dex$^{-1}$ dependence on abundance,  the distance of a star with [Fe/H] = --1.0 would change by about 6\%. We note that 95\% of our stars have [Fe/H] $> -1.0$.

\citet{Piet2003} looked at the red clump magnitude dependence on age and [Fe/H] in J and K and found only a low or no dependence of K magnitude on [Fe/H] over the wide metallicity range of the LMC, SMC and the Carina and Fornax dwarf galaxies. They concluded that the mean K band magnitude of the red clump is an appropriate distance indicator for a wide range in [Fe/H] and age. \citet{vanHels2007} similarly found that the absolute magnitude is an appropriate distance indicator in the range $0.4 >$  [Fe/H] $ > Ð0.5$, although they adopted a slightly fainter value of $M_K = -1.57$ for the mean absolute magnitude of the clump.

We note that theory predicts a metallicity dependence of the absolute magnitude of the clump \citep[i.e.][]{Girardi2001} and this has been supported by earlier observations \citep[i.e.][]{Sara1999, Twarog1999}.  \citet{Udalski1998} found a weak metallicity dependence in the I-band in the range  $Ð0.6 < $ [Fe/H]  $< 0.2$ using red clump stars in 15 Magellanic cloud clusters. However, he concluded that the absolute I-band magnitude of the red clump is independent of age for ages 2 - 10 Gyrs.   In summary, we will adopt $M_K = -1.61$ for the clump, independent of age and abundance.


\begin{thebibliography}{99}


\bibitem[\protect\citeauthoryear{Abadi et al.}{2003}]{Abadi2003} 
Abadi M.~G., Navarro J.~F., Steinmetz M., Eke V.~R., 2003, ApJ, 591, 499 


\bibitem[\protect\citeauthoryear{Alonso, Arribas, 
\& Mart{\'{\i}}nez-Roger}{1999}]{Alonso1999} Alonso A., Arribas S., Mart{\'{\i}}nez-Roger C., 1999, A\&AS, 140, 261 

\bibitem[\protect\citeauthoryear{Alves--Brito et 
al.}{2010}]{AlvesBrito2010} Alves--Brito A., Mel{\'e}ndez J., Asplund M., Ram{\'{\i}}rez I., Yong D., 2010, A\&A, 513, A35 

\bibitem[\protect\citeauthoryear{Alves}{2000}]{Alves2000} Alves 
D.~R., 2000, ApJ, 539, 732 

\bibitem[\protect\citeauthoryear{Arce 
\& Goodman}{1999}]{ArceGoodman1999} Arce H.~G., Goodman A.~A., 1999, ApJ, 512, L135 

\bibitem[\protect\citeauthoryear{Athanassoula}{2005}]{Athanassoula2005} 
Athanassoula E., 2005a, MNRAS, 358, 1477 

\bibitem[\protect\citeauthoryear{Athanassoula}{2005b}]{Athanassoula2005b} 
Athanassoula E., 2005b, AIPC, 804, 333 

\bibitem[\protect\citeauthoryear{Athanassoula}{2008}]{Athanassoula2008} 
Athanassoula E., 2008, arXiv, arXiv:0802.0151 

\bibitem[\protect\citeauthoryear{Babusiaux et 
al.}{2010}]{Babusiaux2010} Babusiaux C., et al., 2010, A\&A, 519, A77 

\bibitem[\protect\citeauthoryear{Bekki 
\& Tsujimoto}{2011}]{Bekki2011} Bekki K., Tsujimoto T., 2011, MNRAS, 416, L60 

\bibitem[\protect\citeauthoryear{Bensby et al.}{2010}]{Bensby2010} 
Bensby T., et al., 2010, IAUS, 265, 346 

\bibitem[\protect\citeauthoryear{Bensby et 
al.}{2011}]{Bensby2011} Bensby T., et al., 2011, A\&A, 533, A134 

\bibitem[\protect\citeauthoryear{Bessell, Castelli, 
\& Plez}{1998}]{Bessell1998} Bessell M.~S., Castelli F., Plez B., 1998, A\&A, 333, 231 
 
 \bibitem[\protect\citeauthoryear{Binney, Merrifield, 
\& Wegner}{2000}]{BinneyMerri} Binney J., Merrifield M., Wegner G.~A., 2000, AmJPh, 68, 95 

\bibitem[\protect\citeauthoryear{Blitz 
\& Spergel}{1991}]{Blitz1991} Blitz L., Spergel D.~N., 1991, ApJ, 379, 631 

 \bibitem[\protect\citeauthoryear{Brook et al.}{2007}]{Brook2007} 
Brook C.~B., Kawata D., Scannapieco E., Martel H., Gibson B.~K., 2007, ApJ, 
661, 10 

\bibitem[\protect\citeauthoryear{Bureau 
\& Freeman}{1999}]{Bureau1999} Bureau M., Freeman K.~C., 1999, AJ, 118, 126 

\bibitem[\protect\citeauthoryear{Cabrera-Lavers et 
al.}{2008}]{CL2008} Cabrera-Lavers A., Gonz{\'a}lez-Fern{\'a}ndez C., Garz{\'o}n F., Hammersley P.~L., L{\'o}pez-Corredoira M., 2008, A\&A, 491, 781 

\bibitem[\protect\citeauthoryear{Cambr{\'e}sy, Jarrett, 
\& Beichman}{2005}]{Cambresy2005} Cambr{\'e}sy L., Jarrett T.~H., Beichman C.~A., 2005, A\&A, 435, 131 


\bibitem[\protect\citeauthoryear{Carollo et 
al.}{2010}]{Carollo2010} Carollo D., et al., 2010, ApJ, 712, 692 

\bibitem[\protect\citeauthoryear{Cassisi et 
al.}{2006}]{Cassisi2006} Cassisi S., Pietrinferni A., Salaris M., 
Castelli F., Cordier D., Castellani M., 2006, MmSAI, 77, 71 

\bibitem[\protect\citeauthoryear{Cassisi}{2010}]{Cassisi2010} 
Cassisi S., 2010, IAUS, 262, 13 

\bibitem[\protect\citeauthoryear{Chiba 
\& Beers}{2000}]{ChibaBeers2000} Chiba M., Beers T.~C., 2000, AJ, 119, 2843 

\bibitem[\protect\citeauthoryear{Combes 
\& Sanders}{1981}]{Combes1981} Combes F., Sanders R.~H., 1981, A\&A, 96, 164 

\bibitem[\protect\citeauthoryear{Diemand, Madau, 
\& Moore}{2005}]{Diemand2005} Diemand J., Madau P., Moore B., 2005, MNRAS, 364, 367 

\bibitem[\protect\citeauthoryear{Dwek et al.}{1995}]{Dwek1995} 
Dwek E., et al., 1995, ApJ, 445, 716 

\bibitem[\protect\citeauthoryear{Falc{\'o}n--Barroso et 
al.}{2006}]{FalconBarroso2006} Falc{\'o}n--Barroso J., et al., 2006, 
MNRAS, 369, 529 

\bibitem[\protect\citeauthoryear{Freeman et 
al.}{2012}]{Freeman2012} Freeman K., et al., 2012, submitted MNRAS

\bibitem[\protect\citeauthoryear{Fulbright, McWilliam, 
\& Rich}{2007}]{Fulbright2007} Fulbright J.~P., McWilliam A., Rich R.~M., 2007, ApJ, 661, 1152 

\bibitem[\protect\citeauthoryear{Freeman}{2008}]{Freeman2008} 
Freeman K.~C., 2008, ASPC, 396, 3 

\bibitem[\protect\citeauthoryear{Gerhard}{2002}]{Gerhard2002} 
Gerhard O., 2002, ASPC, 273, 73 

\bibitem[\protect\citeauthoryear{Gilmore, Wyse \& Jones}{1995}]{Gilmore1995} Gilmore G., Wyse R.~F.~G., Jones J.~B., 1995, AJ, 109, 1095 

\bibitem[\protect\citeauthoryear{Girardi 
\& Salaris}{2001}]{Girardi2001} Girardi L., Salaris M., 2001, MNRAS, 323, 109 

\bibitem[\protect\citeauthoryear{Gonzalez et 
al.}{2011}]{Gonzalez2011} Gonzalez O.~A., Rejkuba M., Zoccali M., Valenti E., Minniti D., 2011, A\&A, 534, A3

\bibitem[\protect\citeauthoryear{Hammersley et 
al.}{1994}]{Hammer1994} Hammersley P.~L., Garzon F., Mahoney T., 
Calbet X., 1994, MNRAS, 269, 753 

\bibitem[\protect\citeauthoryear{Haywood}{2008}]{Haywood2008} 
Haywood M., 2008, MNRAS, 388, 1175 

\bibitem[\protect\citeauthoryear{Hill et 
al.}{2011}]{Hill2011} Hill V., et al., 2011, A\&A, 534, A80 

\bibitem[\protect\citeauthoryear{Howard et al.}{2009}]{Howard2009} 
Howard C.~D., et al., 2009, ApJ, 702, L153 

\bibitem[\protect\citeauthoryear{Kirby, Guhathakurta, 
\& Sneden}{2008}]{Kirby2008} Kirby E.~N., Guhathakurta P., Sneden C., 2008, ApJ, 682, 1217 

\bibitem[\protect\citeauthoryear{Kobayashi 
\& Nakasato}{2011}]{Kobayashi2011} Kobayashi C., Nakasato N., 2011, ApJ, 729, 16 

\bibitem[\protect\citeauthoryear{Laney, Joner, 
\& Pietrzy{\'n}ski}{2012}]{Laney2012} Laney C.~D., Joner M.~D., Pietrzy{\'n}ski G., 2012, MNRAS, 419, 1637 


\bibitem[\protect\citeauthoryear{Laney, Joner, 
\& Pietrzy{\'n}ski}{2012}]{Laney2012} Laney C.~D., Joner M.~D., Pietrzy{\'n}ski G., 2012, MNRAS, 419, 1637 


\bibitem[\protect\citeauthoryear{Lecureur et 
al.}{2007}]{Lecureur2007} Lecureur A., Hill V., Zoccali M., Barbuy B., G{\'o}mez A., Minniti D., Ortolani S., Renzini A., 2007, A\&A, 465, 799 

\bibitem[\protect\citeauthoryear{L{\'o}pez--Corredoira, Cabrera--Lavers, 
\& Gerhard}{2005}]{Lopez2005} L{\'o}pez--Corredoira M., Cabrera--Lavers A., Gerhard O.~E., 2005, A\&A, 439, 107
\bibitem[\protect\citeauthoryear{Luck 
\& Lambert}{2011}]{Luck2011} Luck R.~E., Lambert D.~L., 2011, AJ, 142, 136 

\bibitem[\protect\citeauthoryear{Martinez--Valpuesta 
\& Gerhard}{2011}]{Inma2011} Martinez--Valpuesta I., Gerhard O., 2011, ApJ, 734, L20 

\bibitem[\protect\citeauthoryear{Martinez-Valpuesta, Shlosman, 
\& Heller}{2006}]{Inma2006} Martinez-Valpuesta I., Shlosman I., Heller C., 2006, ApJ, 637, 214 

\bibitem[\protect\citeauthoryear{McWilliam 
\& Rich}{1994}]{McWilliam1994} McWilliam A., Rich R.~M., 1994, ApJS, 91, 749 

\bibitem[\protect\citeauthoryear{McWilliam 
\& Zoccali}{2010}]{McWilliam2010} McWilliam A., Zoccali M., 2010, ApJ, 724, 1491 

 \bibitem[\protect\citeauthoryear{Mel{\'e}ndez et 
al.}{2008}]{Melendez2008} Mel{\'e}ndez J., et al., 2008, A\&A, 484, L21 

\bibitem[\protect\citeauthoryear{Meng \& Rubin}{1993}]{Meng1993} Meng Xiao-Li., Rubin Donald B., 1993,  Biometrika, 80, 267

\bibitem[\protect\citeauthoryear{Ness et al.}{2012a}]{Ness2012a} 
Ness M., et al., 2012a, arXiv, arXiv:1207.0888 

 \bibitem[\protect\citeauthoryear{Ness \& Freeman}{2012}]{proceedings} Ness M., Freeman K., 2012, EPJWC, 19, 6003 

  
\bibitem[\protect\citeauthoryear{Ness et al.}{2012c}]{Ness2012c} Ness M., et al., 2012c. To be submitted to MNRAS


\bibitem[\protect\citeauthoryear{Okuda et al.}{1977}]{Okuda1977} 
Okuda H., Maihara T., Oda N., Sugiyama T., 1977, Natur, 265, 515

\bibitem[\protect\citeauthoryear{Pietrzy{\'n}ski, Gieren, 
\& Udalski}{2003}]{Piet2003} Pietrzy{\'n}ski G., Gieren W., Udalski A., 2003, AJ, 125, 2494 


\bibitem[\protect\citeauthoryear{Raha et al.}{1991}]{Raha1991} 
Raha N., Sellwood J.~A., James R.~A., Kahn F.~D., 1991, Natur, 352, 411 


\bibitem[\protect\citeauthoryear{Rattenbury et 
al.}{2007}]{Rattenbury2007} Rattenbury N.~J., Mao S., Sumi T., Smith 
M.~C., 2007, MNRAS, 378, 1064 


\bibitem[\protect\citeauthoryear{Robin et 
al.}{2012}]{Robin2012} Robin A.~C., Marshall D.~J., Schultheis M., Reyl{\'e} C., 2012, A\&A, 538, A106 

\bibitem[\protect\citeauthoryear{Romero-G{\'o}mez et 
al.}{2011}]{Romero2011} Romero-G{\'o}mez M., Athanassoula E., 
Antoja T., Figueras F., 2011, MNRAS, 418, 1176 

\bibitem[\protect\citeauthoryear{Saguner et 
al.}{2011}]{Saguner2011} Saguner T., Munari U., Fiorucci M., Vallenari A., 2011, A\&A, 527, A40 
 
 \bibitem[\protect\citeauthoryear{Saha, Martinez-Valpuesta, 
\& Gerhard}{2012}]{Saha2012} Saha K., Martinez-Valpuesta I., Gerhard O., 2012, MNRAS, 421, 333 


\bibitem[\protect\citeauthoryear{Sarajedini}{1999}]{Sara1999} 
Sarajedini A., 1999, AJ, 118, 2321 

\bibitem[\protect\citeauthoryear{{Schlegel}, {Finkbeiner}, 
\& {Davis}}{{Schelgel} et~al.}{1998}]{Schlegel} Schlegel D.~J., Finkbeiner D.~P., Davis M., 1998, ApJ, 500, 525 

\bibitem[\protect\citeauthoryear{Skrutskie et 
al.}{2006}]{2MASS} Skrutskie M.~F., et al., 2006, AJ, 131, 
1163 

\bibitem[\protect\citeauthoryear{Smith, Price, 
\& Baker}{2004}]{Smith2004} Smith B.~J., Price S.~D., Baker R.~I., 2004, ApJS, 154, 673 

\bibitem[\protect\citeauthoryear{Soto, Rich, 
\& Kuijken}{2007}]{Soto2007} Soto M., Rich R.~M., Kuijken K., 2007, ApJ, 665, L31 

\bibitem[\protect\citeauthoryear{Tumlinson}{2010}]{Tumlinson2010} 
Tumlinson J., 2010, ApJ, 708, 1398 


\bibitem[\protect\citeauthoryear{Twarog, Anthony-Twarog, 
\& Bricker}{1999}]{Twarog1999} Twarog B.~A., Anthony-Twarog B.~J., Bricker A.~R., 1999, AJ, 117, 1816 

\bibitem[\protect\citeauthoryear{Udalski}{1998}]{Udalski1998} 
Udalski A., 1998, AcA, 48, 383 


\bibitem[\protect\citeauthoryear{Uttenthaler et 
al.}{2012}]{Uttenthaler2012} Uttenthaler S., Schultheis M., Nataf D.~M., Robin A.~C., Lebzelter T., Chen B., 2012, A\&A, 546, A57 

\bibitem[\protect\citeauthoryear{van Helshoecht 
\& Groenewegen}{2007}]{vanHels2007} van Helshoecht V., Groenewegen M.~A.~T., 2007, A\&A, 463, 559 


\bibitem[\protect\citeauthoryear{Valentini 
\& Munari}{2010}]{Valentini2010} Valentini M., Munari U., 2010, A\&A, 522, A79 

\bibitem[\protect\citeauthoryear{Weinberg}{1992}]{Weinberg1992} 
Weinberg M.~D., 1992, ApJ, 384, 81 

\bibitem[\protect\citeauthoryear{Yong, Carney, 
\& Friel}{2012}]{Yong2012} Yong D., Carney B.~W., Friel E.~D., 2012, arXiv, arXiv:1206.6931 

\bibitem[\protect\citeauthoryear{Zhao, Qiu, 
\& Mao}{2001}]{Zhao2001} Zhao G., Qiu H.~M., Mao S., 2001, ApJ, 551, L85 

\bibitem[\protect\citeauthoryear{Zoccali et 
al.}{2003}]{Zoccali2003} Zoccali M., et al., 2003, A\&A, 399, 931 

\bibitem[\protect\citeauthoryear{Zoccali et 
al.}{2008}]{Zoccali2008} Zoccali M., Hill V., Lecureur A., Barbuy B., Renzini A., Minniti D., G{\'o}mez A., Ortolani S., 2008, A\&A, 486, 177 

\bibitem[\protect\citeauthoryear{Zoccali et 
al.}{2008}]{Zoccali2008alpha} Zoccali M., Lecureur A., Hill V., Barbuy 
B., Renzini A., Minniti D., G{\'o}mez A., Ortolani S., 2008, MmSAI, 79, 503 

\end{thebibliography}
\end{document}